\documentclass[twocolumn]{aastex61}
\usepackage{booktabs}
\usepackage{hyperref}
\hypersetup{
    colorlinks=true,
    linkcolor=blue,
    filecolor=magenta,      
    urlcolor=cyan,
}
\usepackage{rotating}
\usepackage{multirow}
\usepackage{float}
\usepackage{longtable} % para tablas largas
\usepackage{bbding}
\usepackage{amsmath}
\usepackage{blindtext}
\usepackage{lipsum}
\usepackage{graphicx}

\usepackage{lineno}
\newenvironment{Contfigure*}{%
\renewcommand{\thefigure}{\arabic{figure} (Cont.)}%
\addtocounter{figure}{-1}%
\begin{figure*}}{%
\renewcommand{\thefigure}{\arabic{figure}}%
\end{figure*}}

\newenvironment{Contfigure}{%
\renewcommand{\thefigure}{\arabic{figure} (Cont.)}%
\addtocounter{figure}{-1}%
\begin{figure}}{%
\renewcommand{\thefigure}{\arabic{figure}}%
\end{figure}}

\def\arcsec{\ifmmode^{\prime\prime}\;\else$^{\prime\prime}\;$\fi}
\def\arcmin{\ifmmode^{\prime}}

\received{??} %July 1, 2018
\revised{??}%June 27, 2018
\accepted{??}%\today
\submitjournal{ApJ}

\shorttitle{Testing physical scenarios for the reflection features of type-1 AGN}
\shortauthors{C. Victoria-Ceballos}

\begin{document}
\title{Testing physical scenarios for the reflection features of type-1 AGN using XMM-Newton and NuSTAR simultaneous observations}

\correspondingauthor{C. Victoria-Ceballos, PhD student}
\email{c.victoria@irya.unam.mx}

\author{C\'esar Ivan Victoria-Ceballos}
\affil{Instituto de Radioastronom\'ia y Astrof\'isica (IRyA-UNAM), 3-72 (Xangari), 8701, Morelia, Mexico}
\author{Omaira Gonz\'alez-Mart\'in}
\affil{Instituto de Radioastronom\'ia y Astrof\'isica (IRyA-UNAM), 3-72 (Xangari), 8701, Morelia, Mexico}
\author{Josefa Masegosa}
\affil{Instituto de Astrofísica de Andalucía (IAA-CSIC), Glorieta de la Astronomía s/n E-18008, Granada, Spain}
\author{Anna Lia Longinotti}
\affil{Instituto de Astronomía, (IA-UNAM), Circuito Exterior, Ciudad Universitaria, Ciudad de México 04510, Mexico}
\author{Donaji Esparza-Arredondo}
\affil{Instituto de Astrofísica de Canarias, Calle Vía Láctea, s/n, E-38205 La Laguna, Tenerife, Spain}
\affil{Departamento de Astrofísica, Universidad de La Laguna, E-38206 La Laguna, Tenerife, Spain}
\author{Natalia Osorio-Clavijo}
\affil{Instituto de Radioastronom\'ia y Astrof\'isica (IRyA-UNAM), 3-72 (Xangari), 8701, Morelia, Mexico}

\begin{abstract}

Above $\sim$3 keV, the X-ray spectrum of the active galactic nuclei (AGN) is characterized by the intrinsic continuum and compton reflection features. For type-1 AGN, several regions could contribute to the reflection. In order to investigate the nature of the reflecting medium, we perform a systematic analysis of the reflector using \emph{XMM}-Newton and \emph{NuSTAR} observations of a sample of 22 type-1 AGN. We create a baseline model which includes Galactic absorption and an intrinsically absorbed power-law plus a reflection model. We test a set of nine reflection models in a sub-sample of five objects. Based on these results, we select three models to be tested on the entire sample, accounting for distinct physical scenarios: neutral/distant reflection, ionized/relativistic reflection, and neutral/distant+ionized/relativistic reflection, namely hybrid model. We find that 18 sources require the reflection component to fit their spectra. Among them, 67$\%$ prefer the hybrid model. Neutral and ionized models are equally preferred by three sources. We conclude that both the neutral/distant reflector most probably associated  with the inner edges of the torus and the ionized/relativistic reflector associated with the accretion disk are required to describe the reflection in type-1 AGN. %Although we only found two sources preferring the ionized/relativistic reflector, it is worth to notice they show the highest photon indices. Thus, we find tentative evidence on intrinsic difference on the reflector for type-1 AGN according to its Eddington ratio.

\end{abstract}

\section{Introduction}\label{sec:intro}

%{\bf According to the unified model of active galactic nuclei \citep[AGN][]{Antonucci93, Urry-Padovani95}, these objects present different structures around the super massive black hole (SMBH), with different physical and chemical properties. The accretion disk is the closest structure to the SMBH, formed by ionized material that spirals down towards the SMBH. On the SMBH there is a plasma of hot electrons, which re-emits the electrons coming from the accretion disk as X-ray photons, which is why this structure is known as the X-ray corona. \citep{Haardt91}. Part of the material that falls towards the SMBH is collimated from its poles at relativistic velocities, forming the so-called relativistic jets \citep{Blandford19}. The broad line region (BLR), with a size between $10^{-4}$ and 0.1 pc is formed by clouds of ionized gas revoling around the SMBH at velocities close to the velocity of light \citep{}. The narrow line region (NLR) are non-ionized clouds that rotate at slower speeds, at a distance of approximately the SMBH. Finally, the structure farthest from the SMBH is made of cold dust distributed in a toroidal shape around the central engine, this is located a few parsecs from the SMBH, commonly called dusty torus. \citep[see][for a review]{Hagai15}.}

The typical X-ray spectrum above $\sim$ 3 keV of active galactic nuclei (AGN) is composed by the intrinsic continuum and reflection features. The primary X-ray emission originates in the corona of relativistic electrons close to the super massive black hole (SMBH), where UV and optical photons from accretion disk are Compton upscattered \citep[][]{Haardt93}. \cite{Pounds90} find that the intrinsic continuum could be modelled with a power law as a function of the photon energy of photon index $\Gamma$ $\sim$ 1.9. \cite{Zdziarski95} find that this power-law continuum has an exponential cut-off of order of several hundred keV. %The reason for this cut-off energy is because the electrons in the corona are in thermal equilibrium and have a Maxwellian distribution with temperature $\rm{kT_e}$ $\sim$ 100 keV.   

X-ray photons in the corona are emitted in all directions, and they can reach and be reflected by the surrounding medium, such the accretion disk, the broad line region (BLR), and/or the torus \citep[][]{Matsuoka90}. Some distinctive features can be observed in the X-ray spectra when the X-ray photons are reflected by one or more of these components. Such features are a hump-like continuum peaking $\sim$ 30 keV (product of the electron down scattering of high-energy photons and photoelectric absorption of low-energy photons) and several fluorescent emission lines (most notably the $\rm{FeK\alpha}$ emission line at 6.4 keV) \citep[][]{George91}.

The reflecting medium could arise from several AGN components around the SMBH. The resulting reflection features depend on the dynamical, geometrical, and chemical properties. For instance, the accretion disk and the BLR are the closest regions to the SMBH and are composed of ionized gas, while the obscuring torus, much more distant from the SMBH, is composed of neutral gas. When the $\rm{FeK\alpha}$ line originates from neutral material, it is observed with a narrow profile. However, it can be broad and blurred by relativistic effects, when it is emitted close to the SMBH \citep{Laor91}.% By other hand, the reflecting medium that produces the compton hump remains as an object of debate.}

%Since the different reflecting structures around the SMBH have different characteristics (eg. the accretion disk and the BLR are the closest regions to the SMBH and are composed of ionized gas, while the obscuring torus, much more distant from the SMBH, is composed of neutral material), the reflection features shows different characteristics.   
%Otherwise is narrow when it is originated from more distant material \citep{Liu16}.

Different origins have been proposed to explain the hard X-ray spectrum of the AGN, among which is relativistic reflection, and distant reflectors \citep[][]{Nardini11, Patrick11, Mehdipour15}. In the same way, there are different models, which are used throughout the fitting of the spectral energy distribution (SED) of the AGN. For instance, {\sc pexrav} \citep{Magdziarz95} assumes optically thick cold material distributed in a slab, {\sc relxill} \citep{Dauser10, Garcia14} models irradiation of accretion by a broken power law emissivity, {\sc reflionx} \citep{Ross99, Ross05} assume an optically-thick atmosphere, and it adds fluorecense lines.

Due to the reflection features in the X-ray spectrum of the AGN are present in a wide range, the best way to study the nature of the reflection component is by using high-energy and -quality observations. The Nuclear Spectroscopic Telescope Array \citep[\emph{NuSTAR}][]{Harrison13} has an unprecedented sensitivity to hard X-ray photons, while, \emph{XMM}-Newton \citep[][]{Jansen01} provides an excellent resolution below 10 keV and in particular around $\rm{FeK\alpha}$ emission line. Different authors have exploit the advantages offered by these two telescopes conducting studies with simultaneous observations to study the hard X-ray spectrum of various AGN \citep[e.g.][]{Porquet18, Liu20, Diaz20, Traina21, Marchesi22}.

While the nature of the Compton reflector is well established for type-2 AGN, which are mostly dominated by reflection in the distant and cold torus \citep{Brightman11, Ricci15, Marchesi18}, type-1 AGN are the ideal laboratories to explore the contribution of the disk and BLR to this reflection \citep{Falocco14, Panagiotou19}.

The aim of this work is to investigate the reflection medium of the primary X-ray radiation in a sample of type-1 AGN. For this, we used simultaneous observations from \emph{XMM}-Newton and \emph{NuSTAR} satellites, covering a spectral range from 3 keV up to $\sim$ 70 keV, in which it is expected to detect the reflection features. The paper is organized as follows. Section \ref{sec:observations and data reduction} describes the sample selection and the data reduction. Section \ref{sec:x-ray_models} describes the X-ray models tested and the fitting procedure. We present our main results and we discuss them in Section \ref{sec:Results} and \ref{sec:Discussion} respectively.

\begin{table*}[ht]
%\tiny
\scriptsize
\renewcommand{\tabcolsep}{0.1cm}
\begin{center}
\begin{tabular}{lccccccccccccccc}\hline
& RA & Dec & z & Type & \multicolumn{5}{c}{\emph{XMM}-Newton} & \multicolumn{5}{c}{\emph{NuSTAR}} \\\hline
& J2000.0 & J2000.0 & & & Obs ID & Obs. date & Exp. & Counts & Bins & Obs ID & Obs. date & Exp. & Counts & Bins\\
&&&&&&& (ks) &&&&& (ks) && \\
(1) & (2) & (3) & (4) & (5) & (6) & (7) & (8) & (9) & (10) & (11) & (12) & (13) & (14) & (15) \\ \hline
Mrk\,335 & 00 06 25.7 & +20 13 31 & 0.035 & Sy1.2 & 0780500301 & 18-07-11 & 114.5 & 6892.7 & 55 & 80201001002 & 18-07-10 & 82.26 & 3359.5 & 138 \\
Fairall\,9 & 01 23 45.3 & -58 46 46 & 0.047 & Sy1.2 & 0741330101 & 14-05-09 & 141.4 & 116551 & 155 & 60001130002 & 14-05-09 & 49.21 & 24661.9 & 352 \\
Mrk\,1040 & 02 28 17.1 & +31 20 07 & 0.011 & Sy1.5 & 0760530301 & 15-08-15 & 84.6 & 106701 & 466 & 60101002004 & 15-08-15 & 64.24 & 43260.9 & 499 \\
Mrk\,1044 & 02 30 07.8 & -08 58 13 & 0.016 & Sy1 & 0824080301 & 18-08-03 & 140.7 & 103577 & 155 & 60401005002 & 18-08-03 & 267.08 & 92680.2 & 560 \\
NGC\,1365 & 03 33 32.5 & -36 09 34 & 0.004 & Sy1.8 & 0692840401 & 13-01-23 & 133.62 & 141235 & 466 & 60002046007 & 13-01-23 & 73.65 & 54529 & 542 \\
Ark\,120 & 05 16 07.0 & -00 10 16 & 0.032 & Sy1 & 0721600401 & 14-03-22 & 133.3 & 211387 & 466 & 60001044004 & 14-03-22 & 65.45 & 62593.4 & 511 \\
Mrk\,382 & 07 55 30.2 & +39 12 37 & 0.027 & Sy1 & 0843020801 & 19-10-30 & 34.5 & 3586.6 & 69 & 60501008002 & 19-10-29 & 52.36 & 1934.1 & 80 \\
NGC\,3227 & 10 23 37.2 & 19 52 38 & 0.004 & Sy1.5 & 0782520201 & 16-11-09 & 92 & 71597 & 466 & 60202002002 & 16-11-09 & 49.8 & 49452.5 & 515 \\
NGC\,3783 & 11 39 08.1 & -37 43 39 & 0.011 & Sy1 & 0780860901 & 16-12-11 & 115 & 93018 & 174 & 80202006002 & 16-12-11 & 25.66 & 24209 & 406 \\
NGC\,4051 & 12 03 04.7 & +44 33 30 & 0.003 & Sy1.2 & 0830430201 & 18-11-07 & 83.2 & 48490 & 699 & 60401009002 & 18-11-04 & 311.14 & 130991 & 758 \\
NGC\,4151 & 12 10 43.7 & +39 24 06 & 0.002 & Sy1.5 & 0679780301 & 12-11-14 & 12.21 & 37955 & 699 & 60001111005 & 12-11-14 & 61.53 & 393011 & 1024 \\
Mrk\,766 & 12 18 18.6 & +29 48 01 & 0.013 & Sy1.5 & 0763790401 & 15-07-05 & 29.3 & 20186 & 199 & 60101022002 & 15-07-05 & 23.57 & 10318.9 & 238 \\
NGC\,4593 & 12 39 45.0 & -05 20 06 & 0.008 & Sy1 & 0740920201 & 14-12-29 & 26 & 21054 & 279 & 60001149002 & 14-12-29 & 23.32 & 14279.4 & 296 \\
IRAS\,13197-1627 & 13 22 30.6  & -16 43 14 & 0.020 & Sy1.8 & 0763220201 & 16-01-17 & 142.5 & 12003 & 233 & 60101020002 & 16-01-17 & 78.5 & 9925.79 & 284 \\ 
IRAS\,13224-3809 & 13 25 13.2 & -38 25 20 & 0.066 & NLSy1 & 0792180301 & 16-08-01 & 140.5 & 2704 & 22 & 60202001012 & 16-08-01 & 171.65 & 2328.62 & 50 \\
MCG\,-06-30-15 & 13 36 02.0 & -34 17 10 & 0.008 & Sy1.2 & 0693781401 & 13-02-02 & 48.92 & 50651 & 155 & 60001047005 & 13-02-02 & 29.65 & 23126.7 & 360 \\
NGC\,5548 & 14 17 52.3 & +25 05 17 & 0.025 & Sy1.5 & 0720111001 & 13-07-23 & 57 & 71133 & 466 & 60002044005 & 13-07-23 & 49.52 & 46097.7 & 510 \\
Mrk\,841 & 15 03 54.3  & +10 25 17  & 0.036 & Sy1.5 & 0763790501 & 15-07-14 & 29.5 & 14721 & 155 & 60101023002 & 15-07-14 & 23.42 & 10118.8 & 249 \\
IGRJ\,19378-0617 & 19 37 30.9 & -06 14 15 & 0.010 & Sy1 & 0761870201 & 15-10-01 & 141.4 & 131342 & 349 & 60101003002 & 15-10-01 & 65.52 & 32697.6 & 364 \\
Mrk\,915 & 22 36 43.4 & -12 34 11  & 0.024 & Sy1 & 0744490401 & 14-12-02 & 135 & 38862 & 279 & 60002060002 & 14-12-02 & 52.98 & 12994.4 & 298 \\
MR\,2251-178 & 22 54 09.9 & -17 33 03 & 0.064 & Sy1 & 0763920601 & 15-06-17 & 38.9 & 52739 & 349 & 60102025004 & 15-06-17 & 23.19 & 32048.2 & 399 \\
NGC\,7469 & 23 03 12.4 & +08 50 48 & 0.014 & Sy1.2 & 0760350501 & 15-12-23 & 90.9 & 92397 & 266 & 60101001008 & 15-12-22 & 23.48 & 17163.1 & 315 \\
\hline
\end{tabular}
\end{center}
\caption{Observational parameters for the sample. (1) Name of the source; (2) right ascension (in hours, minutes, and seconds); (3) declination (in degrees, minutes of arc, and seconds of arc); (4) Redshift; (5) AGN classification; (6)-(10) observation ID, date of the observation, exposure time of the observation in ksec, number of counts in the 3–10 keV band background-substracted, and number of bins in the 3–10 keV band in \emph{XMM}-Newton; (11)-(15) observation ID, date of the observation, exposure time of the observation in ksec, number of counts in the 3–60 keV band background-substracted, and number of bins in the 3–60 keV band in \emph{NuSTAR}.}
\label{tab:sample_data}
\end{table*}

\section{Sample and data reduction} \label{sec:observations and data reduction}

\subsection{Sample selection} \label{sec:sample_selection}

We select our sample according two criteria. First, in order to account the variability of the sources, we select a sample of AGNs with simultaneous observations of \emph{XMM}-Newton and \emph{NuSTAR}. We consider simultaneous those observations in which, given the exposure time, they overlap even though they started on different days. Our second criterion is for those sources with two or more simultaneous observations, we select those with the longest exposure time, which allows us to have the best quality data. According to these two criteria, we obtain a sample of 63 AGNs: 26 Sy1-Sy1.8, 20 Sy2, 5 NLSy1, 9 QSO, 2 Blazar, and 1 unclassified AGN. Among all these sources, the type-1 AGN are the main target of our investigation, therefore we discard the Sy2 sources, the Blazar objects, the QSOs (because their high redshift), the radio sources (because their flat spectra), and the unclassified AGN. Also, we discard 4 NLSy1 and 5 Seyfert galaxies due to their noisy \emph{NuSTAR} spectrum. Our final sample contains 22 sources: 1 NLSy1, 19 Sy1-Sy1.5, and 2 Sy1.8. Table\,\ref{tab:sample_data} shows the final list of objects with the observation details of the two observatories.

%{\bf Finally we select the type-1 AGN, which are the main target of our investigation. Our final sample contains 22 sources: 1 NLSy1, 19 Sy1-Sy1.5, and 2 Sy1.8. Table\,\ref{tab:sample_data} shows the final list of objects with the observation details of the two observatories. (Note that we discard 4 NLSy1 and 5 Seyfert galaxies due to their noisy \emph{NuSTAR} spectrum).}

\subsection{Data reduction} \label{sec:data_reduction}

\emph{XMM}-Newton and \emph{NuSTAR} data reduction was performed with the \emph{XMM}-Newton Science Analysis System (SAS)\footnote{https://www.cosmos.esa.int/web/xmm-newton/sas}, and the \emph{NuSTAR} Data Analysis Software (NuSTARDAS)\footnote{https://heasarc.gsfc.nasa.gov/docs/nustar/analysis/} package, respectively. For \emph{XMM}-Newton we use data from the EPIC pn camera \citep{Struder01} because of the higher count rate and lower distortion due to pile-up. We use circular regions with 40 arcsec radii to extract the spectra. The background events were selected from a source-free circular region with 40 arcsec radii on the same CCD as the source. For \emph{NuSTAR} data we use 65 arcsec radius circular extraction regions for both source and background spectra. The background region was selected from a region on the same chip, uncontaminated with source photons.

\section{X-ray spectral models and fitting} \label{sec:x-ray_models}

\subsection{Baseline model}

In general terms, the X-ray spectra of AGN above $\sim$ 3 keV shows two main components: the intrinsic continuum modelled by a cutoff power law, and the reflection component that can be associated to an ionized or neutral medium. We create a baseline model considering these features: 

\begin{equation}
M = Cte * Abs_{Gal} * (absorber * intrinsic + reflection)
\end{equation}

\noindent Where $\rm{Cte}$ is a multiplicative constant to account of \emph{NuSTAR} and \emph{XMM}-Newton cross-calibration issues. $\rm{Abs_{Gal}}$ accounts for the Galactic absorption \citep[using the NH tool within FTOOLS, which is fixed to the HI maps of][]{Kalberla05}. $\rm{absorber*intrinsic}$ represents the intrinsic continuum absorbed by the material along the LOS to the observer, which we model with a cutoff power-law affected by a neutral absorber. The last component in the baseline model, accounts for the reflection. Several model are used (see below) depending on the material producing this reflection (geometry and composition). Free and fixed parameters for the reflection component depend on the model used.\\
Note that if we consider the emission from ionised/relativistic reflection by the accretion disk, it should be affected by the same column density that affects the intrinsic emission, however, this effect is observed at low energies (below 3 keV). The absorber affecting the ionized reflection will be consider when an analysis of the data below 3 keV is performed.

\begin{figure}[!t]
\begin{center}
\includegraphics[width=1.0\columnwidth]{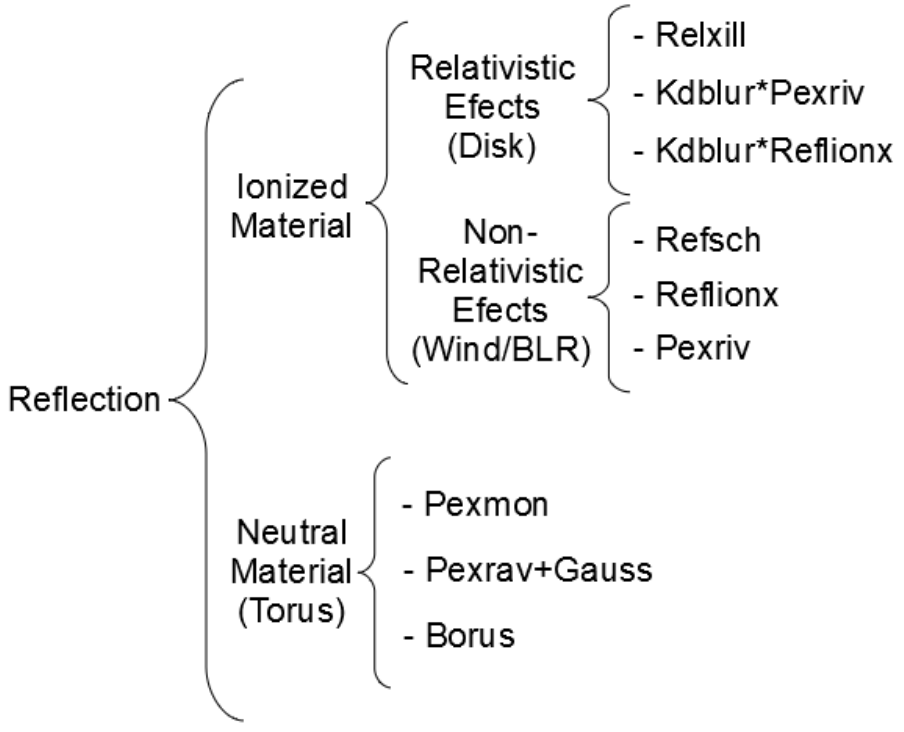}
\caption{Scheme of reflection models used, accounting the scenarios described in Section\,\ref{sec:reflection_models}.}
\label{fig:models}
\end{center}
\end{figure}

\begin{figure*}[!t]
\begin{center}
\includegraphics[width=0.75\columnwidth, clip, trim=10 1 0 60]{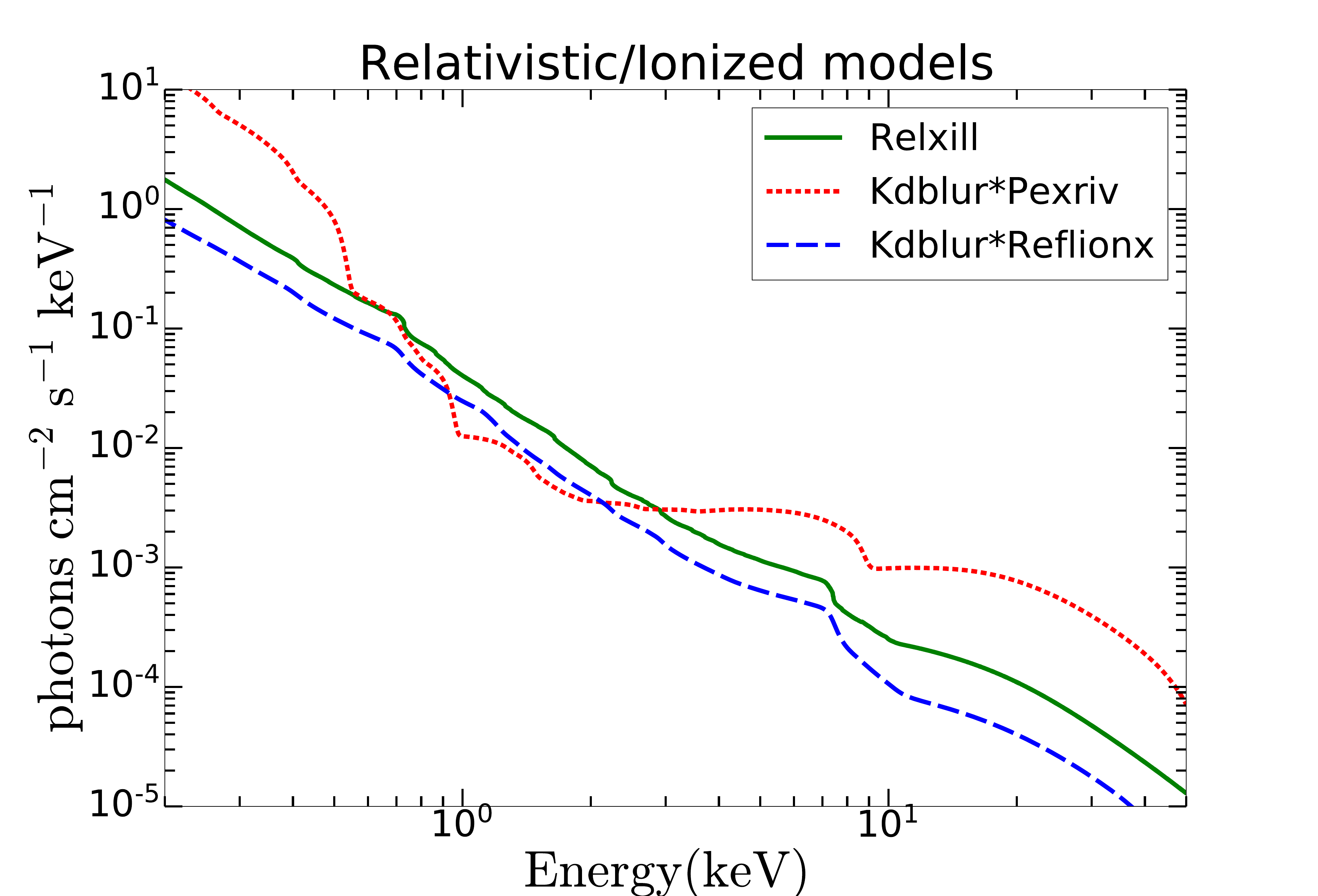}
\includegraphics[width=0.66\columnwidth, clip, trim=223 1 0 60]{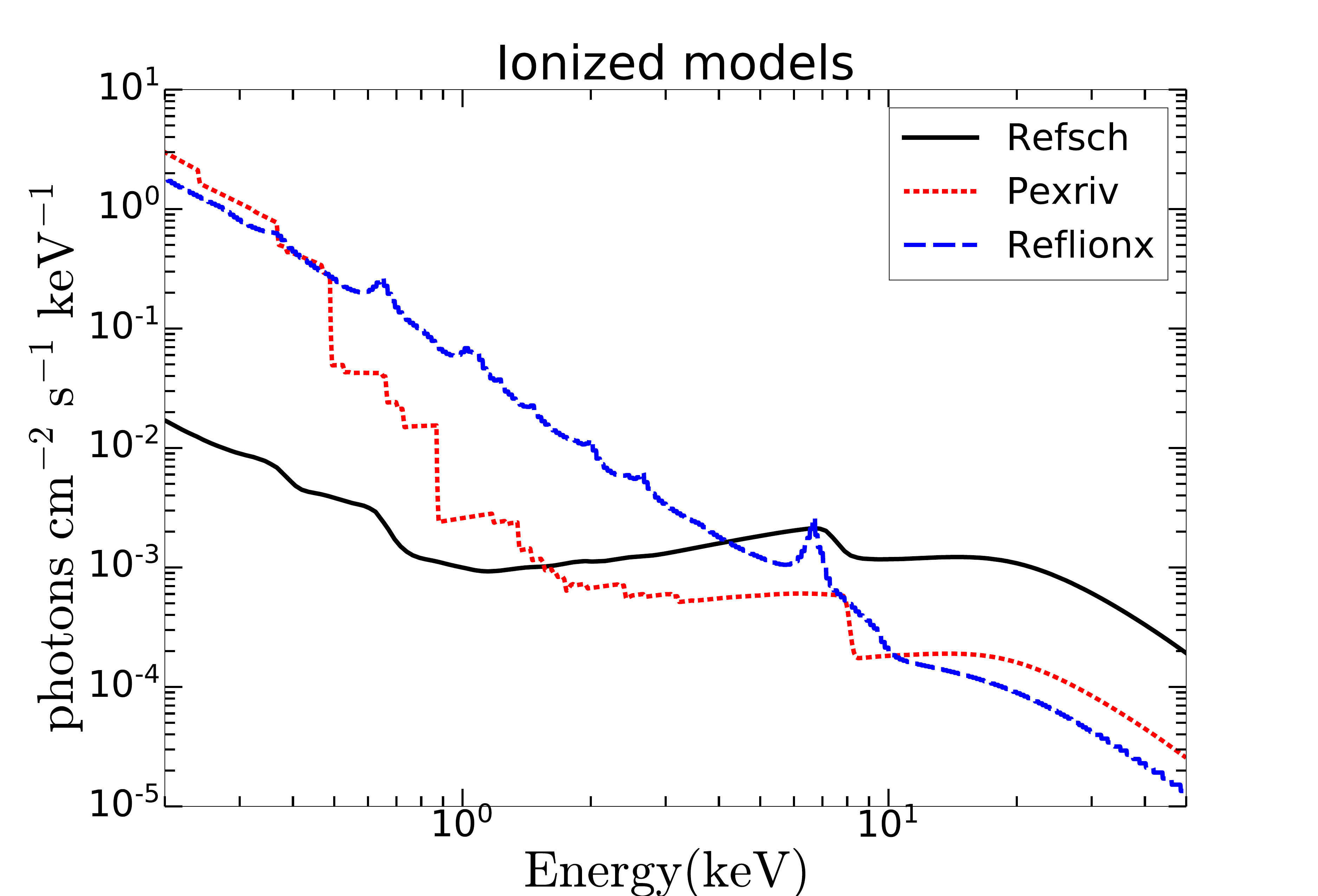}
\includegraphics[width=0.675\columnwidth, clip, trim=223 1 0 60]{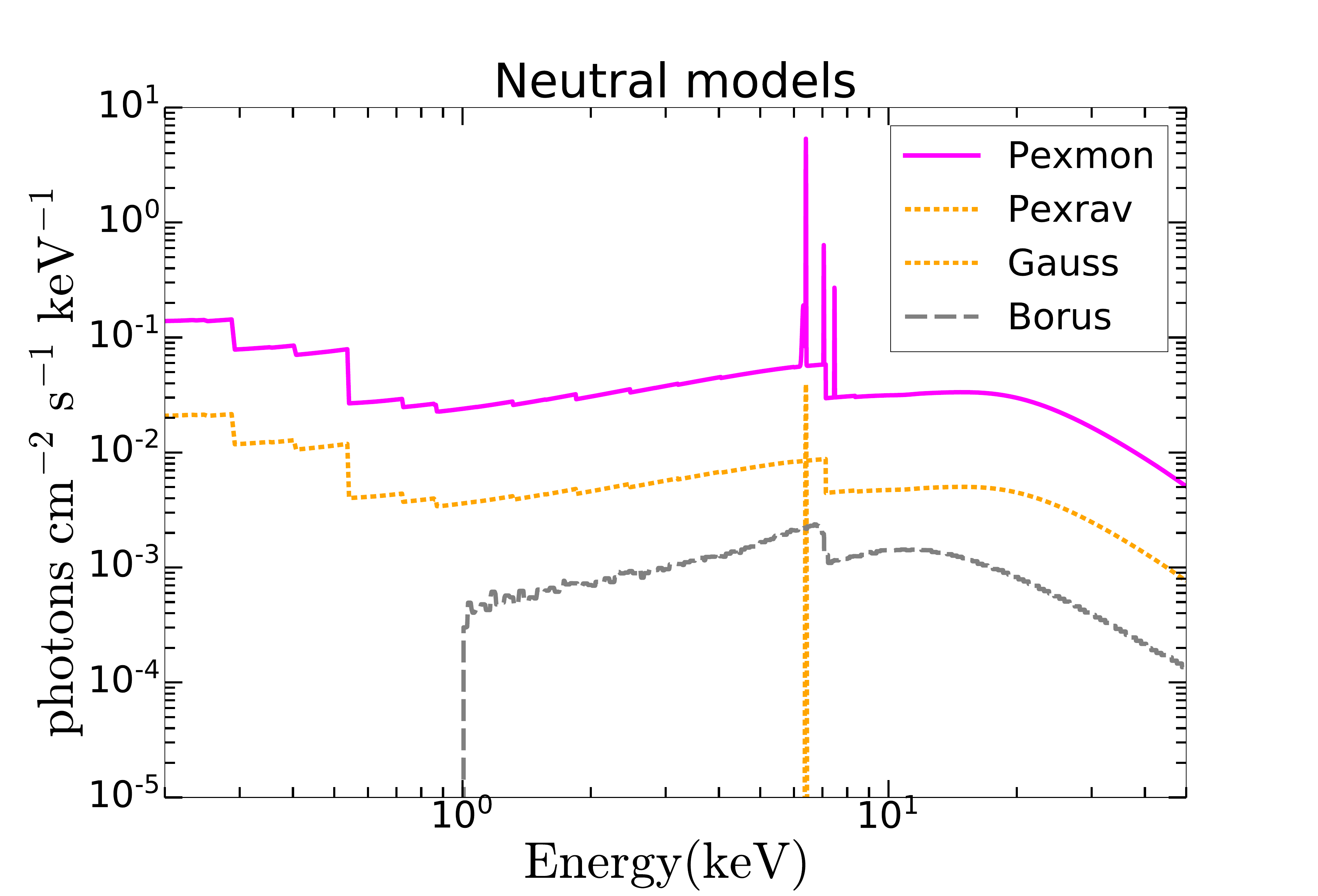}
\caption{Example of the resulting X-ray spectra of the reflection models used in this analysis (see Section\,\ref{sec:reflection_models}). Ionized and relativistic, ionized, and neutral models are shown in the left, middle, and right panels, respectively. For all models, the photon index $\rm{\Gamma}$= 2, the high energy cut-off $\rm{E_{cut}}$= 300, the abundance $\rm{Z}$=$\rm{Z_{\odot}}$, the iron abundance $\rm{Z_{Fe}}$= $\rm{Z_{\odot}}$, the viewing angle towards the system i=45$\rm{^o}$, the inner radius of the disk $\rm{R_{in}}$=2$\rm{r_{g}}$, the outer radius of the disk $\rm{R_{out}}$=100$\rm{r_{g}}$, the disk temperature $\rm{T_{disk}}$=30000 K, the index emissivity Index=3, and the ionization parameter $\rm{\xi}$= 1000, when the parameter is inside the model. Also, for {\sc borus}, the column density NHTor=$10^{24}$ $\rm cm^{-2}$, the angular size thTor=45$\rm{^o}$; for {\sc gauss}, the center and the width of the emission line are $\rm{E_{line}}$=6.4 keV and $\sigma$=0.01 keV, respectively; for {\sc relxill} the spin of the black-hole is set to $\rm{a}$=0.998.}
\label{fig:models_groups}
\end{center}
\end{figure*}

\subsection{Reflection models} \label{sec:reflection_models}

In order to explore the different reflection possibilities, we consider the physical scenarios showed in Figure \ref{fig:models}:

$\rm{\bullet}$ \underline{Reflection in an ionized medium.} We contemplate four ionized reflection models to account for the reflection in a medium such as the accretion disk, wind or the Broad Line Region (BLR): (1) {\sc pexriv} \citep{Magdziarz95}, (2) {\sc reflionx} \citep{Ross99, Ross05}, (3) {\sc relxill} \citep{Dauser10, Garcia14} and (4) {\sc refsch} \citep{Fabian89, Magdziarz95}. We also use the convolution model {\sc kdblur} in order to account for the relativistic effects when it is needed.  

$\rm{\bullet}$ \underline{Reflection in a neutral medium.} For the reflection in an neutral/distant medium such as the molecular torus we use three models: (1) {\sc pexmon} \citep{Nandra07}, (2) {\sc pexrav} \citep{Magdziarz95} and (3) {\sc borus02} \citep{Balokovic18}. For the {\sc pexrav} model, we also add a Gaussian line component in order to account for the $\rm{FeK\alpha}$ emission line feature. We named this model {\sc pexrav+gauss}. Note that {\sc pexrav} and {\sc pexmon} models are made by using physical approximations on the geometry, assuming the reflecting medium as a semi-infinite plane-parallel surface. Note also that these models consider the reflection medium to be neutral. It is possible to attribute these characteristics to the torus since it is composed of neutral material, and, due to the distance between the x-ray corona and the torus, it is reasonable to think of the torus as a semi-infinite plane-parallel surface, seen from the corona. Different authors have previously used these models to represent the putative torus, among them are \cite{Ricci11}, \cite{Brightman12}, \cite{Reis12}, \cite{Bauer15} \cite{Brightman15}, \cite{Laha21}, \cite{Osorio-Clavijo22}, \cite{Inaba22}.

Figure\,\ref{fig:models_groups} shows an example of the spectra produced for each of these nine X-ray reflection models. For all of them, we link parameters to the same value when there are common in several models. The main difference between neutral, ionized, and relativistic/ionized models is that the last two constantly grow from high to low energies, except {\sc pexriv} and {\sc refsch}; while the neutral models show a decrease below $\sim$ 6 keV. On the other hand, neutral models show a narrow $\rm{FeK\alpha}$ emission line compared to the ionized or ionized/relativistic models. Finally, note that despite the fact that the {\sc pexmon} and {\sc pexrav} models do not assume a toroidal geometry, they naturally reproduce the reflection by the torus because the physical approximation by which they are built is valid for the torus. This is highlighted in Figure\,\ref{fig:models_groups} where the spectral shape produced by {\sc borus}\footnote{The sharp decrease of flux shown for {\sc borus} model at 1\,keV (Figure\,\ref{fig:models_groups}, right) is probably associated to the energy range where this model is evaluated. Note that this does not affect our analysis because we use data above 3\,keV.} model is consistent with that of {\sc pexmon} and {\sc pexrav} models.

We will fit these nine models on a sub-sample of five test objects in order to obtain the models that give the best fits, and then use them to fit the full sample of 22 sources (see Section\,\ref{sec:Spectral fitting procedure}). Table\,\ref{tab:models_parameters} shows the parameters of each tested reflection model.  Note that, for the power-law we fixed the energy cut-off to a high value of 1000 keV. In all reflection models, we linked the energy cut-off and the photon index to such of the power-law, when these parameters are required.

%{\sc relxill} and {\sc kdblur*reflionx} are very similar along the full range. {\sc kdblur*pexriv} shows a wider compton hump.

\begin{table*}[ht!]
\scriptsize 
\begin{center}
\begin{tabular}{cccccccccccccccccccccccccc} \hline
& {\sc pexmon} & {\sc pexrav+gauss} & {\sc borus} & {\sc pexriv} & {\sc reflionx} & {\sc refsch} & {\sc relxill} & {\sc kdblur} \\ \hline
$\rm{\Gamma}$ & $\rm{\Gamma_{PL}}$ & $\rm{\Gamma_{PL}}$ & $\rm{\Gamma_{PL}}$ & $\rm{\Gamma_{PL}}$ & $\rm{\Gamma_{PL}}$ & $\rm{\Gamma_{PL}}$ & $\rm{\Gamma_{PL}}$ & - \\
$\rm{E_{cut}}$ & $\rm{E_{cut.PL}}$ & $\rm{E_{cut.PL}}$ & $\rm{E_{cut.PL}}$ & $\rm{E_{cut.PL}}$ & - & - & - & - \\
$\rm{rel_{refl}}$ & -1* & -1* & - & -1* & - & -1* & -1* & - \\
$\rm{Z}$ & 0.5 - 5 & 0.5 - 5 & - & 0.5 - 5 & - & 0.5 - 5 & - & - \\
$\rm{Z_{Fe}}$ & 0.5 - 5 & 0.5 - 5 & 0.5 - 5 & 0.5 - 5 & 0.5 - 5 & 0.5 - 5 & 0.5 - 5 & - \\
$\rm{Incl}$ & 5 - 85 & 18 - 87 & 20 - 85 & 18 - 87 & - & 20 - 85 & 5 - 80 & 20 - 85 \\
$\rm{E_{line}}$ & - & 6.3 - 6.6 & - & - & - & - & - & - \\
$\rm{\sigma}$ & - & 0.002 - 0.1 & - & - & - & - & - & - \\
$\rm{NHTor}$ & - & - & $10^{22}$ - $10^{25}$ & - & - & - & - & - \\
$\rm{thTor}$ & - & - & 5 - 80 & - & - & - & - & - \\
$\rm{T_{disk}}$ & - & - & - & $10^{4}$ - $10^{6}$ & - & $10^{4}$ - $10^{6}$ & - & - \\
$\rm{\xi}$ & - & - & - & $10^{-3}$ - $10^{3}$ & 10 - $10^{4}$ & 0.1 - $10^{3}$ & 0.01 - 4.7 & - \\
$\rm{R_{in}}$ & - & - & - & - & - & 10* & 1* & 10* \\
$\rm{R_{out}}$ & - & - & - & - & - & 400* & 100* & 400* \\
$\rm{Index1}$ & - & - & - & - & - & - & =Index2 & - \\
$\rm{Index2}$ & - & - & - & - & - & - & -9.8 - 9.8 & - \\
$\rm{R_{br}}$ & - & - & - & - & - & - & (($\rm{R_{out}}$-$\rm{R_{in}}$)/2)+$\rm{R_{in}}$* & - \\
$\rm{a}$ & - & - & - & - & - & - & 0.01 - 0.998 & - \\
\hline
\end{tabular}
\end{center}
\caption{Parameters of the reflection models tested in Section\,\ref{sec:reflection_models}. Values represent the range in which the parameter is evaluated. $\rm{\Gamma}$: photon index; $\rm{E_{cut}}$: cut-off energy (in keV); $\rm{rel_{refl}}$: strength of the reflection characterized by the reflection scaling factor; $\rm{Z}$ and $\rm{Z_{Fe}}$: abundance of elements heavier than He, and iron abundance, respectively (relative to solar); $\rm{E_{line}}$: gaussian line energy (in keV); $\rm{\sigma}$: gaussian line width (in keV); $\rm{Incl}$: inclination angle (in degrees); $\rm{NHTor}$ and $\rm{thTor}$: column density {\bf (in $\rm cm^{-2}$)}, and angular size (in degrees) of the torus, respectively; $\rm{T_{disk}}$: disk temperature (in K); $\rm{\xi}$: ionization parameter (in ergcm/s and log for {\sc relxill}); $\rm{R_{in}}$ and $\rm{R_{out}}$: inner and outer radius of the disk, respectively (ISCO for {\sc relxill} and $\rm{R_{g}}$ for all other models); $\rm{Index1}$, $\rm{Index2}$ and $\rm{R_{br}}$: index emissivity between $\rm{R_{in}}$ and $\rm{R_{br}}$, and index emissivity between $\rm{R_{br}}$ and $\rm{R_{out}}$, respectively; $\rm{a}$: spin of the black hole. Asterisks indicate when the values were fixed. Note that $\rm{\Gamma}$ and $\rm{E_{cut}}$ were linked to the corresponding value of the power-law.}
\label{tab:models_parameters}
\end{table*}

\subsection{Considerations for the simultaneous fitting of \emph{XMM}-Newton and \emph{NuSTAR} spectra} \label{sec:Considerations for the simultaneous fitting}

\emph{XMM}-Newton spectra have, in general, a larger number of counts compared to the \emph{NuSTAR} spectra (see Table\,\ref{tab:sample_data}. This could yield to an over-weight of the \emph{XMM}-Newton compared to the \emph{NuSTAR} spectra because the former have systematically the largest number of bins in the composed \emph{NuSTAR}+\emph{XMM}-Newton spectra. In order to validate the use of $\chi^{2}$ techniques allowing both spectra to account similarly in the resulting spectral fit, we re-binned the \emph{XMM}-Newton spectra according with \emph{NuSTAR} spectra, so that we have the same spectral bins in each spectral file in the common spectral band (3-10 keV). For this, we use the {\sc grppha} task within {\sc ftools}, which group the spectrum from a lower to an upper channel with grouping a certain amount of channels per bin. In order to estimate the amount of channels per bin, we computed the number of channels in \emph{NuSTAR} that correspond to an spectral bin in \emph{XMM}-Newton in the 3-10 keV energy band, using the lower and upper channel of \emph{XMM}-Newton equivalent to that of \emph{NuSTAR} for this energy range.
In order to determine the confidence intervals of the parameters in the models, we estimate the 1-$\sigma$ errors, which corresponds to a probability of 68$\%$. The 1-$\sigma$ error is calculated through the {\sc error} command within {\sc xspec}. For this we set $\rm{\Delta\chi^2}$=1, and the parameter and all other non frozen parameters are varied.

%each indicated parameter is varied until the value of the fit statistic, minimized by allowing all the other non-frozen parameters to vary, is equal to the last value of fit statistic determined by delta fit statistic = 1.}\footnote{https://heasarc.gsfc.nasa.gov/xanadu/xspec/manual/Xspec Manual.html}

%with as many bins as there are channels in each group. For this, we first compared th. 

%We then computed the number of channels per bin in the \emph{XMM}-Newton data to achieve the same number of spectral bins in both satellites.]

%Finally, the number of bins is obtained by dividing the number of channesls in the \emph{XMM}-Newton spectrum by the number channels in the \emph{NuSTAR} spectrum.}

\subsection{Statistical techniques} \label{sec:Statistical techniques}

In order to determine the best model, we use the $\rm{\chi^2}$ statistics through the standard $\rm{\chi^2_r = \chi^2/dof}$, where dof are the degrees of freedom. To compare between two models where one of them come from adding an extra model component to the previous one, we use the F-statistics. When the F-test probability is low (\rm{$\leq 10^{-4}$}) the model significantly improves the fit. We also use the Akaike Information Criterion (AIC) to evaluate to what extent a model is better than another. To this end, we use the eq.\,5 in \cite{Emmanoulopoulos16} to calculate the $\rm{AIC_c}$

\begin{equation}
    AIC_c = 2k - 2C_L + \chi^2 + \frac{2k(k+1)}{N-k-1}
\end{equation}

\noindent where $C_L$ is the constant likelihood of the true hypothetical model, $k$ is the number of free model parameters, and $N$ is the number of data points. We then calculate the difference between two different models, $\rm{\Delta[AIC_{c}]}$

\begin{equation}
    \Delta[AIC_{c}] = AIC_{c,2} - AIC_{c,1}
\end{equation}

Finally, we estimate the evidence ratio, $\rm{\epsilon}$,

\begin{equation}
    \epsilon = e^-\frac{\Delta[AIC_{c}]}{2}
\end{equation}

\noindent which is a measure of the relative likelihood of one versus another model. When the evidence ratio is almost 300, the model is the best among the alternatives.

\begin{table}[!t]
\scriptsize 
\renewcommand{\tabcolsep}{0.1cm}
\begin{center}
\begin{tabular}{lccc}\hline
Source & {\sc power-law} & {\sc relxill} & F-test \\
& \multicolumn{2}{c}{$\chi^{2}$/d.o.f. = $\chi^{2}_{r}$} & \\\hline
Mrk\,335 & 443.8/188=2.36 & 255/183=1.4 & \checkmark \\
Fairall\,9 & 766.4/502=1.53 & 561/497=1.13 & \checkmark \\ 
Mrk\,1040 & 1237.8/960=1.29 & 998.3/955=1.05 & \checkmark \\ 
Mrk\,1044 & 1111/710=1.56 & 768.2/705=1.09 & \checkmark \\ 
NGC\,1365 & 3719.3/1003=3.71 & 1739/998=1.74 & \checkmark \\ 
Ark\,120 & 1344.1/972=1.38 & 1060.4/967=1.1 & \checkmark \\ 
{\bf Mrk\,382} & 161.7/143=1.13 & 151.1/138=1.1 & 0.09 \\
NGC\,3227 & 1425.4/976=1.46 & 1146.3/971=1.18 & \checkmark \\ 
NGC\,3783 & 1755/575=3.05 & 1121.3/570=1.97 & \checkmark \\ 
NGC\,4051 & 1857.1/1452=1.28 & 1457.9/1447=1.01 & \checkmark \\ 
NGC\,4151 & 3036.9/1718=1.77 & 3291.8/1713=1.92 &  \\ 
Mrk\,766 & 470.1/432=1.09 & 443.7/427=1.04 & \checkmark \\ 
NGC\,4593 & 676.6/570=1.19 & 569.8/565=1.01 & \checkmark \\ 
IRAS\,13197-1627 & 1697.2/512=3.31 & 665.7/507=1.31 & \checkmark \\ 
{\bf IRAS\,13224-3809} & 314.2/245=1.28 & 301.5/240=1.26 & 0.07 \\
MCG\,-06-30-15 & 691.2/510=1.36 & 497.5/505=0.99 & \checkmark \\
NGC\,5548 & 1203.2/971=1.24 & 1079.1/966=1.12 & \checkmark \\ 
{\bf Mrk\,841} & 413.8/398=1.04 & 402.7/393=1.02 & 0.05 \\
IGRJ\,19378-0617 & 973.4/708=1.37 & 742.7/703=1.06 & \checkmark \\ 
Mrk\,915 & 694.2/572=1.21 & 595.5/567=1.05 & \checkmark \\ 
{\bf MR\,2251-178} & 720/742=0.97 & 717.6/737=0.97 & 0.78 \\
NGC\,7469 & 707.9/465=1.52 & 489.8/460=1.06 & \checkmark \\ 
\hline
\end{tabular}
\end{center}
\caption{F-statistics applied to the sample comparing a simplest power-law (absorption*power-law) and a reflection (absorption*power-law + {\sc relxill}) model. Column 1 shows the name of the soure. Columns 2 and 3 shows the \rm{$\chi^{2}$} statistic for the power-law and reflection models respectively. Column 4 shows the F-test obtained. Sources in bold face indicate those that do no require reflection component, according to F-test.}
\label{tab:models_PL}
\end{table}

\subsection{Spectral fitting procedure} \label{sec:Spectral fitting procedure}

We start our analysis by fitting the spectra to our baseline model in order to check the presence of the reflection component. To do this, we compare the power-law fit with a model which account the reflection component ({\sc relxill}). In order to determine the best model, we use the $\rm{\chi^2}$ statistics through the standard $\rm{\chi^2_r = \chi^2/dof}$, where dof are the degrees of freedom. To compare between two models where one of them come from adding an extra model component to the previous one, we use the F-statistics. When the F-test probability is low (\rm{$\leq 10^{-4}$}) the model significantly improves the fit. We also use the Akaike Information Criterion (AIC) to evaluate to what extent a model is better than another. Note that, we use AIC when comparing models not related to each other, and F-test when comparing nested models. Objects for which the reflection component is not statistically needed were discarded in the following analysis.\\
Our aim is to test as many models as possible that can be used as baseline model. However, this is time consuming considering for the nine baseline models, which makes the analysis unpractical for a large sample of objects as the one analysed here.

Thus, we select a subsample of five objects to which we fit our set of the nine reflection models described in Section\,\ref{sec:reflection_models}. Our criteria for the selection of this subsample was to choose sources that have previously been fitted with different models, with the intention of not biasing the subsample towards objects that clearly prefer some model. Note that the selection of the test objects was not through the number of counts or signal-to-noise ratio to avoid biases in the selection of the models preferred by the objects, since, the model preferred by sources with a high number of counts may not be the best model for sources with smaller number of counts. The five test objects are NGC\,3783, MCG\,-06-30-15, Fairall\,9, Mrk\,1044, and Mrk\,335. X-ray spectrum of NGC\,3783 and Fairall\,9 are well fitted by using absorption components \citep{Blustin02, Krongold03, Krongold05}, and through reflection of the accretion disk \citep{Gondoin01, Emmanoulopoulos11}, respectively. MCG\,-06-30-15 and Mrk\,335 has been well fitted through relativistic reflection \citep[][]{Chiang11, Longinotti07} and absorption models \citep{Miller09, ONeill07}. The X-ray spectrum of Mrk\,1044 has been fitted by using relativistic and distant reflection \citep{Mallick18} and also by adding wind components \citep{Dewangan07, Krongold21}. In this way, we cover the different reflection scenarios, two objects extensively fitted through disk reflection or only absorption scenario, and three objects fitted with equal success using one or the other model. The results from these objects determine a sub-set of models that we apply to the overall sample of type-1 AGN.

Note that, in order to corroborate that the choice of the subsample does not affect the result on the best models, we choose three other objects (in addition to the first five), namely Mrk\,1040, Mrk\,915, and Ark\,120, to which we also fit the nine initial models. For these three new objects we find that the preferred models are {\sc pexrav+gauss} and {\sc relxill}. These results strongly suggest that the selection of objects for the subsample does not alter the results on the models selected for testing on the full sample.

%Mrk\,335 has been fitted through relativistic accretion disk \citep{Longinotti07} and partial covering models \citep{ONeill07}. 

%Finally, since the spectral fitting described previously was performed keeping fixed the cut-off energy parameter in the power-law associated with the primary X-ray radiation, we test the effect of the energy cut-off allowing the cut-off energy parameter to vary in the best fit obtained of each source.

\subsection{Intrinsic parameters} \label{sec:Intrinsic parameters}

We also study the relation between the final preferred models and the SMBH mass, luminosity and the Eddington ratio of the sources (see Table\,\ref{tab:sample_Lums}). For this purpose, we compute the 2–10 keV luminosity for each component of the spectral fit using the {\sc cluminosity} command within {\sc xspec}. From the luminosity of the intrinsic continuum, we calculate the bolometric luminosity, $\rm{L_{bol}}$, from X-rays following the relation from \cite{Marconi04}:

\begin{equation}
log[L_{bol}/L_{2-10keV}] = 1.54+0.24\mathcal{L}+0.012\mathcal{L}^2-0.0015\mathcal{L}^3
\end{equation}

\noindent where $\mathcal{L}$ = log($\rm{L_{bol}}$-12), and $\rm{L_{bol}}$ is in units of $L_{\odot}$. Then, we calculated the Eddington ratio, $\lambda$, defined as $\lambda = L_{bol}/L_{edd}$, where $L_{edd}$ is the Eddington luminosity. 

\begin{table*}[ht]
\scriptsize 
\begin{center}
\begin{tabular}{cccccc}\hline
Model & NGC\,3783 & MCG\,-6-30-15 & Fairall\,9 & Mrk\,1044 & Mrk\,335 \\
& \multicolumn{5}{c}{$\chi^{2}$/d.o.f.=$\chi^{2}_{r}$ / AIC} \\ \hline

\multirow{2}{*}{{\sc relxill}} & 1121.3/570 = 1.97 & \bf{497.5/505 = 0.99} & 561/497 = 1.13 & \bf{768.2/705 = 1.09} & 255/183 = 1.4 \\
& $5.3\times 10^{35}$ & 1 & 70 & 1 & $3.1\times 10^{7}$ \\
\multirow{2}{*}{{\sc kdblur*pexriv}} & 1462.9/569 = 2.57 & 571.1/504 = 1.13 & 796/496 = 1.6 & 851.4/704 = 1.21 & 379.3/182 = 2.08 \\
& $1.4\times 10^{110}$ & $1.7\times 10^{16}$ & $1.3\times 10^{53}$ & $1.9\times 10^{18}$ & $ 5.5\times 10^{34}$ \\
\multirow{2}{*}{{\sc kdblur*reflionx}} & 1236.4/571 = 2.17 & 507.6/506 = 1.0 & 603.1/498 = 1.21 & 775.1/706 = 1.1 & 302.5/184 = 1.64 \\
& $3.1\times 10^{60}$ & 91 & $5.7\times 10^{10}$ & 19 & $ 3.5\times 10^{17}$ \\
\multirow{2}{*}{{\sc refsch}} & 1366.4/569 = 2.4 & 538.4/504 = 1.07 & 701.8/496 = 1.41 & 825/704 = 1.17 & 391.9/182 = 2.15 \\
& $1.5\times 10^{89}$ & $1.3\times 10^{9}$ & $4.6\times 10^{32}$ & $3.7\times 10^{12}$ & $ 3.1\times 10^{37}$ \\
\multirow{2}{*}{{\sc reflionx}} & 1325.8/573 = 2.31 & 538.3/508 = 1.06 & 614.7/500 = 1.23 & 807/708 = 1.14 & 296.1/186 = 1.59 \\
& $2.8\times 10^{79}$ & $1.5\times 10^{8}$ & $6.5\times 10^{12}$ & $5.5\times 10^{7}$ & $ 4.3\times 10^{15}$ \\ 
\multirow{2}{*}{{\sc pexriv}} & 1648.4/570 = 2.89 & 575.9/505 = 1.14 & 795.8/497 = 1.6 & 905.2/705 = 1.28 & 382/183 = 2.09 \\
& $1.5\times 10^{150}$ & $1.1\times 10^{17}$ & $6.8\times 10^{52}$ & $5.6\times 10^{29}$ & $ 1.2\times 10^{35}$ \\ 
\multirow{2}{*}{{\sc pexmon}} & 1120.8/572 = 1.96 & 541.9/507 = 1.07 & 598/499 = 1.2 & 1028.7/707 = 1.46 & 248.3/185 = 1.34 \\
& $1.4\times 10^{35}$ & $1.5\times 10^{9}$ & $2.6\times 10^{9}$ & $1.3\times 10^{56}$ & $ 3.2\times 10^{5}$ \\
\multirow{2}{*}{{\sc pexrav+gauss}} & \bf{955.7/569 = 1.68} & 507.6/504 = 1.01 & \bf{551.4/496 = 1.11} & 960/704 = 1.36 & \bf{219.3/182 = 1.2} \\
& 1 & 268 & 1 & $7.6\times 10^{41}$ & 1 \\
\multirow{2}{*}{{\sc borus}} & 1057.4/571 = 1.85 & 542.6/506 = 1.07 & 556.8/498 = 1.12 & 1019.2/706 = 1.44 & 270.9/184 = 1.47 \\ 
& $4.2\times 10^{21}$ & $3.6\times 10^{9}$ & 5 & $1.9\times 10^{54}$ & $ 4.7\times 10^{10}$ \\
\hline
\end{tabular}
\end{center}
\caption{Models applied to the test objects described in Section\,\ref{sec:Spectral fitting procedure}. First and second rows correspond to $\rm{\chi^2}/d.o.f.$ and AIC statistics respectively. We show in bold face the best model according to AIC. Number one in AIC rows indicates the best model, other numbers indicate how many times the model is worse than the best model. Note that the model name represents the reflection component in our baseline model: absorber*intrinsic + reflection.}
\label{tab:test_models}
\end{table*}

\section{Results}\label{sec:Results}

\subsection{The existence of the reflection component} \label{sec:The existence of the reflection component}

\begin{figure}[!t]
\begin{center}
\includegraphics[width=1.0\columnwidth]{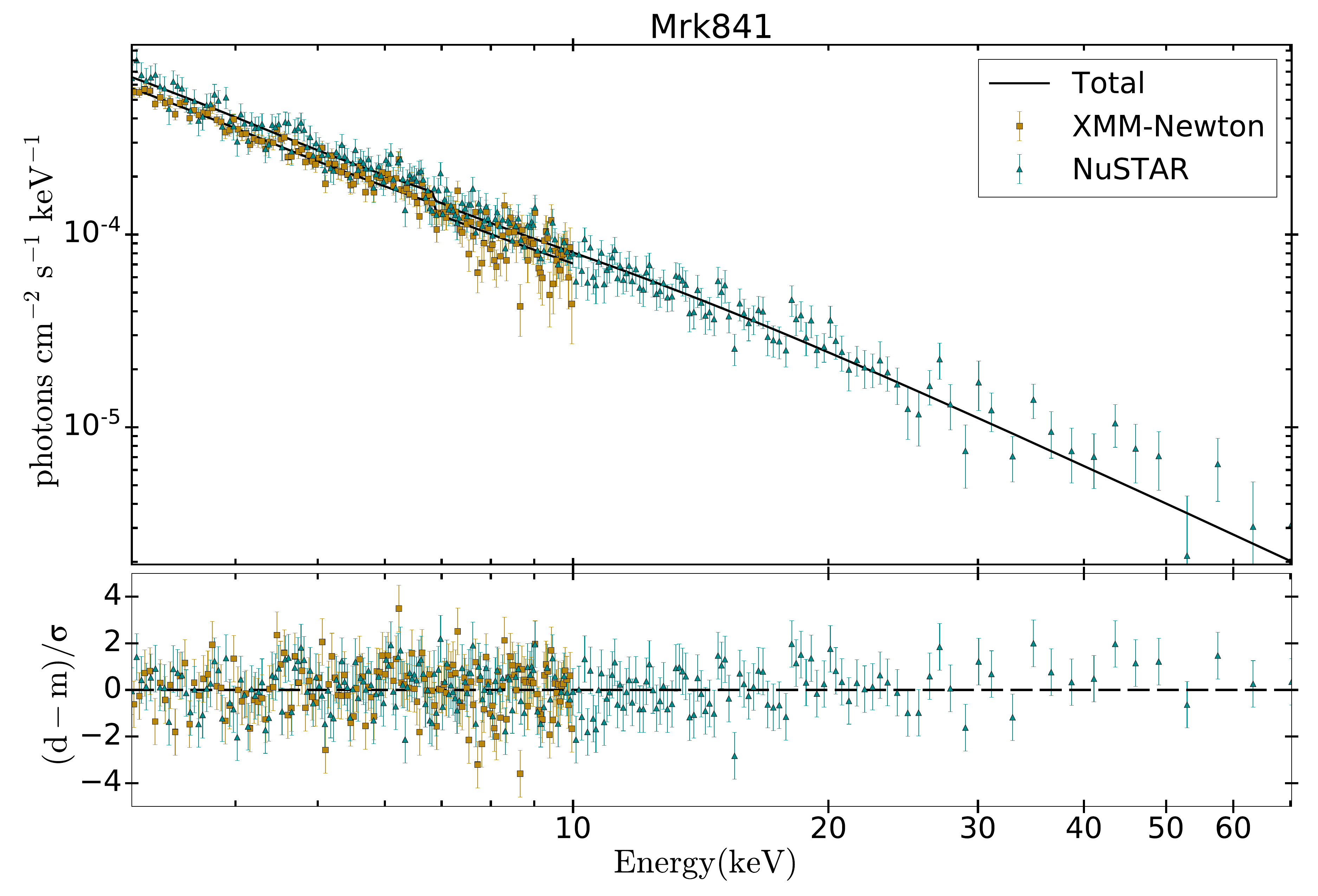}
\caption{Power-law fit to Mrk\,841. We show the best fit (black solid line) to the data in the top panel and the ratio between model and data in the bottom panel. The gold dots and dark cyan diamonds show the data from \emph{XMM}-Newton and \emph{NuSTAR} respectively. Top panel shows the result of plotting {\sc ufspec} in the {\sc xspec} language. The data points plotted are calculated by D*UM/FM, where D is the observed data, UM is the theoretical model integrated over the plot bin, and FM is the model times the response.}
\label{fig:Mrk841_non_reflection}
\end{center}
\end{figure}

Table\,\ref{tab:models_PL} shows the $\chi^2$ results of the reflection component test. We corroborate the presence of the reflection component in 18 sources (marked with a check symbol in Table\,\ref{tab:models_PL}). These objects show a F-test probability $\rm{<10^{-4}}$, which confirm the improvement among these two models. Four sources, namely Mrk\,382, IRAS\,13224-3809, Mrk\,841, and MR\,2251-178 show a F-test probability of 0.09, 0.07, 0.05, and 0.78, respectively. Therefore, according to the F-statistics, these four objects do not need a reflection component to satisfactorily fit the spectrum.
As an example of objects where the reflection component is not needed, we show in Figure\,\ref{fig:Mrk841_non_reflection} the absorbed power-law fit for Mrk\,841. Note that the spectrum is well fitted with a power-law with photon index of 1.86, which is in good agreement with the photon index of $\sim$ 1.9 expected to the AGN \citep{Zdziarski95}. Note also the absence of the $\rm{FeK\alpha}$ line, and the compton hump, which are the main signatures of the reflection component. A similar behavior is observed in the other three non-reflection objects. The power-law fit of these sources is shown in Appendix\,\ref{sec:Non_reflection_objects}. 

\begin{table*}[ht]
\scriptsize
%\tiny
\renewcommand{\tabcolsep}{0.05cm}
\begin{center}
\setlength{\columnsep}{5 cm}
\begin{tabular}{lccc|lcccc}\hline
Source & {\sc pexrav+gauss} & {\sc relxill} & {\sc hybrid} & Source & {\sc pexrav+gauss} & {\sc relxill} & {\sc hybrid} \\ \hline

Mrk\,335* & 219.3/182=1.2 & 255/183=1.4 & {\bf166.3/176=0.94} & NGC\,4151 & 2543.9/1712=1.49 & 3291.8/1713=1.92 & {\bf 2117.3/1706=1.24} \\
AIC/F-test$\rm {_{mod}}$ & 1 & $3\times10^{7}$ & $6.7\times10^{-9}$ & AIC/F-test$\rm {_{mod}}$ & 1 & $1.5\times10^{162}$ & $8.2\times10^{-64}$ \\
\hline

Fairall\,9 & 556.8/496=1.12 & 561/497=1.13 & {\bf499.6/490=1.02} & Mrk\,766 & {\bf447/426=1.05} & {\bf443.7/427=1.04} & 436.2/420=1.04 \\
AIC/F-test$\rm {_{mod}}$ & 1 & 5 & $1\times10^{-9}$ & AIC/F-test$\rm {_{mod}}$ & 9 & 1 & 0.41 \\
\hline

Mrk\,1040 & 997.4/954=1.05 & 998.3/955=1.05 & {\bf958.2/948=1.01} & NGC\,4593 & {\bf559.4/564=0.99} & {\bf 569.8/565=1.01}  & 550.1/558=0.99 \\
AIC/F-test$\rm {_{mod}}$ & 1 & 1 & $1.1\times10^{-6}$ & AIC/F-test$\rm {_{mod}}$ & 1 & 105 & 0.15 \\
\hline

Mrk\,1044 & 960/704=1.36 & {\bf768.2/705=1.09} & 763.5/698=1.09 & IRAS\,13197-1627 & 643.7/506=1.27 & 665.7/507=1.31 & {\bf 533.2/500=1.07} \\
AIC/F-test$\rm {_{mod}}$ & $7.5\times10^{41}$ & 1 & 0.74 & AIC/F-test$\rm {_{mod}}$ & 1 & $3.5\times10^{4}$ & $3.5\times10^{-18}$ \\
\hline

NGC\,1365* & 2409.3/997=2.42 & 1739/998=1.74 & {\bf 1566.4/991=1.58} & MCG\,-06-30-15 & {\bf 507.6/504=1.01} & {\bf 497.5/505=0.99} & 495.3/498=0.99 \\
AIC/F-test$\rm {_{mod}}$ & $6\times10^{145}$ & 1 & $1.7\times10^{-19}$ & AIC/F-test$\rm {_{mod}}$ & 268 & 1 & 0.946786 \\
\hline

Ark\,120 & 1057.1/966=1.09 & 1060.4/967=1.1 & {\bf 999.3/960=1.04} & NGC\,5548 & 1044/965=1.08 & 1079.1/966=1.12  & {\bf 1010.8/959=1.05}  \\
AIC/F-test$\rm {_{mod}}$ & 1 & 3 & $7.1\times10^{-10}$ & AIC/F-test$\rm {_{mod}}$ & 1 & $2.4\times10^{7}$ & $2.5\times10^{-5}$ \\
\hline

NGC\,3227 & 1143/970=1.18 & 1146.3/971=1.18 & {\bf 1105.3/964=1.15} & IGRJ\,19378-0617 & 946.2/702=1.35 & {\bf 742.7/703=1.06} & 730.9/696=1.05 \\
AIC/F-test$\rm {_{mod}}$ & 1 & 3 & $1.4\times10^{-5}$ & AIC/F-test$\rm {_{mod}}$ & $2.6\times10^{44}$ & 1 & 0.130678 \\
\hline

NGC\,3783* & 955.7/569=1.68 & 1121.3/570=1.97 & {\bf 857.7/563=1.52} & Mrk\,915 & {\bf 583/566=1.03} & 595.5/567=1.05  & 580.4/560=1.04 \\
AIC/F-test$\rm {_{mod}}$ & 1 & $5.3\times10^{35}$ & $2.6\times10^{-11}$ & AIC/F-test$\rm {_{mod}}$ & 1 & 302 & 0.867129 \\
\hline

NGC\,4051 & 1469/1446=1.02 & 1457.9/1447=1.01 & {\bf1418.1/1440=0.98} & NGC\,7469 & 493.2/459=1.07 & 489.8/460=1.06 & {\bf439.5/453=0.97} \\
AIC/F-test$\rm {_{mod}}$ & 430 & 1 & $1.3\times10^{-6}$ & AIC/F-test$\rm {_{mod}}$ & 9 & 1 & $1.9\times10^{-8}$ \\
\hline

\end{tabular}
\end{center}
\caption{Models tested to the full sample. Note that the model name represents the reflection component in our baseline model: absorber*intrinsic + reflection. First and second row of each object shows the $\rm{\chi^2}/d.o.f.$ statistics, and the AIC and F-test results respectively. Preferred model by each object is in bold face. Sources marked with an asterisk are those discarded in the analysis of the intrinsic properties of the sample  due to poor spectral fitting (see Section\,\ref{sec:Intrinsic parameters}).}
\label{tab:models_sample}
\end{table*}

\subsection{Test objects}

As described above, we selected five objects (all of them with clear signs of reflection in the \emph{NuSTAR} observations) to perform an analysis on nine reflection models: three neutral, three ionized and three ionized and relativistic models (see Section\,\ref{sec:reflection_models} and Table\,\ref{fig:models}). We show in Table\,\ref{tab:test_models} the statistics obtained with each model tested to the five test objects. We also show the AIC, in order to compare the goodness between model. We find that the {\sc pexrav+gauss} and {\sc relxill} models are the preferred models by the test objects. {\sc pexrav+gauss} model gives the best fit for three objects: NGC\,3783, Fairall\,9, and Mrk\,335; and {\sc relxill} model gives the best fit for two objects: MCG\,-06-30-15 and Mrk\,1044. The best fits of MCG\,-06-30-15 and Fairall\,9 are obtained through three models ({\sc relxill}, {\sc kdblur*reflionx} and {\sc pexrav+gauss}, and {\sc pexrav+gauss}, {\sc borus} and {\sc relxill}, respectively). For Mrk\,1044 the best fits are obtained with two models ({\sc relxill} and {\sc kdblur*reflionx}). According to these results, we decide to choose the {\sc pexrav+gauss} and {\sc relxill} models, discarding from the subsequent analysis the rest of the models. Note that the {\sc borus} model provide a good fit to Fairall\,9, and {\sc kdblur*reflionx} model provide good fits to MCG\,-06-30-15 and Mrk\,1044. However, we discard them because the physical scenario is properly described by {\sc pexrav+gauss} and {\sc relxill} models, respectively, and they are good fits only in one and two of the five test objects.

%\begin{figure*}[!t]
%\begin{center}
%\includegraphics[width=0.74\columnwidth, clip, trim=10 20 0 70]{figuras/Mrk335_HYBRID_hard_final.pdf}
%\includegraphics[width=0.67\columnwidth, clip, trim=180 20 0 70]{figuras/fairall9_HYBRID_hard_final.pdf}
%\includegraphics[width=0.67\columnwidth, clip, trim=178 20 0 70]{figuras/Mrk1040_HYBRID_hard_final.pdf} \\
%\includegraphics[width=0.74\columnwidth, clip, trim=10 20 0 70]{figuras/NGC1365_HYBRID_hard_final.pdf}
%\includegraphics[width=0.67\columnwidth, clip, trim=180 20 0 70]{figuras/Ark120_HYBRID_hard_final.pdf}
%\includegraphics[width=0.67\columnwidth, clip, trim=178 20 0 70] {figuras/NGC3227_HYBRID_hard_final.pdf} \\ 
%\caption{Best fit to the 12 objects that prefer the hybrid model. We show the best fit (black solid line) to the data in the top panel and the ratio between model and data of the hybrid, neutral and ionized models (marked whit letters H, N, and I respectively) in the bottom panels. Also, we show the reduced $\chi^2$ of the corresponding fit.}
%\label{fig:hard_band_fit_HYBRID}
%\end{center}
%\end{figure*}

\begin{figure*}[!t]
\begin{center}
\includegraphics[width=0.69\columnwidth]{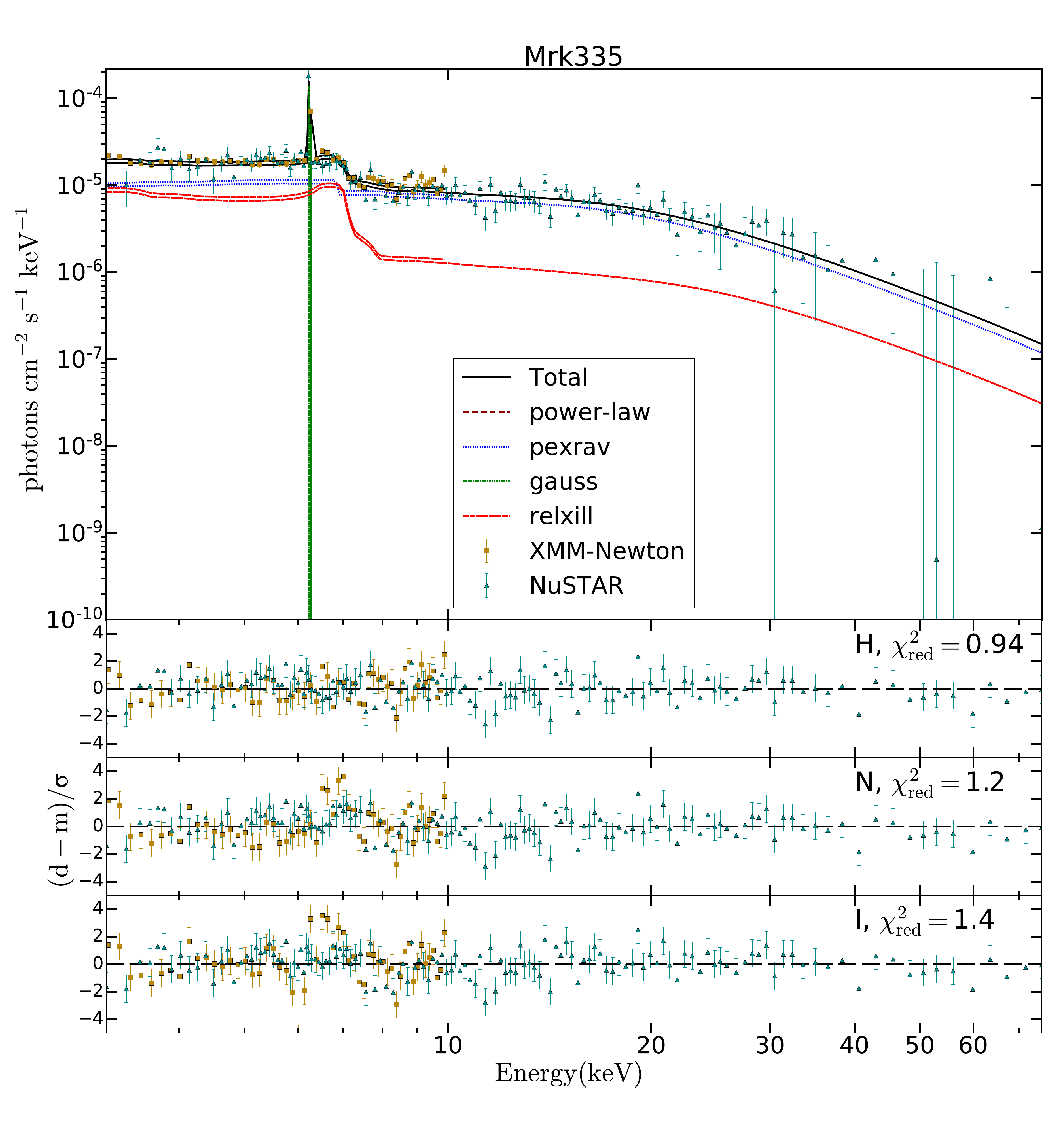}
\includegraphics[width=0.69\columnwidth]{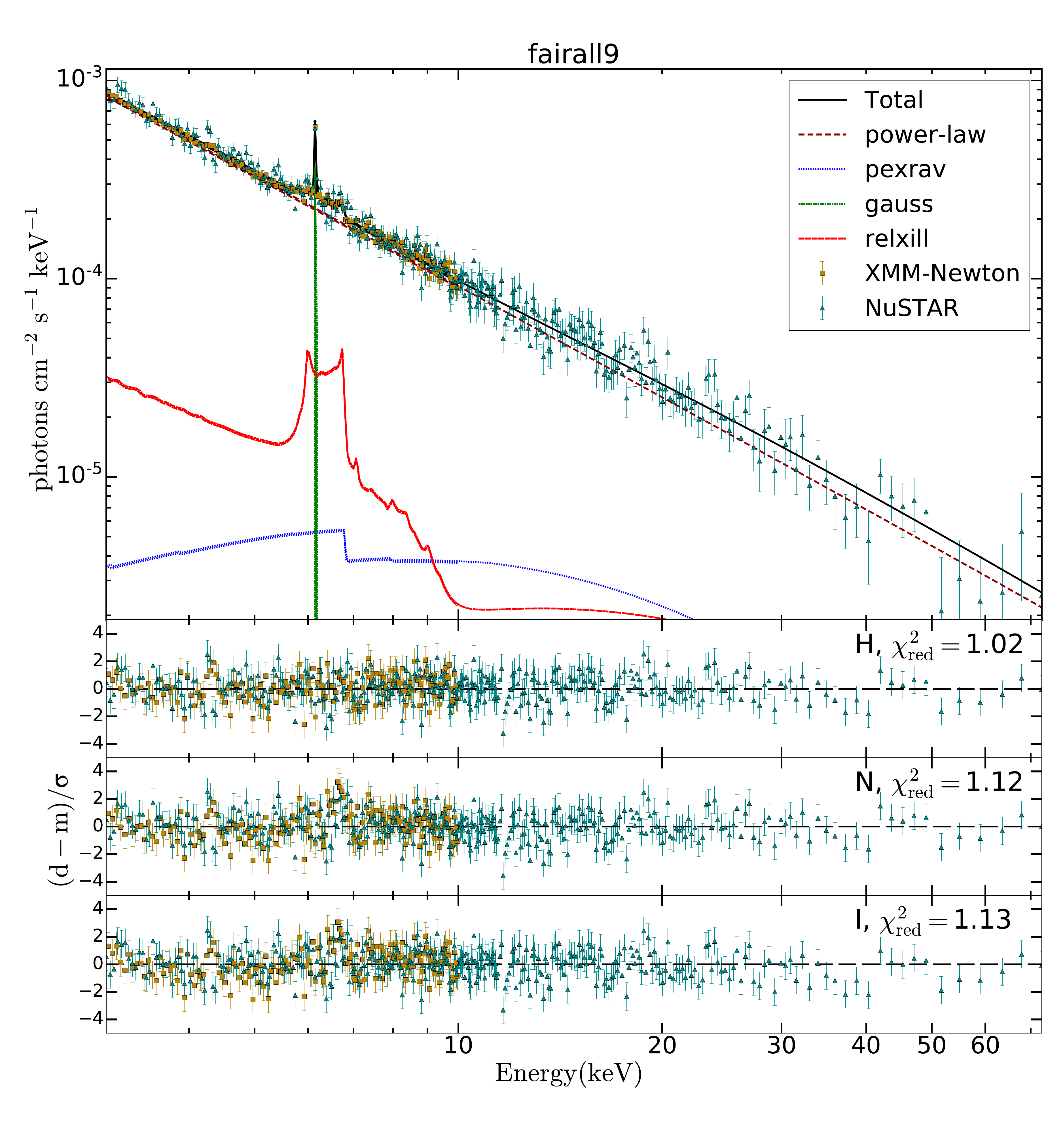}
\includegraphics[width=0.69\columnwidth]{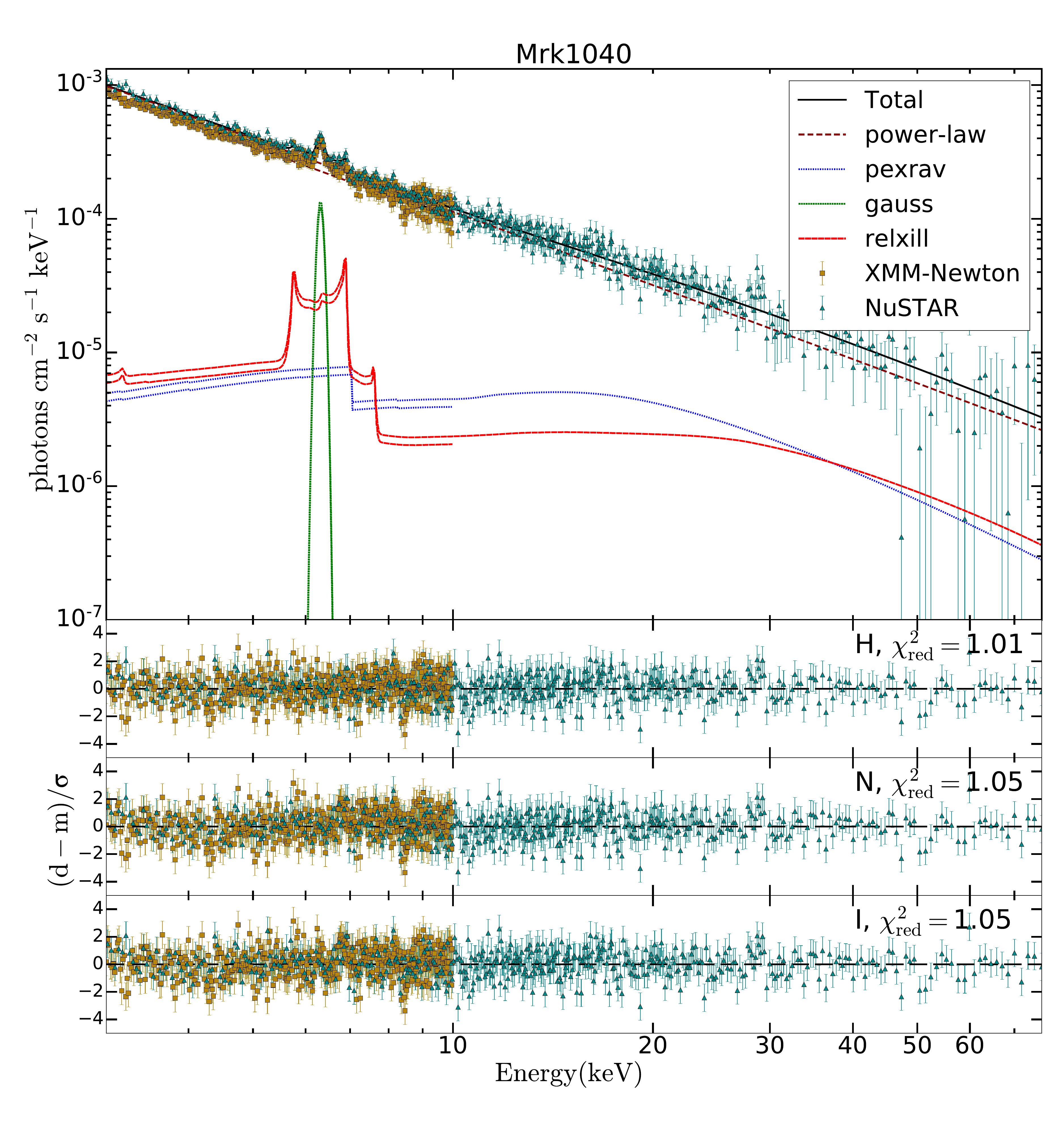} \\
\includegraphics[width=0.69\columnwidth]{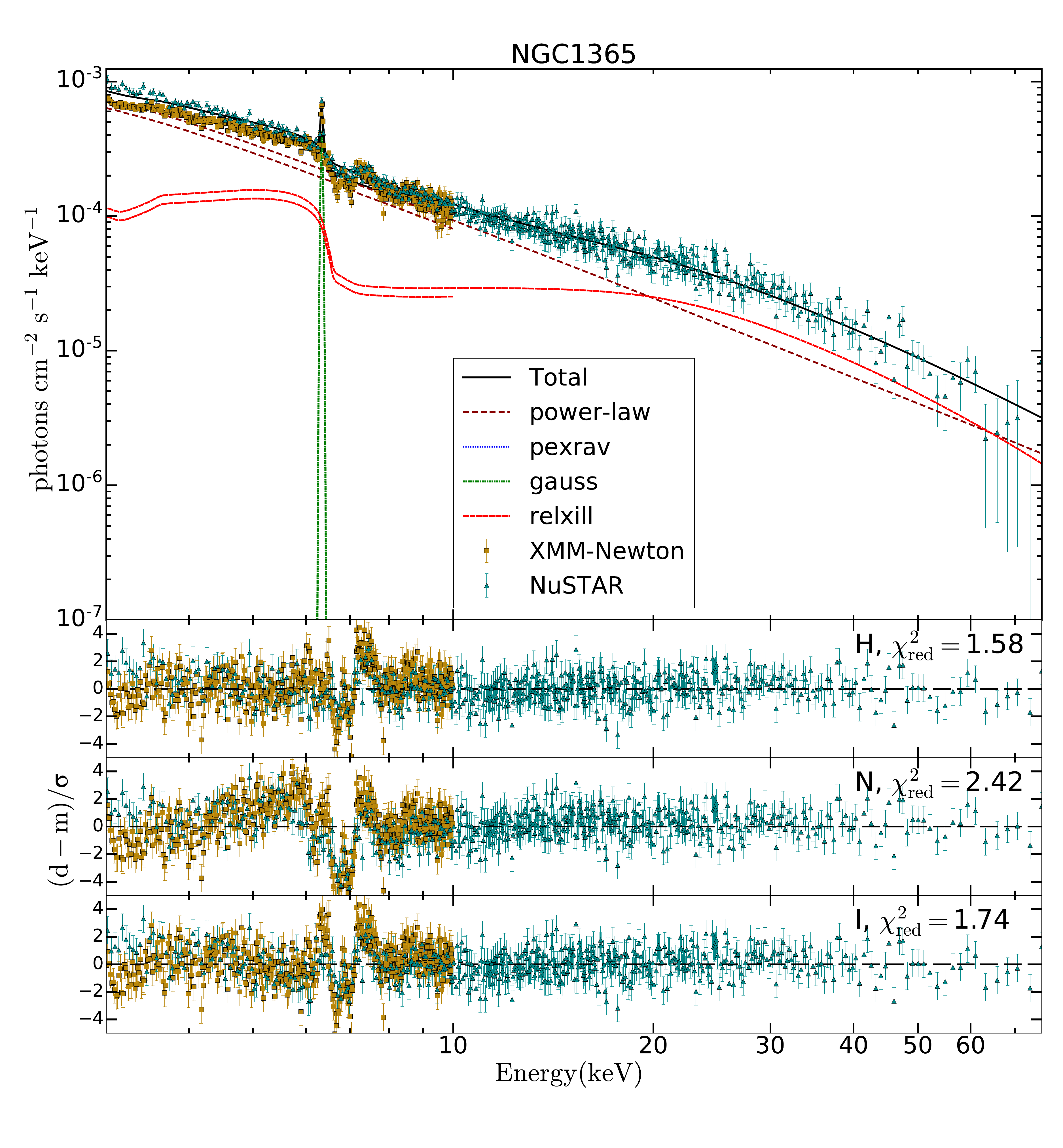}
\includegraphics[width=0.69\columnwidth]{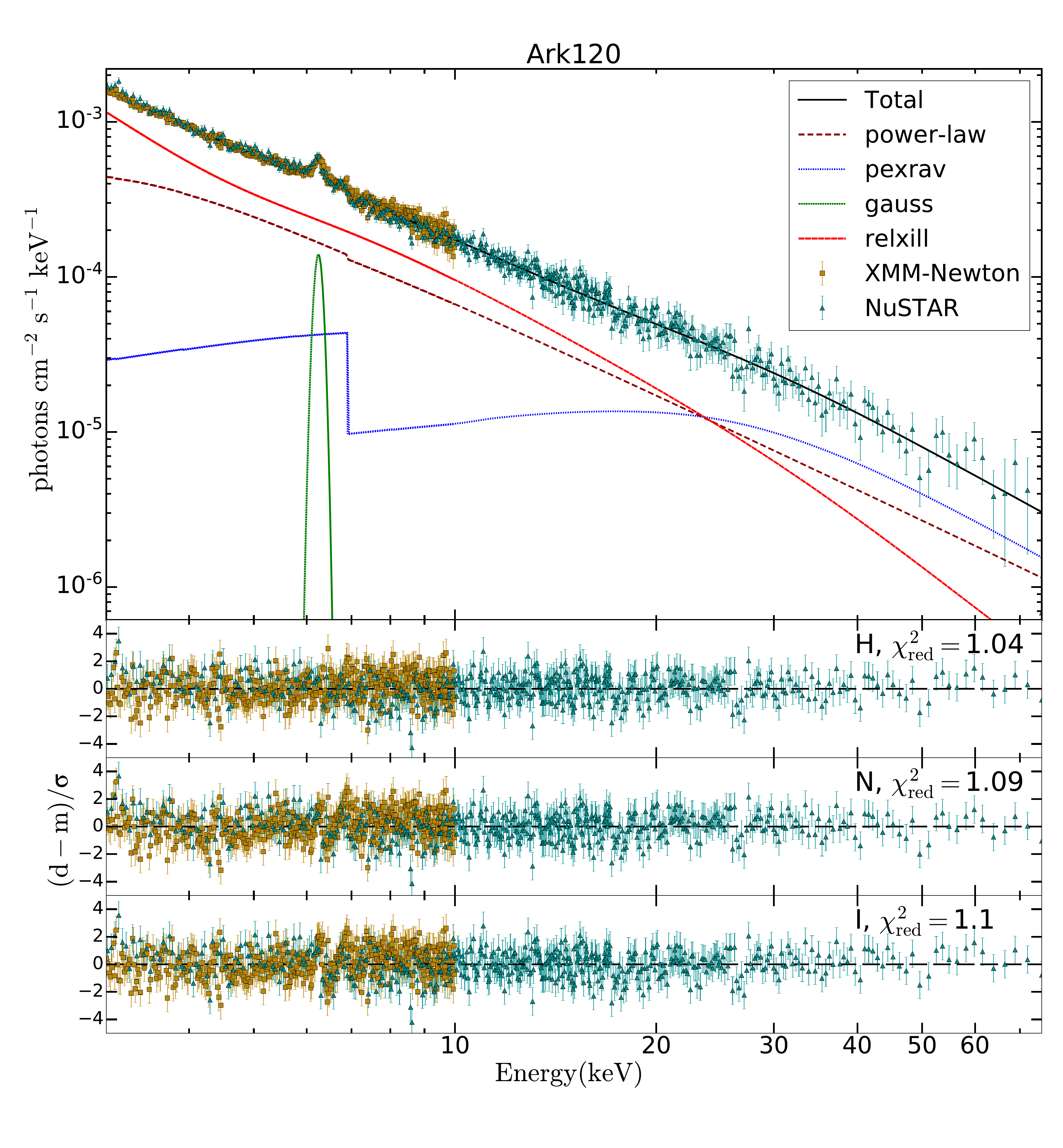}
\includegraphics[width=0.69\columnwidth]{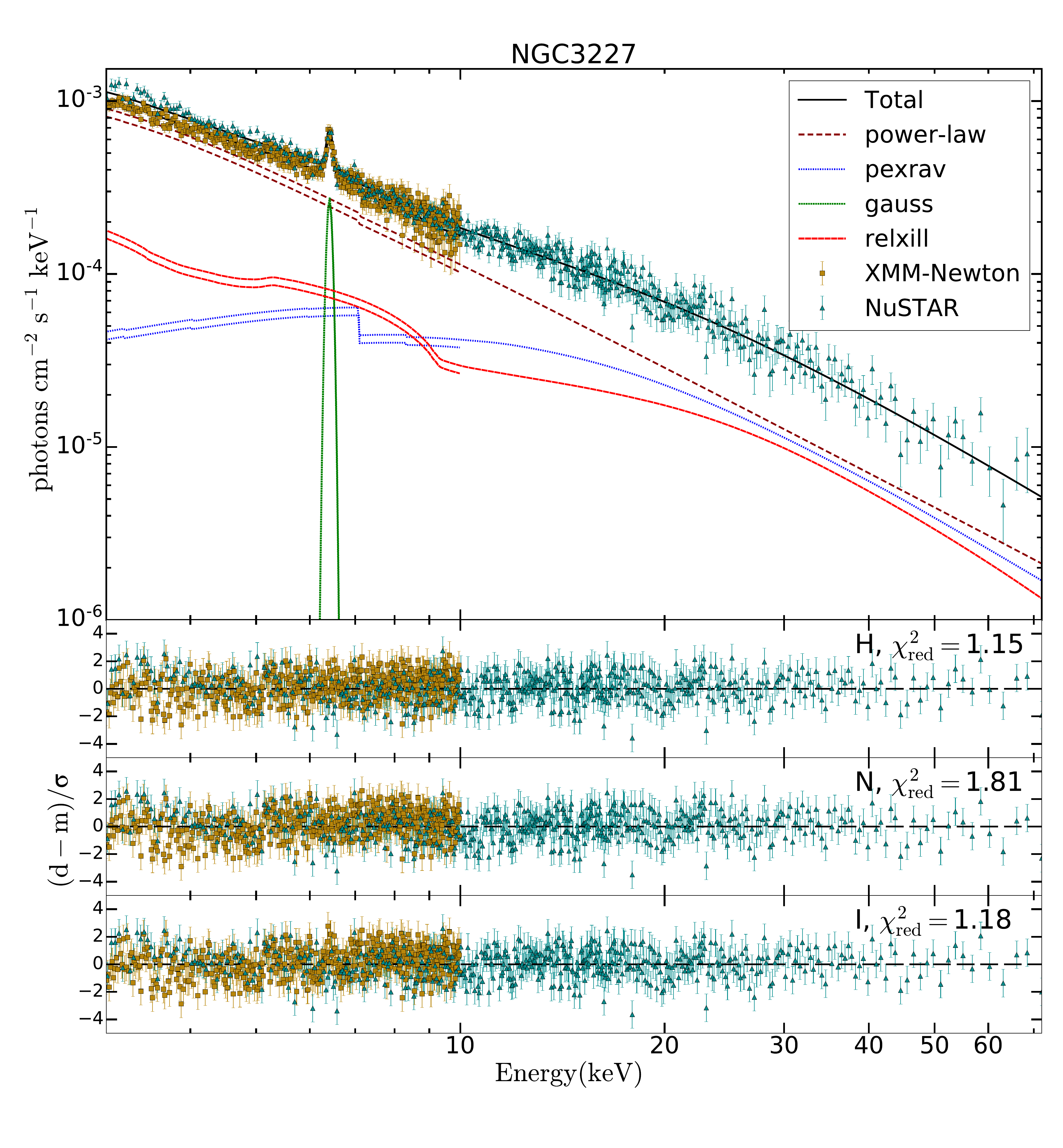} \\ 
\caption{Best fit to the 12 objects that prefer the hybrid model. We show the best fit (black solid line) to the data in the top panel and the ratio between model and data of the hybrid, neutral and ionized models (marked with letters H, N, and I respectively) in the bottom panels. Also, we show the reduced $\chi^2$ of the corresponding fit.}
\label{fig:hard_band_fit_HYBRID}
\end{center}
\end{figure*}

\begin{Contfigure*}[!t]
\begin{center}
\includegraphics[width=0.69\columnwidth]{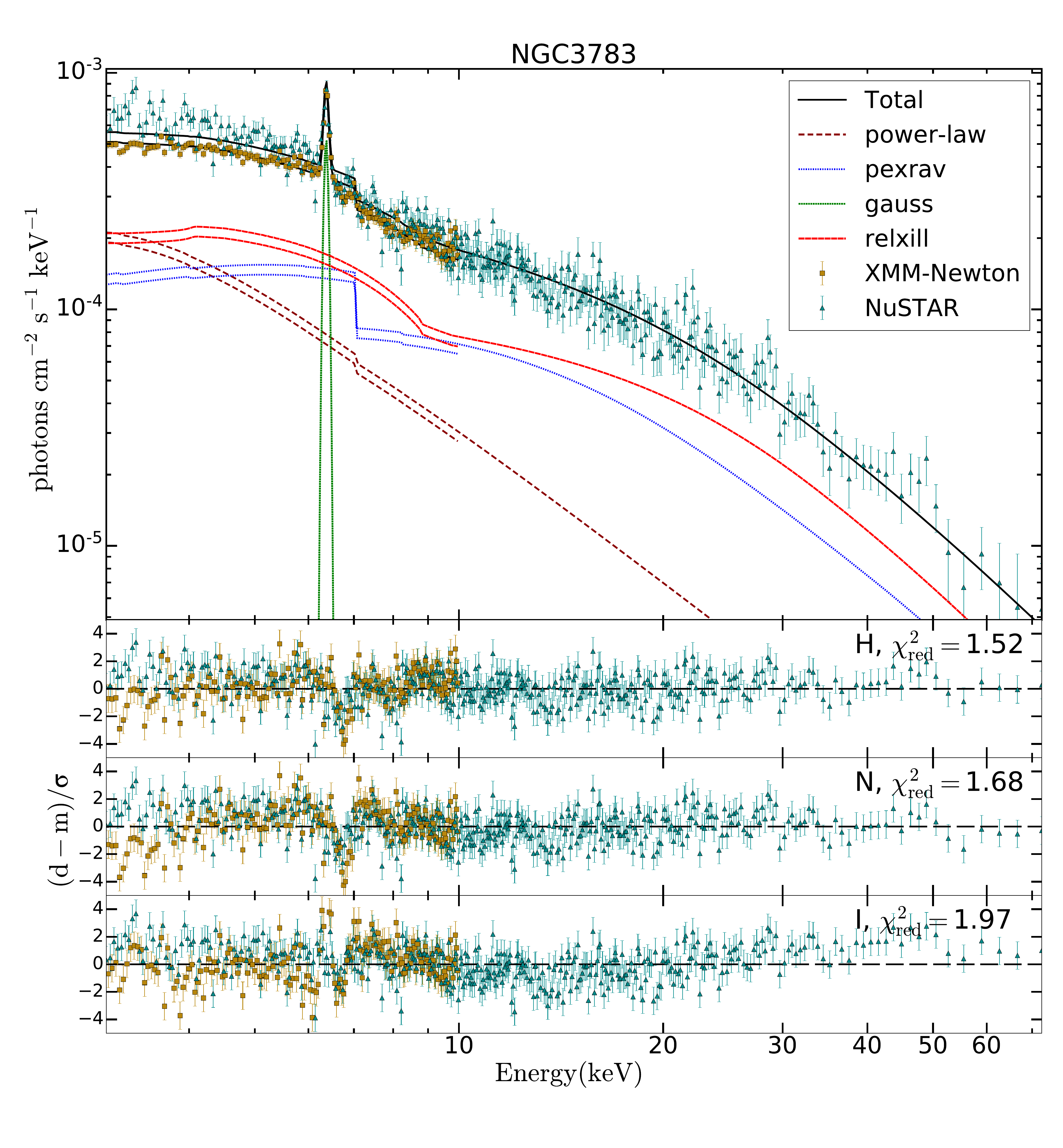}
\includegraphics[width=0.69\columnwidth]{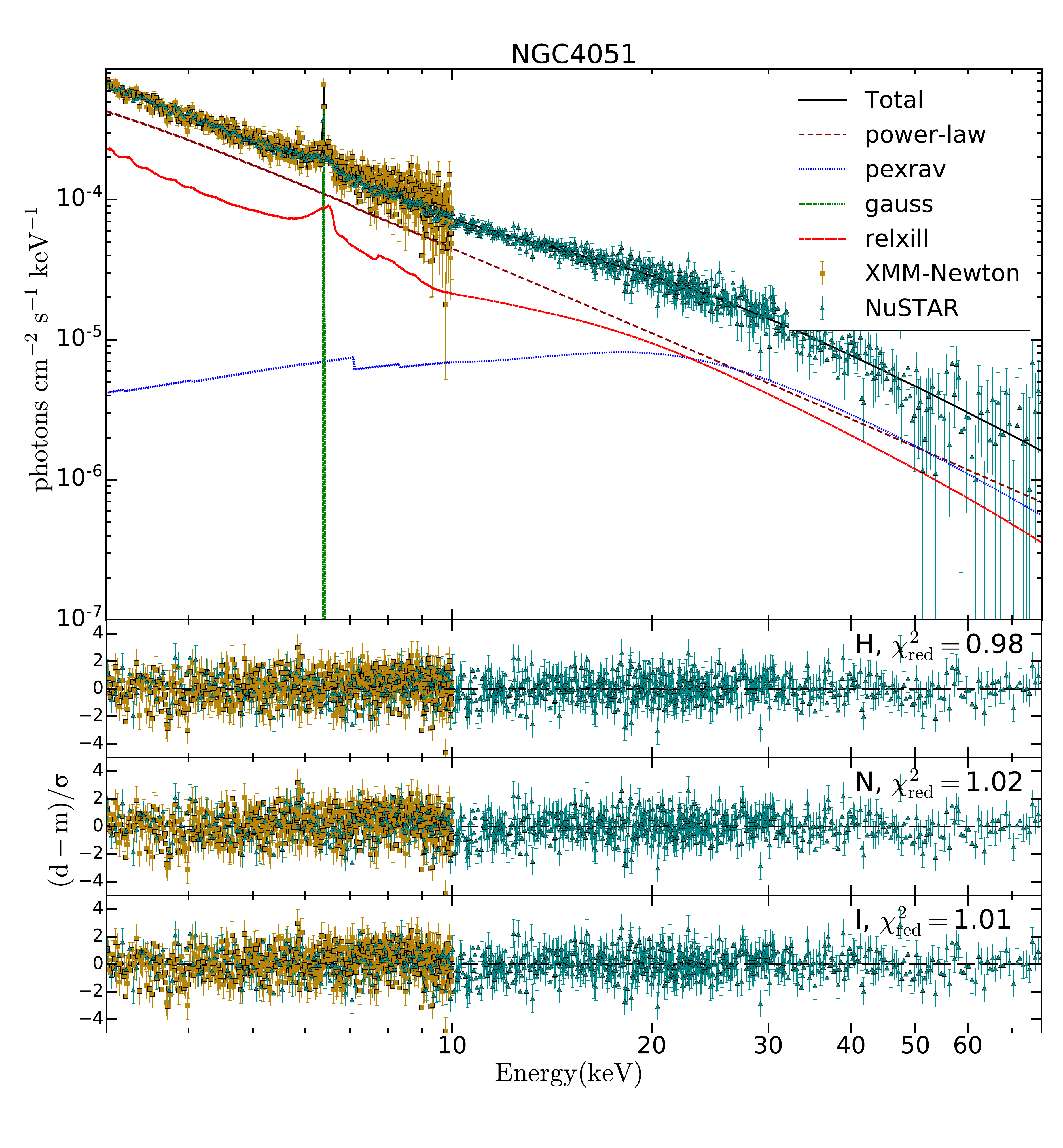}
\includegraphics[width=0.69\columnwidth]{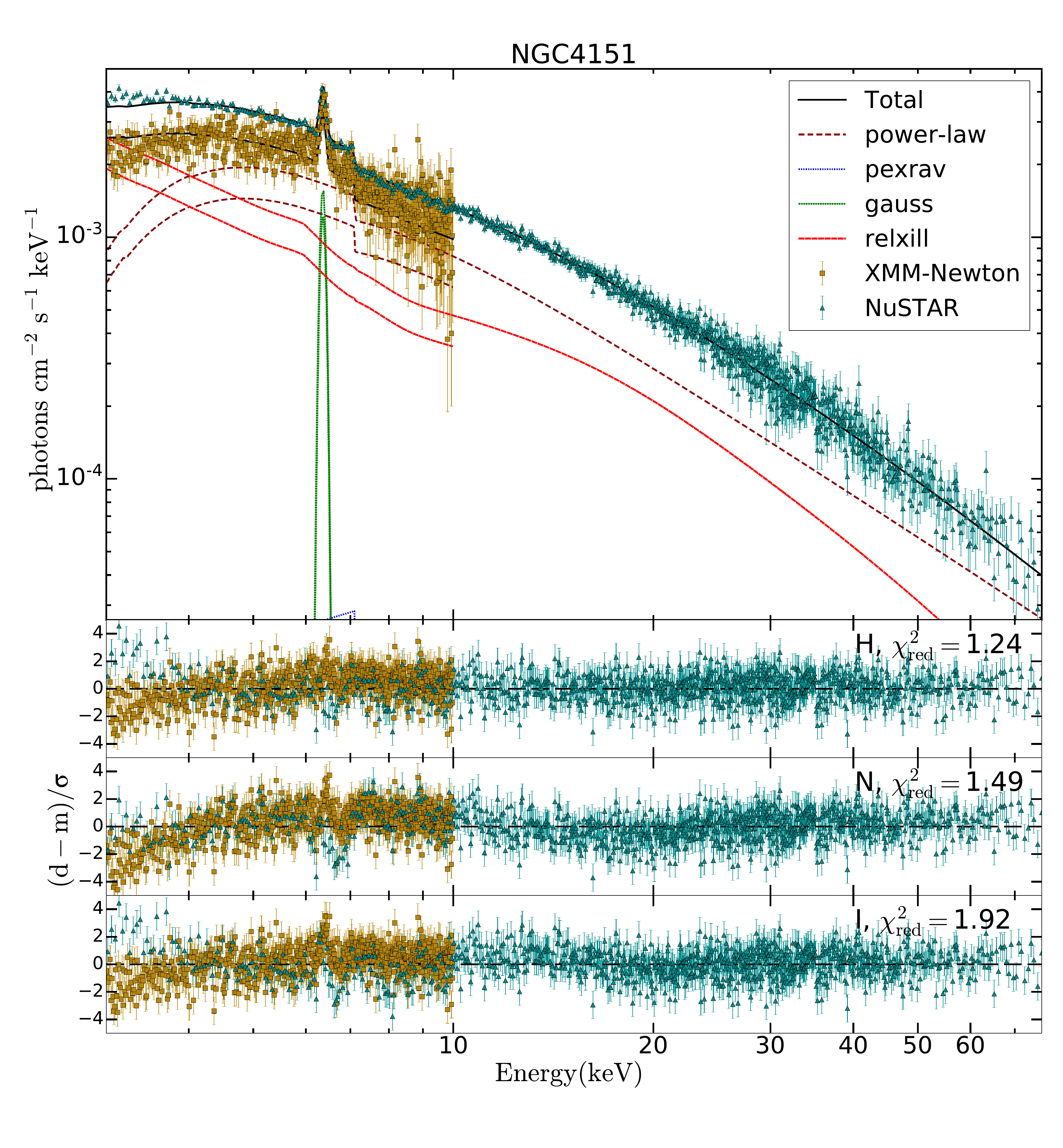} \\
\includegraphics[width=0.69\columnwidth]{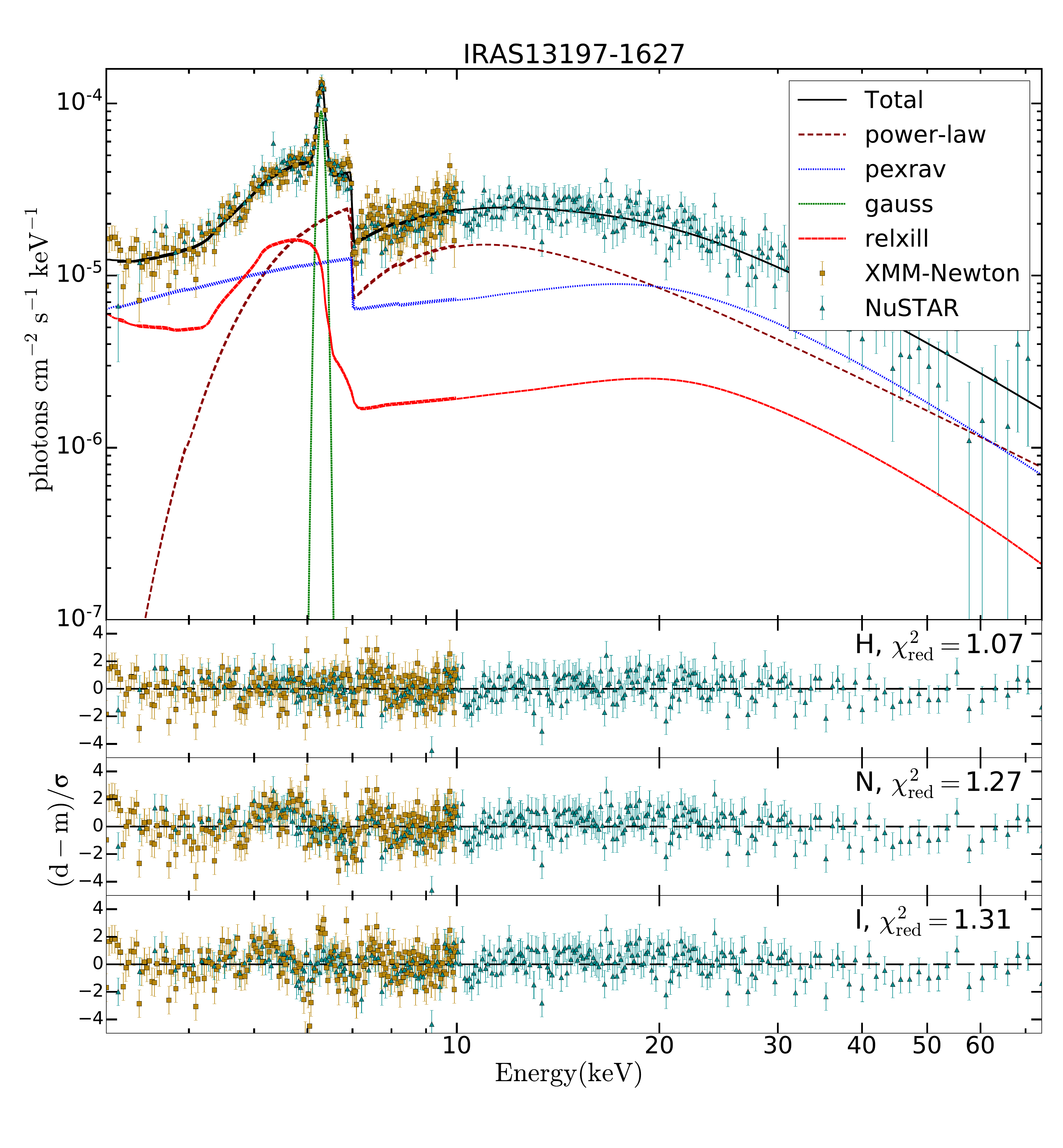}
\includegraphics[width=0.69\columnwidth]{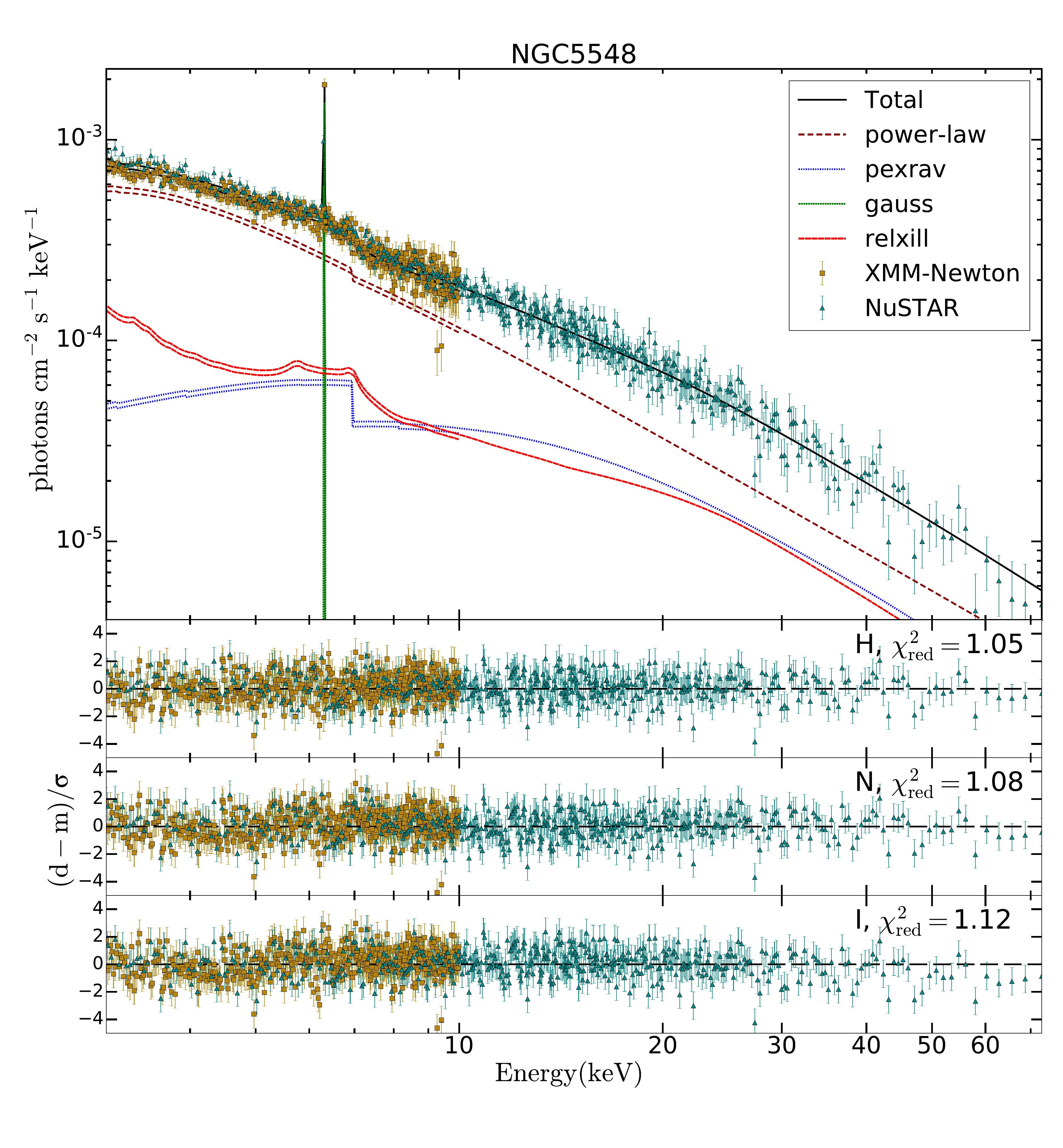}
\includegraphics[width=0.69\columnwidth]{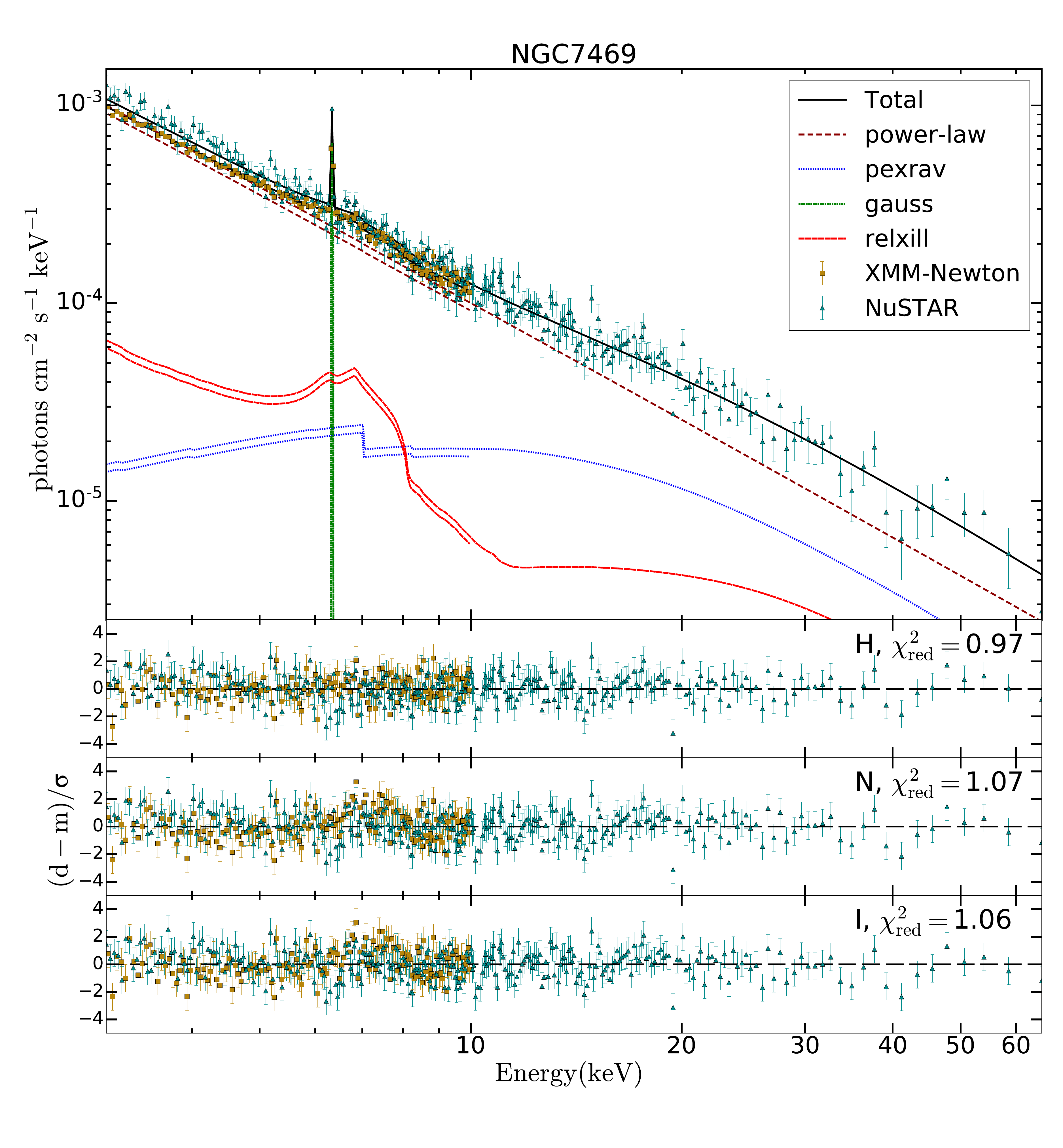}
\caption{}
%\label{fig:hard_band_fit_HYBRID}
\end{center}
\end{Contfigure*}

%\begin{Contfigure*}[!t]
%\begin{center}
%\includegraphics[width=0.74\columnwidth, clip, trim=10 20 0 70]{figuras/NGC3783_HYBRID_hard_final.pdf}
%\includegraphics[width=0.67\columnwidth, clip, trim=180 20 0 70]{figuras/NGC4051_HYBRID_hard_final.pdf}
%\includegraphics[width=0.67\columnwidth, clip, trim=178 20 0 70]{figuras/NGC4151_HYBRID_hard_final.pdf} \\
%\includegraphics[width=0.74\columnwidth, clip, trim=10 20 0 70]{figuras/IRAS13197-1627_HYBRID_hard_final.pdf}
%\includegraphics[width=0.67\columnwidth, clip, trim=180 20 0 70]{figuras/NGC5548_HYBRID_hard_final.pdf}
%\includegraphics[width=0.67\columnwidth, clip, trim=178 20 0 70]{figuras/NGC7469_HYBRID_hard_final.pdf}
%\caption{}
%%\label{fig:hard_band_fit_HYBRID}
%\end{center}
%\end{Contfigure*}

\subsection{Spectral fits for the sample}

Then, we used {\sc pexrav+gauss} and {\sc relxill} models for the 18 objects with evidence of the reflection component. Note that both {\sc pexrav+gauss} and {\sc relxill} models has six free parameters, in addition to the normalization (see Table\,\ref{tab:models_parameters}). Furthermore, we also try a combined model that uses both neutral ({\sc pexrav+gauss}) and ionized relativistic reflector ({\sc relxill}), namelly {\sc hybrid} model ({\sc hybrid} = {\sc power-law} + {\sc pexrav+gauss} + {\sc relxill}). We test these three models in our sample.  In order to determine the best model, first we compare between the two simplest models ({\sc pexrav+gauss} and {\sc relxill}) by using the AIC. Then, we compare the best model given by AIC and the more complex model ({\sc hybrid}) by using the F-test. If the F-test of the {\sc hybrid} model is \rm{$\leq 10^{-4}$} this model is the preferred model by the source. If not, the preferred model is the one indicated by AIC among {\sc pexrav+gauss} and {\sc relxill} models (marked with the number one). Finally, if the AIC of both models give \rm{$\leq 300$}, both models are equally preferred by the analysis. Table\,\ref{tab:models_sample} shows the statistics obtained.

We find that 12 out of the 18 objects (Mrk\,335, Fairall\,9, Mrk\,1040, NGC\,1365, Ark\,120, NGC\,3227, NGC\,3783, NGC\,4051, NGC\,4151, IRAS\,13197-1627, NGC\,5548, and NGC\,7469) prefer the hybrid model (i.e. 67\%), one object (Mrk\,915) prefer the neutral reflector (5\%), two objects (Mrk\,1044, and IGRJ\,19378-0617) prefer the ionized model (11\%), and three objects (Mrk\,766, NGC\,4593, and MCG\,-06-30-15) equally prefer both the neutral and ionized reflection models (17\%).

Figure\,\ref{fig:hard_band_fit_HYBRID} shows the best fit of the 12 objects that prefer the hybrid model, also the residuals of the three models in order to compare them. Residuals of Mrk\,335, Fairall\,9, and NGC\,7469 shows that the main improvement using the hybrid model is between 6-7 keV, through adding the ionized component. Residuals of hybrid model of NGC\,4151 and Ark\,120 shows a better fit above $\sim$ 10 keV and between 6-7 keV, compared to the neutral and ionized models. Note that residuals of NGC\,4151 around 6-7 keV are bigger in {\emph{NuSTAR}} data, possible due to the cross calibration between the satellites. Residuals of hybrid model of IRAS\,13197-1627 shows a better fit below $\sim$ 6 keV compared to the neutral and ionized models. NGC\,1365 and NGC\,3783 have the worst fits in the sample. Residuals of the three models of these two objects shows clear deficiencies of the fit between $\sim$ 6-8 keV. These two objects show possible emission lines below $\sim$ 8-9 keV, which were not considered in the baseline model. In order to obtain a better fit, emission lines are requested to be added to the baseline model. Interestingly, the $\rm{FeK\alpha}$ line is always fitted by the neutral component. Meanwhile, the ionized component is important to fit the edge around 7-9 keV and the range below $\sim$ 6 keV. The compton hump is fitted by the neutral component except for three sources (NGC\,1365, NGC\,3783, and NGC\,4151).

\begin{figure*}[!t]
\begin{center}
\includegraphics[width=1\columnwidth]{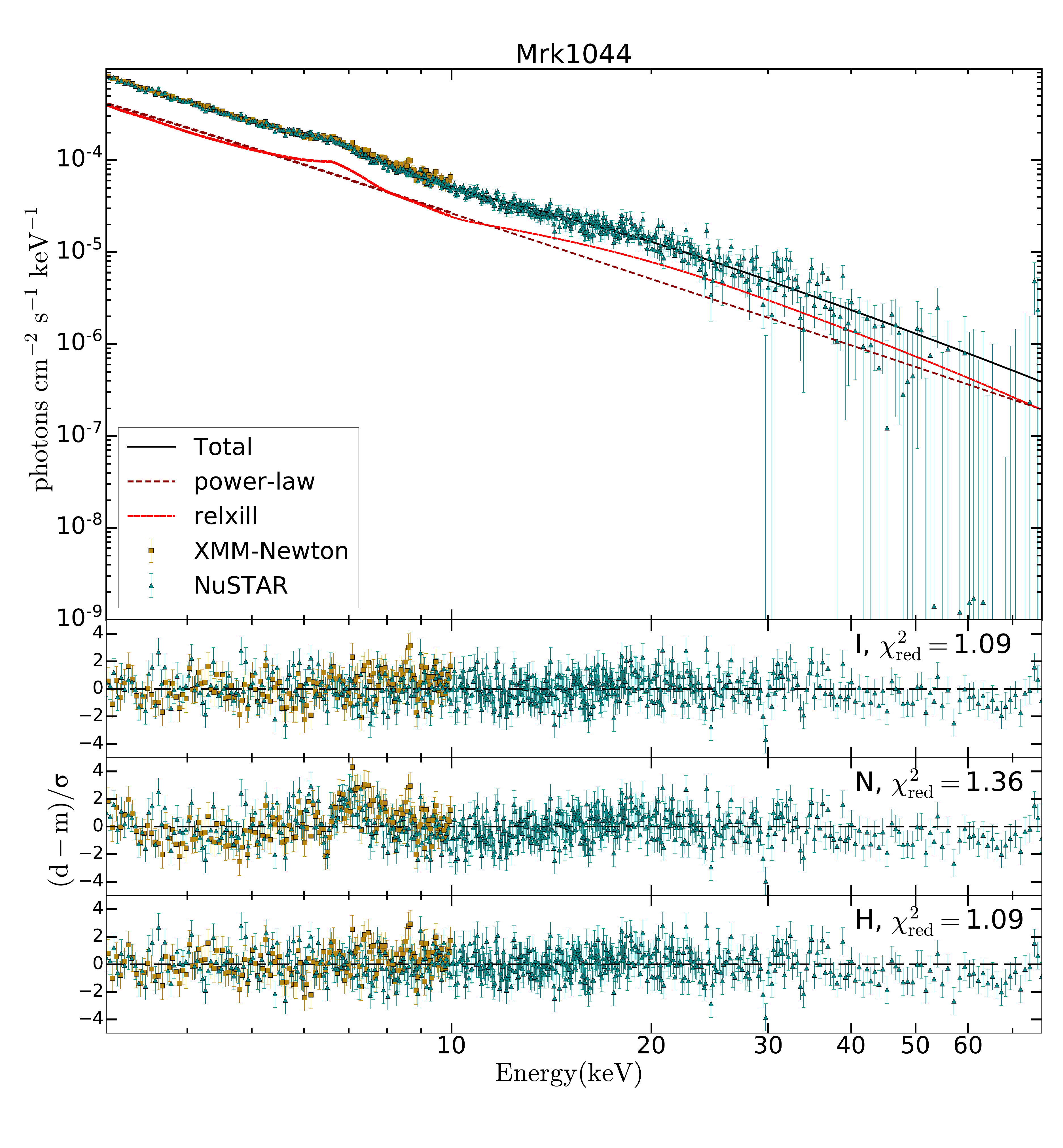}
\includegraphics[width=1\columnwidth]{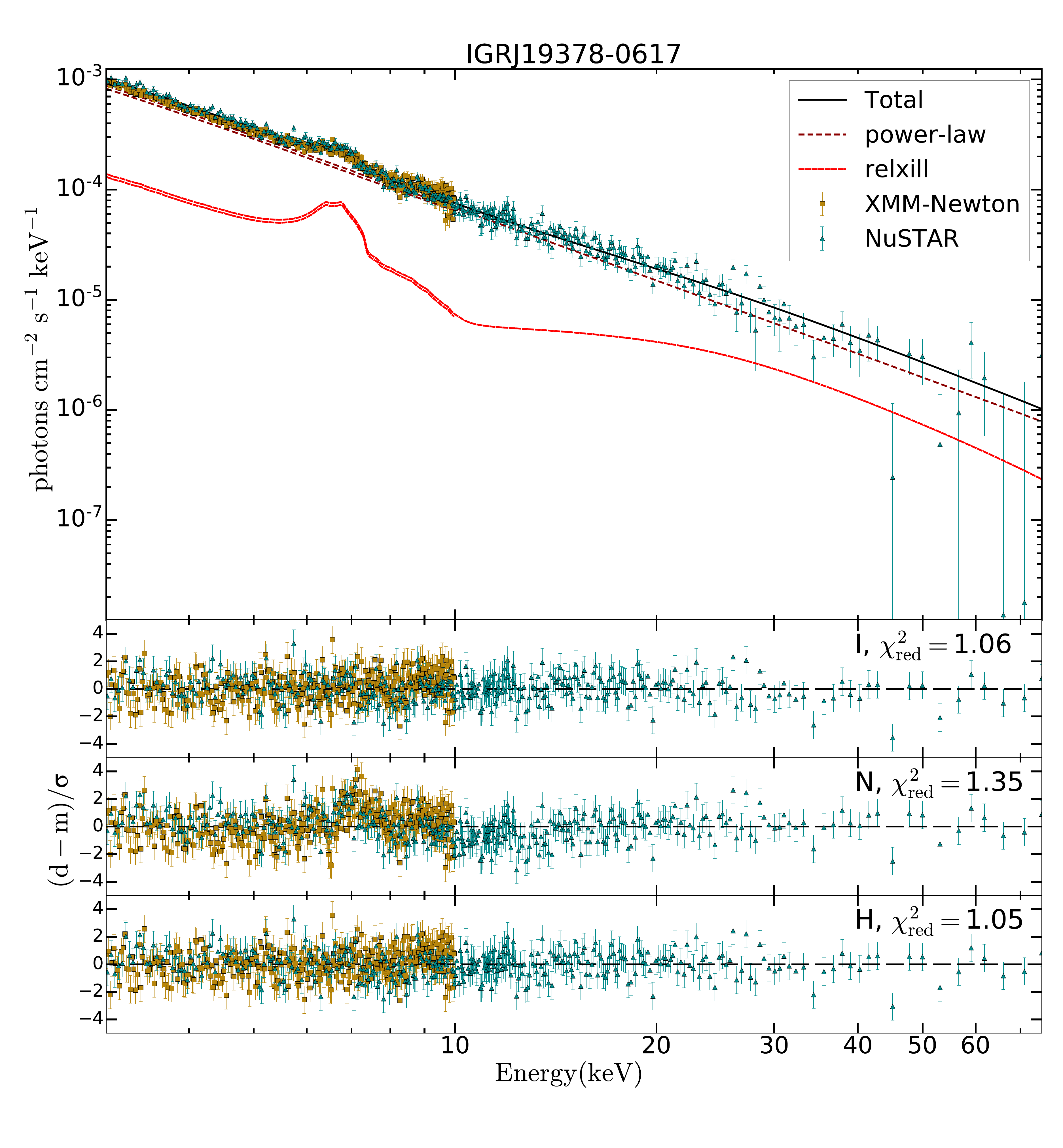}
\caption{Best fit to the two objects that prefer the ionized model. The description is the same as that given in Figure\,\ref{fig:hard_band_fit_HYBRID}.}
\label{fig:hard_band_fit_IONIZED}
\end{center}
\end{figure*}

\begin{figure}[!t]
\begin{center}
\includegraphics[width=0.95\columnwidth]{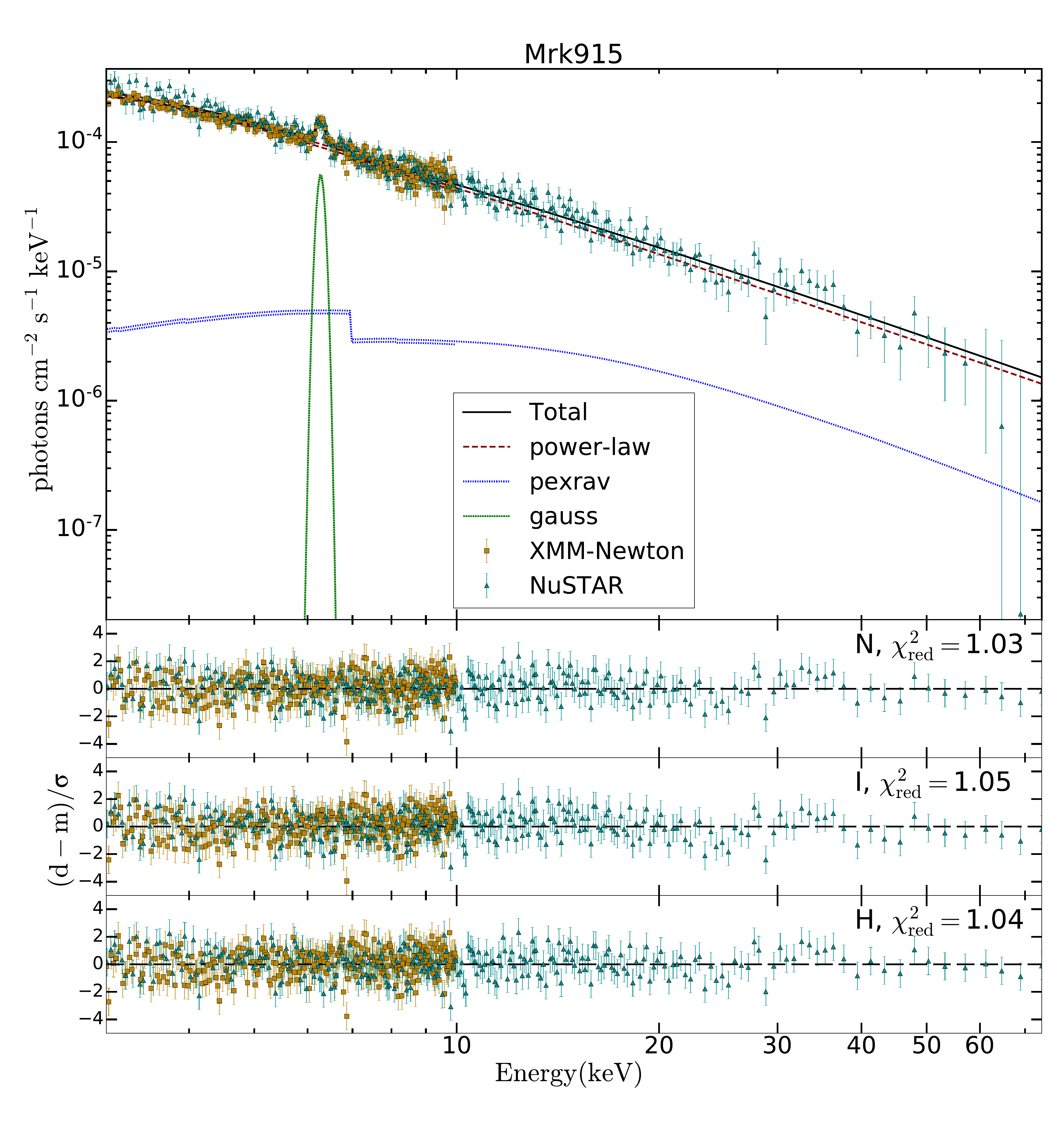}
\caption{Best fit to the object preferring the neutral model. The description is the same as that given in Figure\ref{fig:hard_band_fit_HYBRID}.}
\label{fig:hard_band_fit_NEUTRAL}
\end{center}
\end{figure}

We show in Figure\,\ref{fig:hard_band_fit_NEUTRAL} the spectrum and residuals of Mrk\,915, which is best fitted to the neutral model. Residuals of the three models in this object show minimal differences throughout the spectral range. Figure\,\ref{fig:hard_band_fit_IONIZED} shows the two objects that prefer the ionized model, namely Mrk\,1044 and IGRJ\,19378-0617. In both objects, the neutral model fails to reproduce the data between 6-8\,keV. Also, the ionized component reproduces well the $\rm{FeK\alpha}$ line.%, contrary to the objects that prefer the neutral model when it is fitted by the Gaussian. 

\begin{figure*}[!t]
\begin{center}
\includegraphics[width=1.05\columnwidth, clip, trim=10 885 0 72]{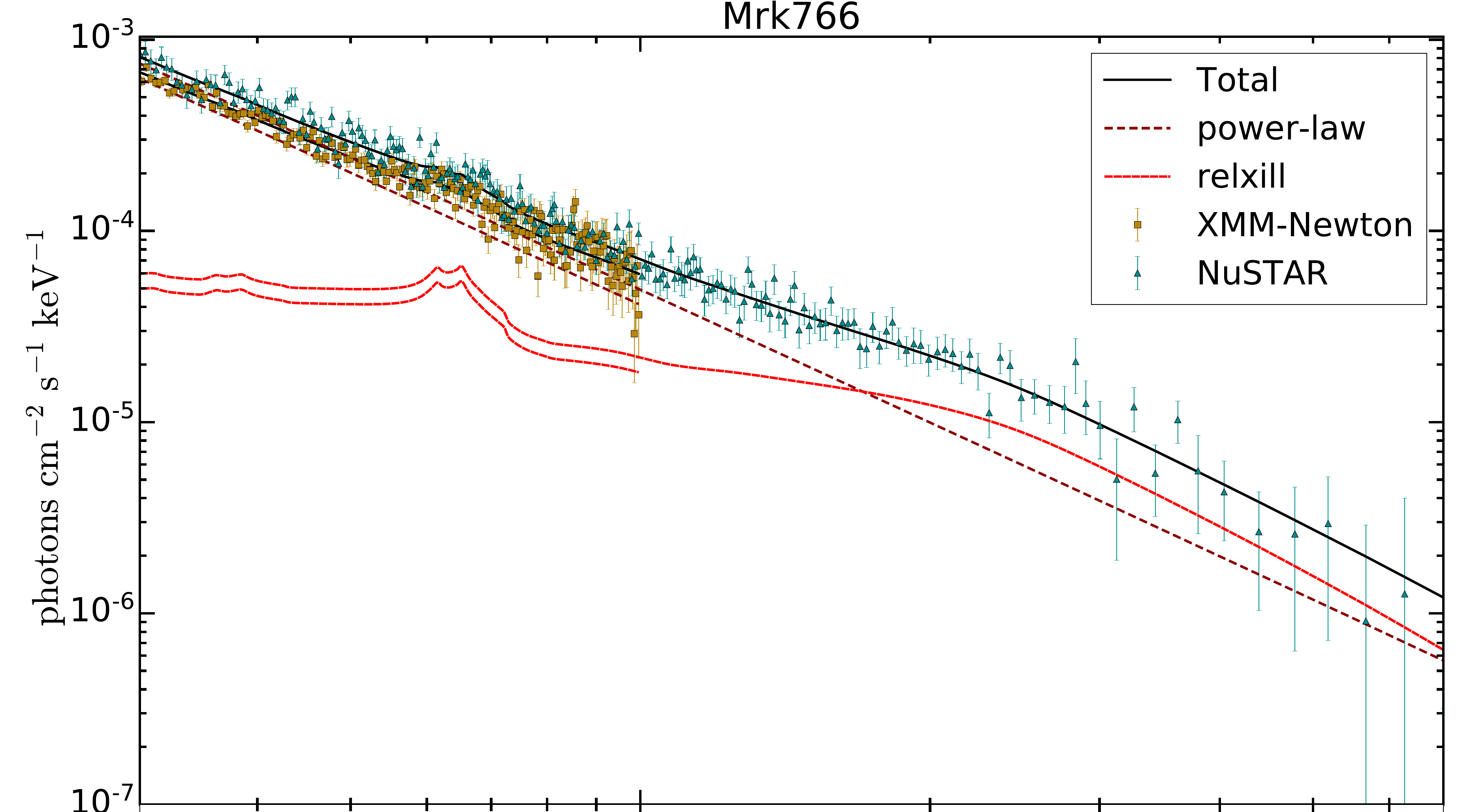}
\includegraphics[width=1.05\columnwidth, clip, trim=10 885 0 72]{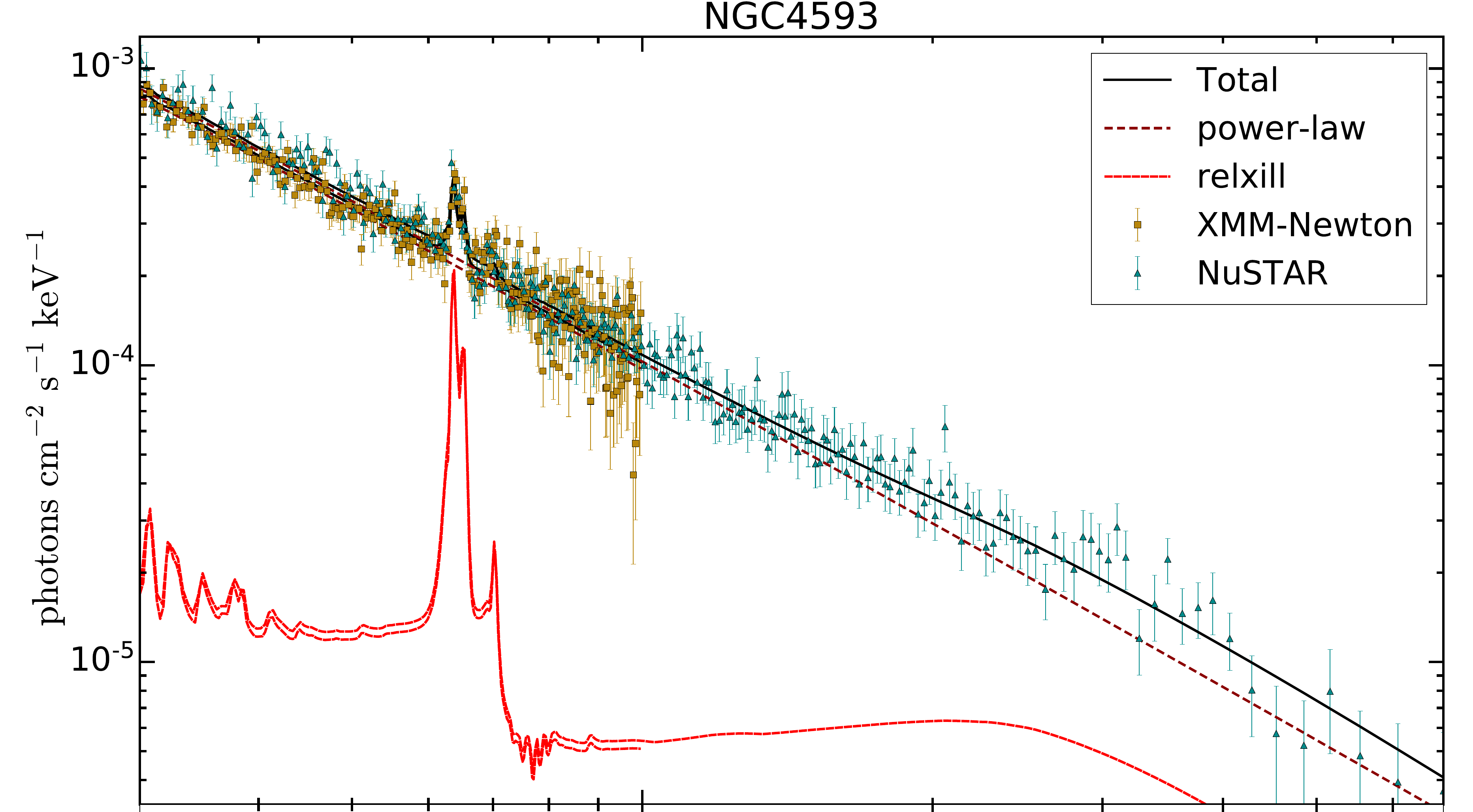}
\includegraphics[width=1.05\columnwidth, clip, trim=10 50 0 112]{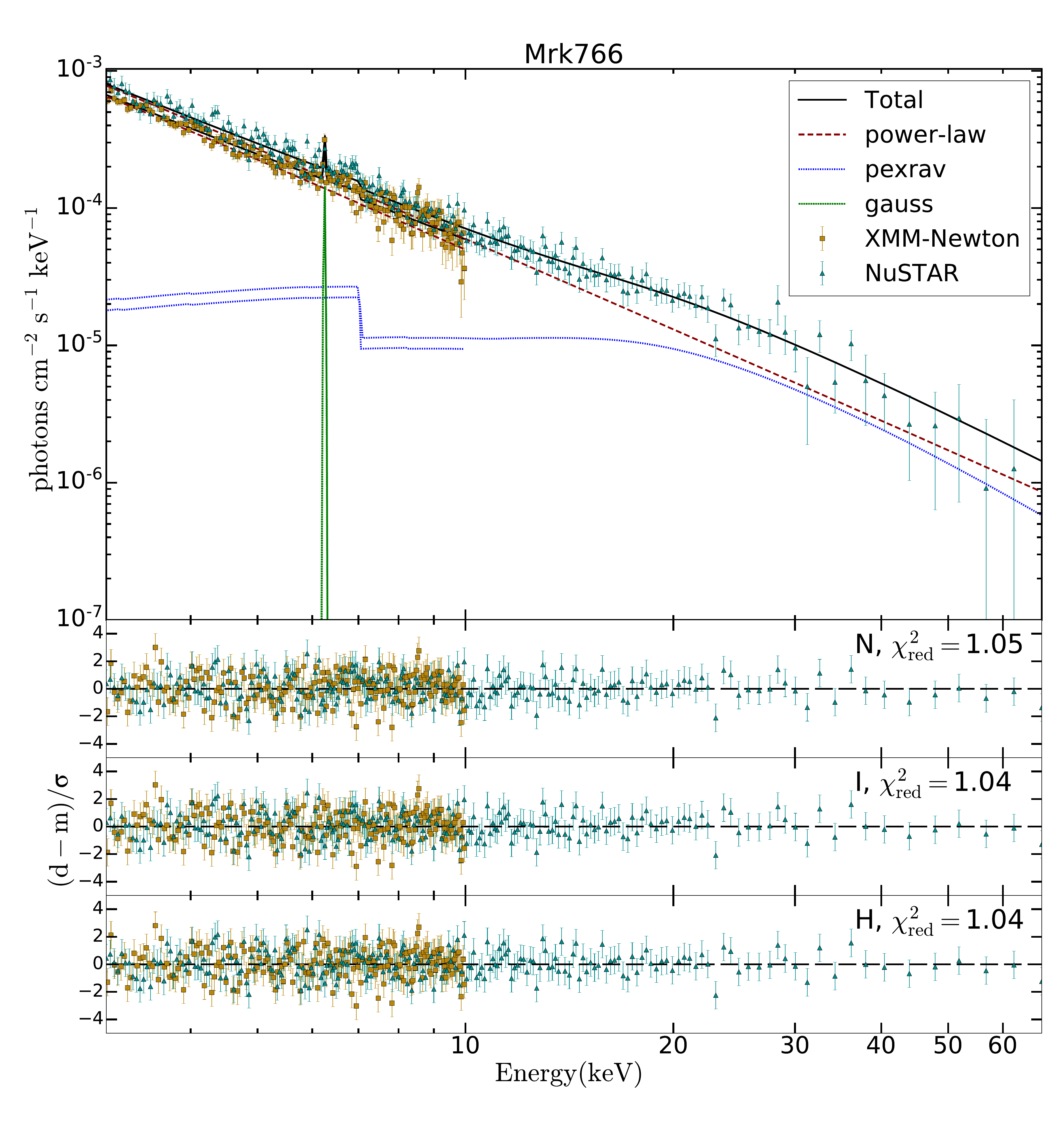}
\includegraphics[width=1.05\columnwidth, clip, trim=10 50 0 112]{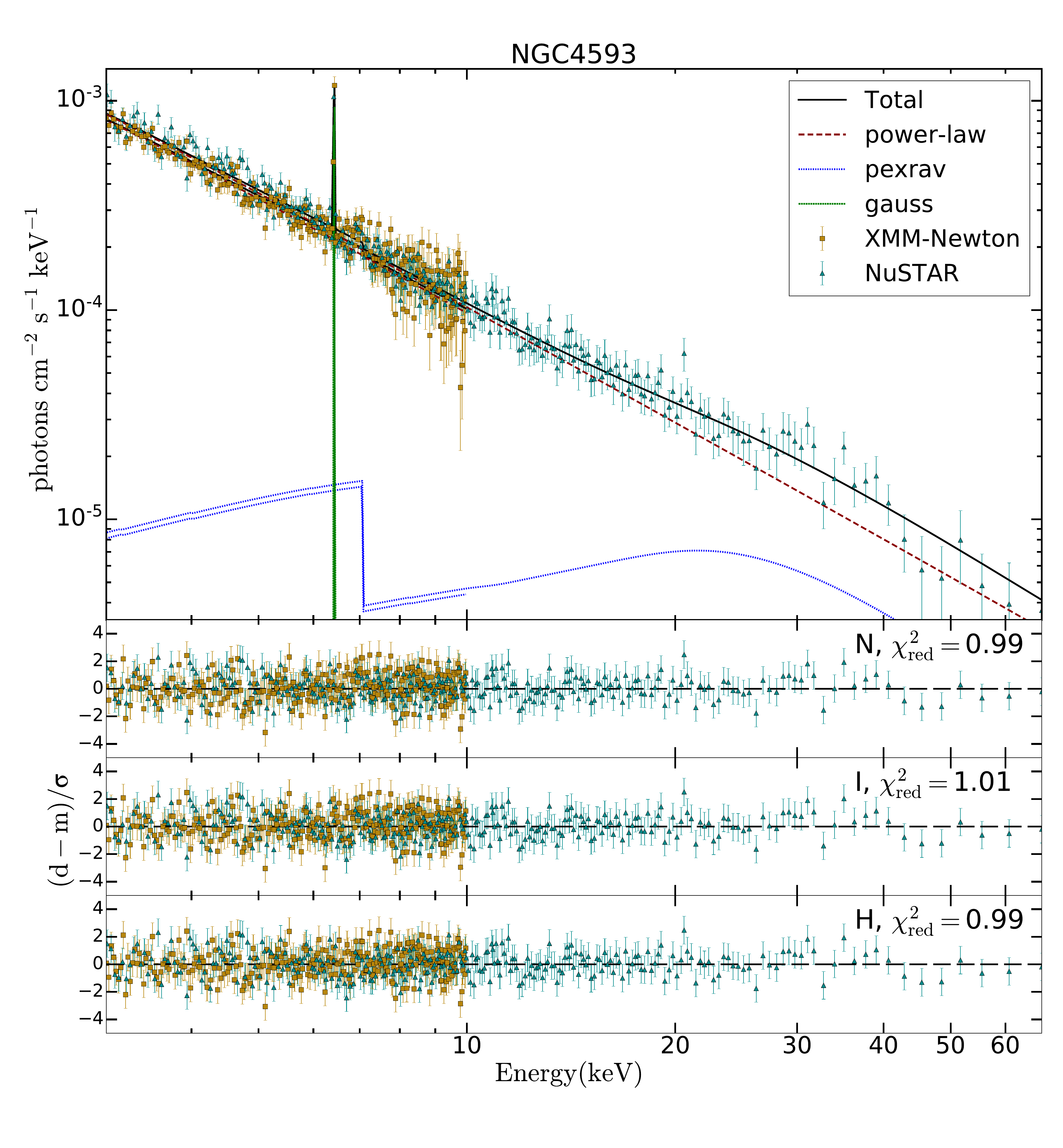}
\caption{Best fit to the three objects that equally prefer the neutral and ionized models. The description is the same as that given in Figure\,\ref{fig:hard_band_fit_HYBRID}.}
\label{fig:hard_band_fit_NEUTRAL_IONIZED}
\end{center}
\end{figure*}

Fits of Mrk\,766, NGC\,4593, and MCG\,-06-30-15 are shown in Figure\,\ref{fig:hard_band_fit_NEUTRAL_IONIZED}. According to F-statistics these objects equally prefer the neutral and ionized models. In both cases the compton hump is well fitted by the {\sc pexrav+gauss} and {\sc relxill} models. For Mrk\,766 the main difference is in the $\rm{FeK\alpha}$ line, which is fitted by a narrow Gaussian line in the neutral model, and with a broad line in the ionized model. For MCG\,-06-30-15, the $\rm{FeK\alpha}$ line is well fitted by a broad line in both models.  

%\begin{figure*}[!t]
%\begin{center}
%\includegraphics[width=1.05\columnwidth, clip, trim=10 50 0 72]{figuras/Mrk1044_IONIZED_hard_final.pdf}
%\includegraphics[width=0.95\columnwidth, clip, trim=178 50 0 72]{figuras/IGRJ19378-0617_IONIZED_hard_final.pdf}
%\caption{Best fit to the two objects that prefer the ionized model. The description is the same as that given in figure \ref{fig:hard_band_fit_HYBRID}.}
%\label{fig:hard_band_fit_IONIZED}
%\end{center}
%\end{figure*}

We show in Table\,\ref{tab:sample_params} the parameters obtained with the preferred model.
Column density and photon index are constrained for almost all objects. For the neutral component, the central energy of the $\rm{FeK\alpha}$ emission line is restricted in all but one object. The inclination angle is restricted for only one source. For the ionized component, ionization parameter is restricted for almost all sources. Iron abundance is the least constrained parameter for this component. Note that we found high Iron abundance for NGC\,3227 ($\rm{2.76\pm^{0.46}_{0.41}}$), NGC\,3783 ($\rm{1.84\pm^{0.12}_{0.09}}$), and NGC\,4593 ($\rm{2.75\pm^{1.51}_{1.06}}$). Similar results has been found previously for the last two sources. \cite{Brenneman11} and \cite{Ursini16} found Iron abundance of $\rm{3.7\pm^{0.9}_{0.9}}$ and $\rm{2.6\pm^{0.2}_{0.4}}$ for NGC\,3783 and NGC\,4593, respectively. Note also, that we do not restrict the iron abundance and inclination of {\sc pexrav+gauss} and {\sc relxill} in {\sc hybrid} model to the same value, since as they are different media (torus and accretion disk, respectively) they are not expected to have the same value.

\begin{Contfigure}[!t]
\begin{center}
\includegraphics[width=0.95\columnwidth, clip, trim=10 885 0 72]{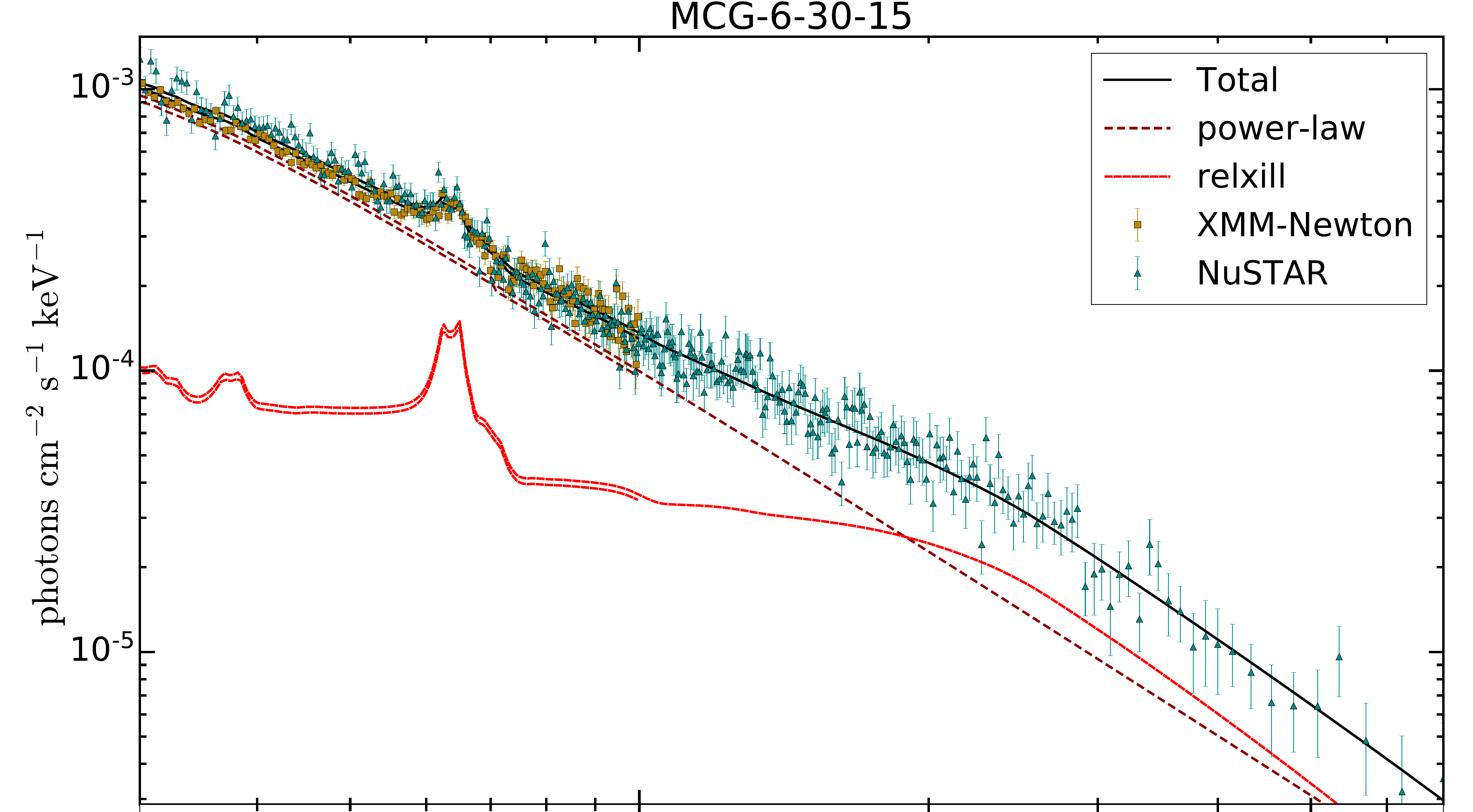}
\includegraphics[width=0.95\columnwidth, clip, trim=10 50 0 112]{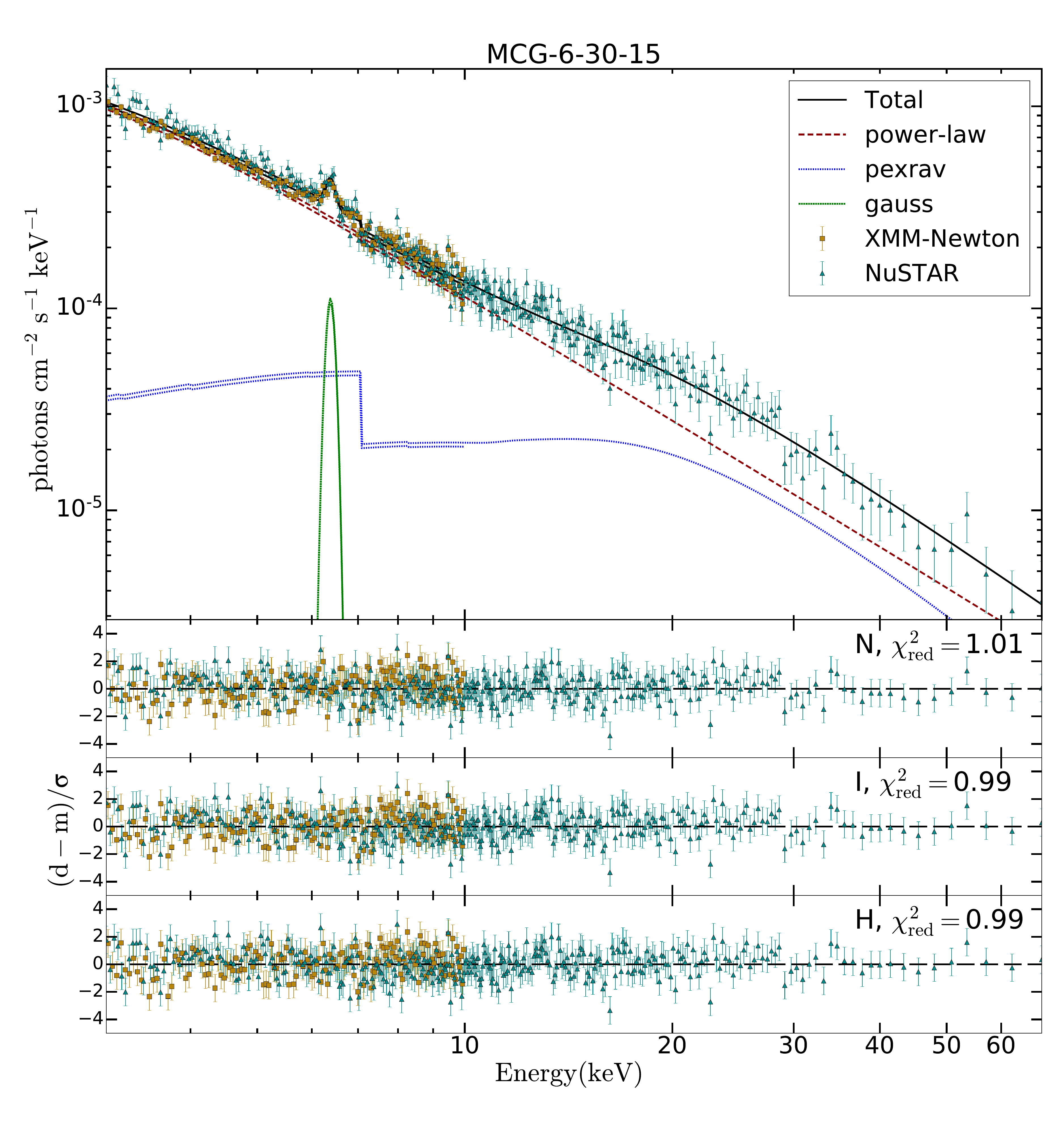}
\caption{}
%\label{fig:hard_band_fit_HYBRID}
\end{center}
\end{Contfigure}

\begin{figure*}[htp]
\begin{center}
\includegraphics[width=0.69\columnwidth, clip, trim=0 115 0 10]{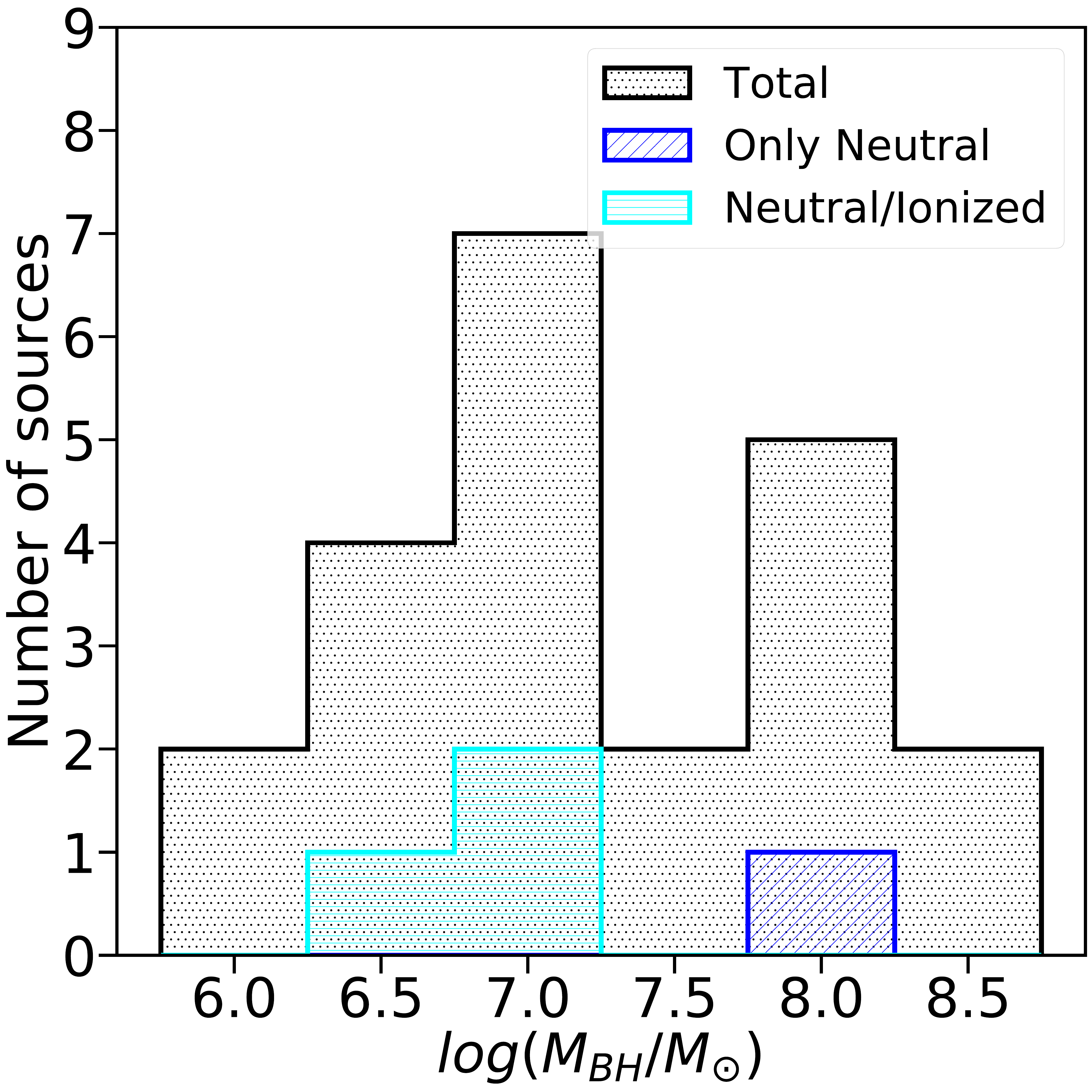} %\hspace{-0.3 cm}
\includegraphics[width=0.63\columnwidth, clip, trim=160 115 0 10]{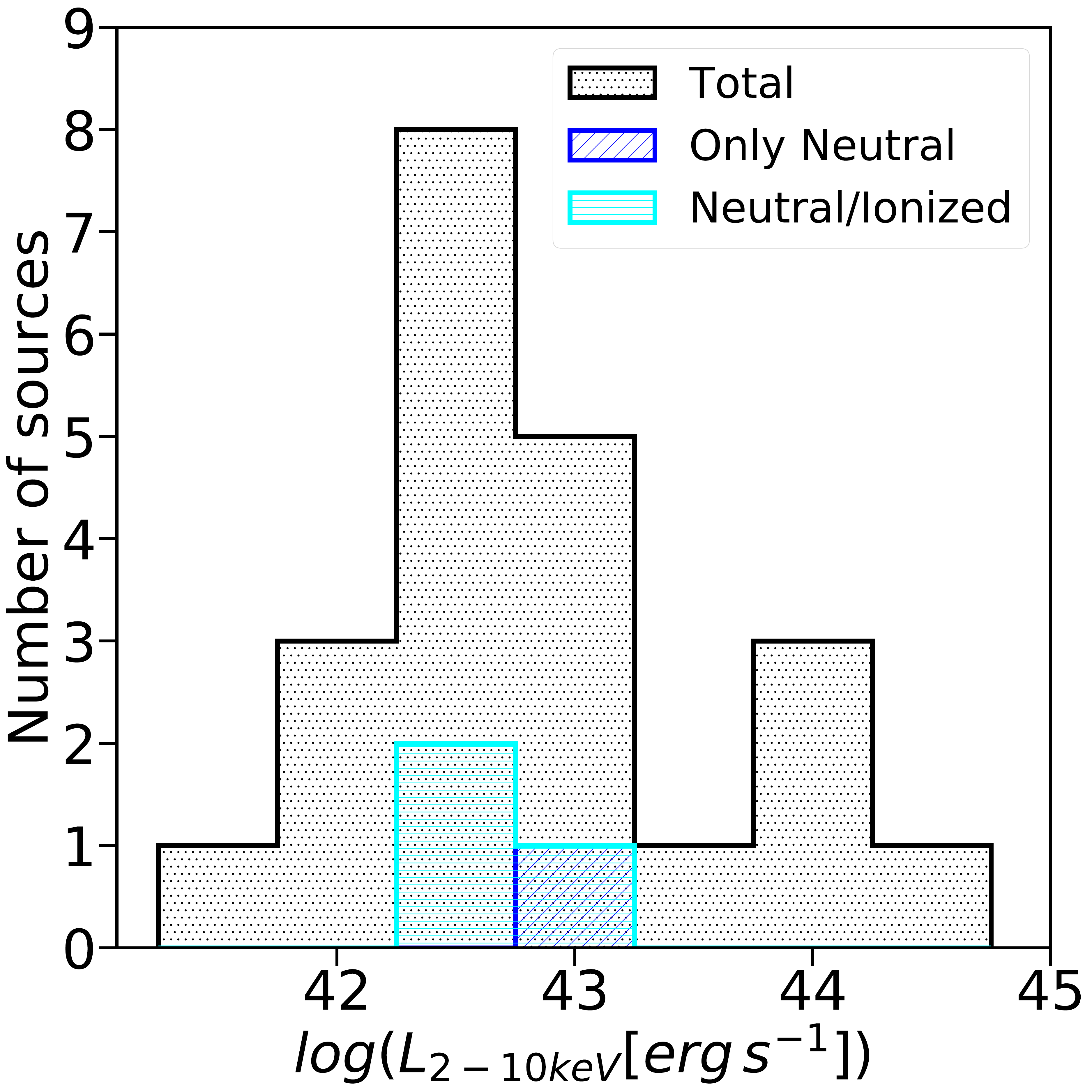} %\hspace{-0.6 cm}
\includegraphics[width=0.63\columnwidth, clip, trim=160 115 0 10]{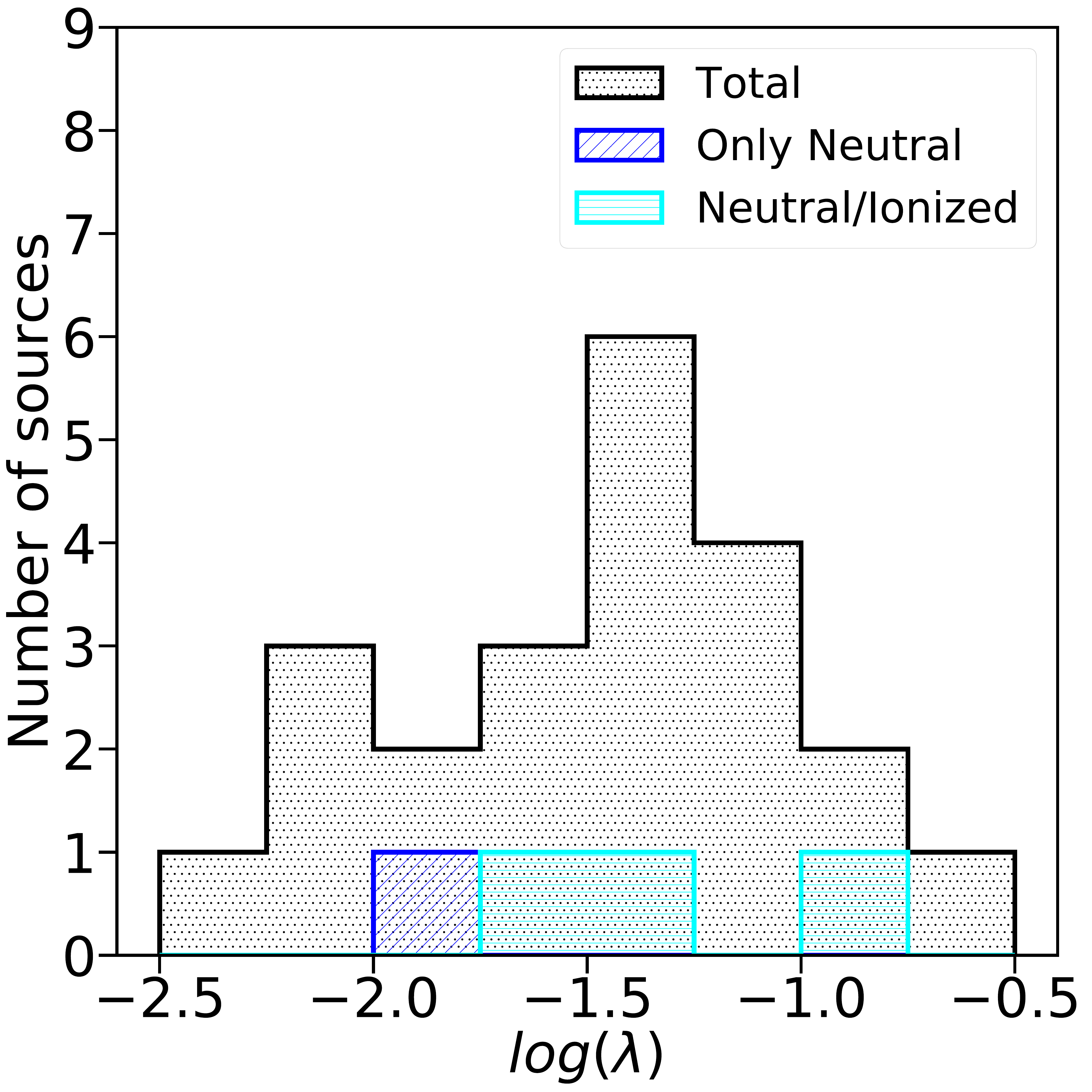} \\ %\hspace{-0.6 cm}
\includegraphics[width=0.69\columnwidth, clip, trim=0 115 0 10]{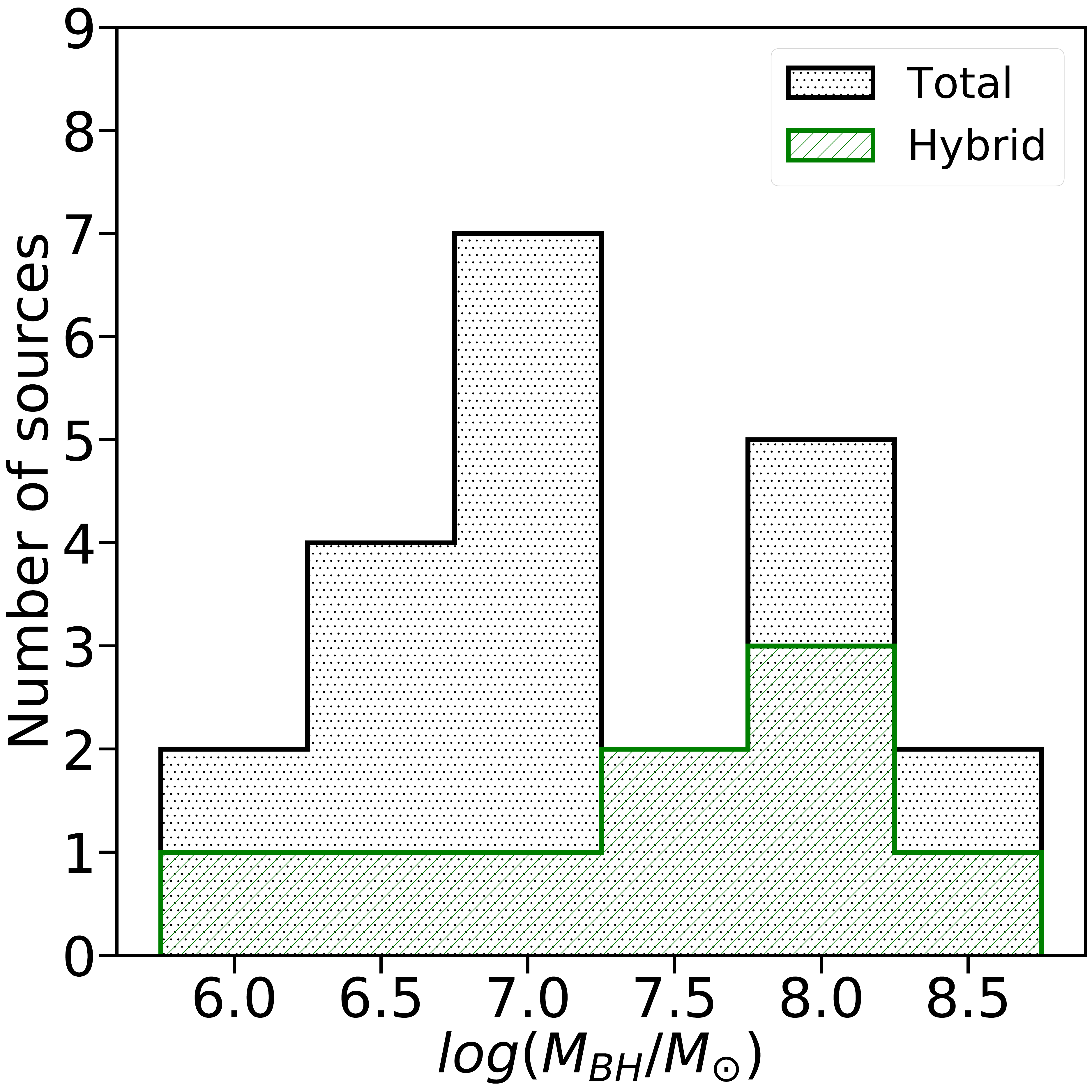} %\hspace{-0.3 cm}
\includegraphics[width=0.63\columnwidth, clip, trim=160 115 0 10]{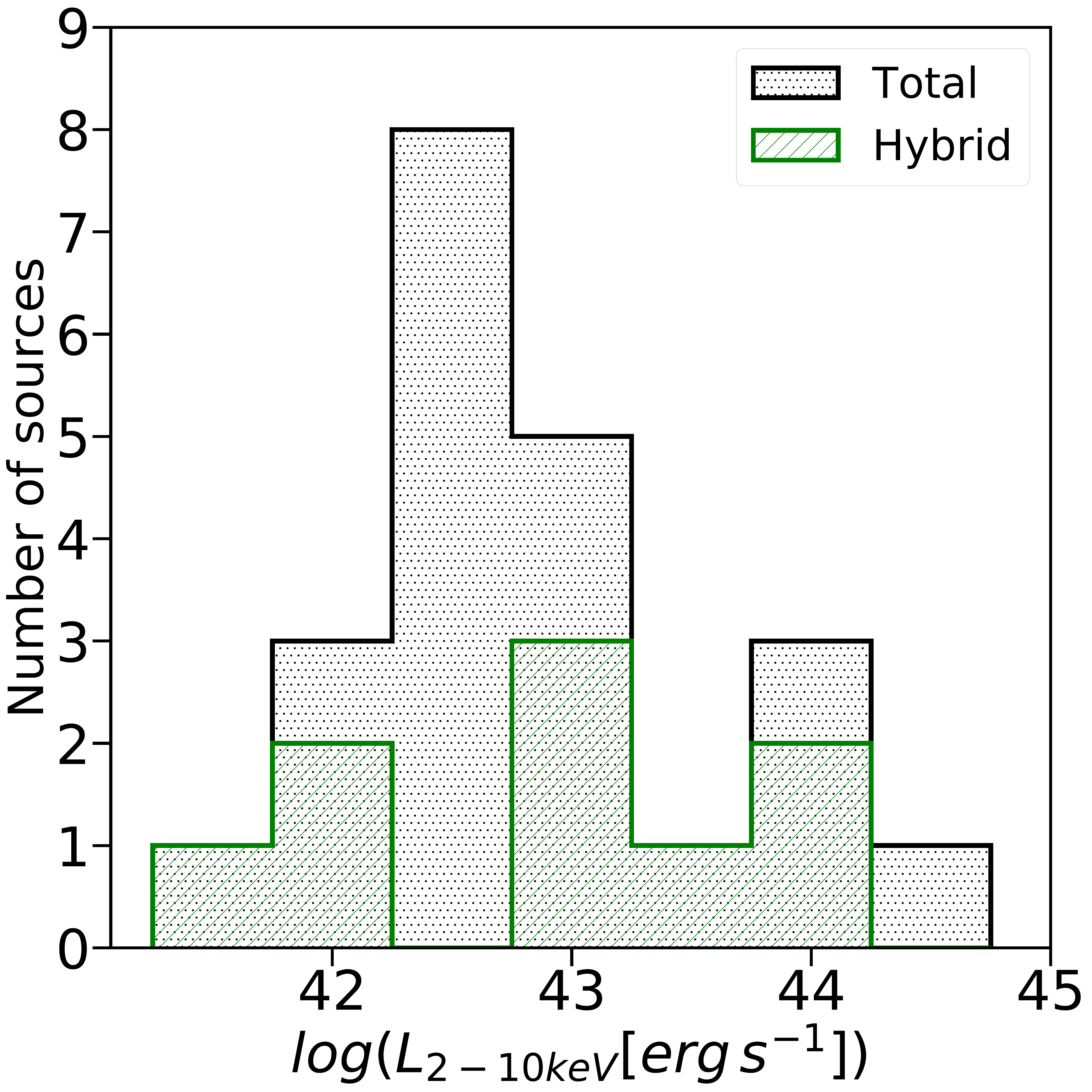} %\hspace{-0.6 cm}
\includegraphics[width=0.63\columnwidth, clip, trim=160 115 0 10]{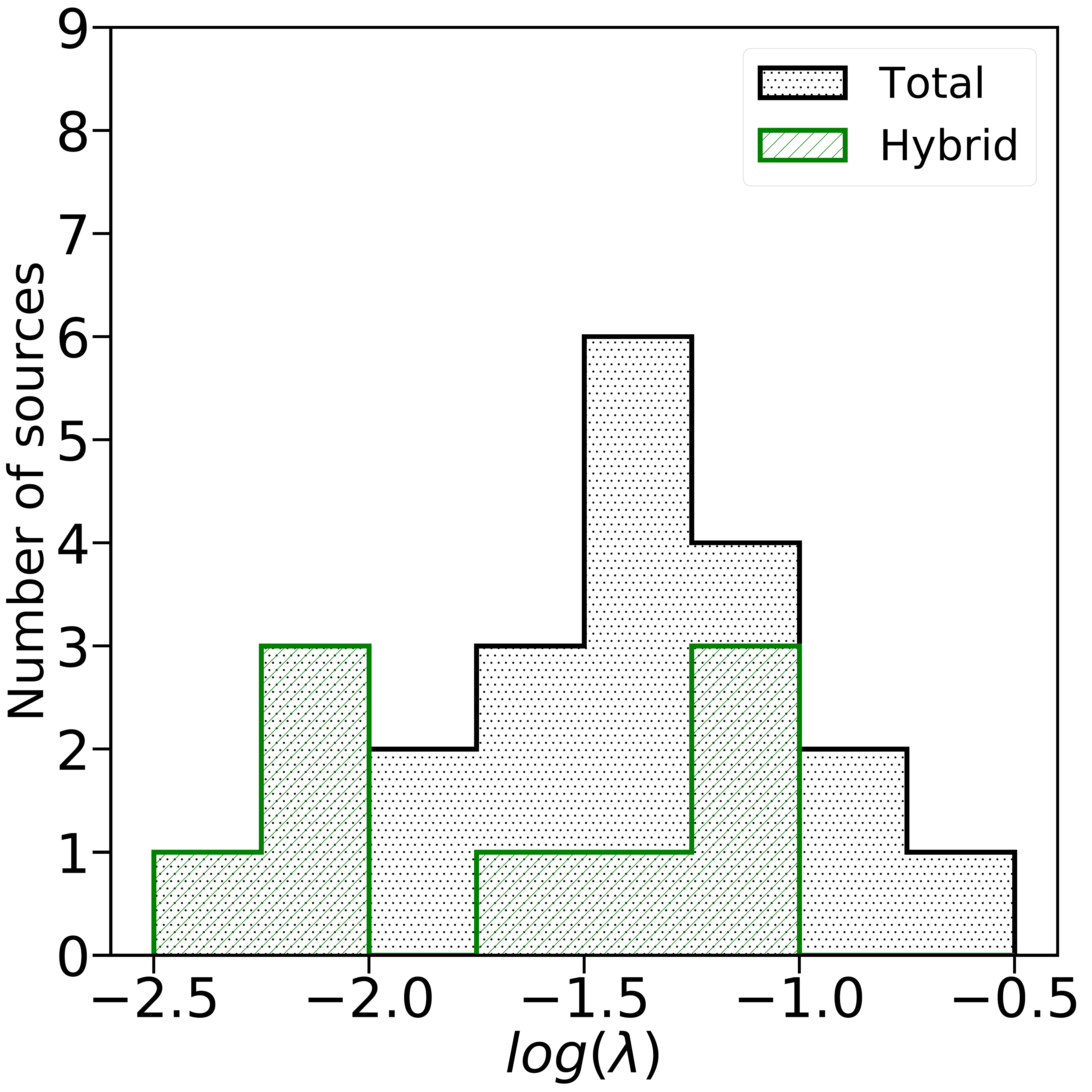} \\ %\hspace{-0.6 cm}
\includegraphics[width=0.69\columnwidth, clip, trim=0 115 0 10]{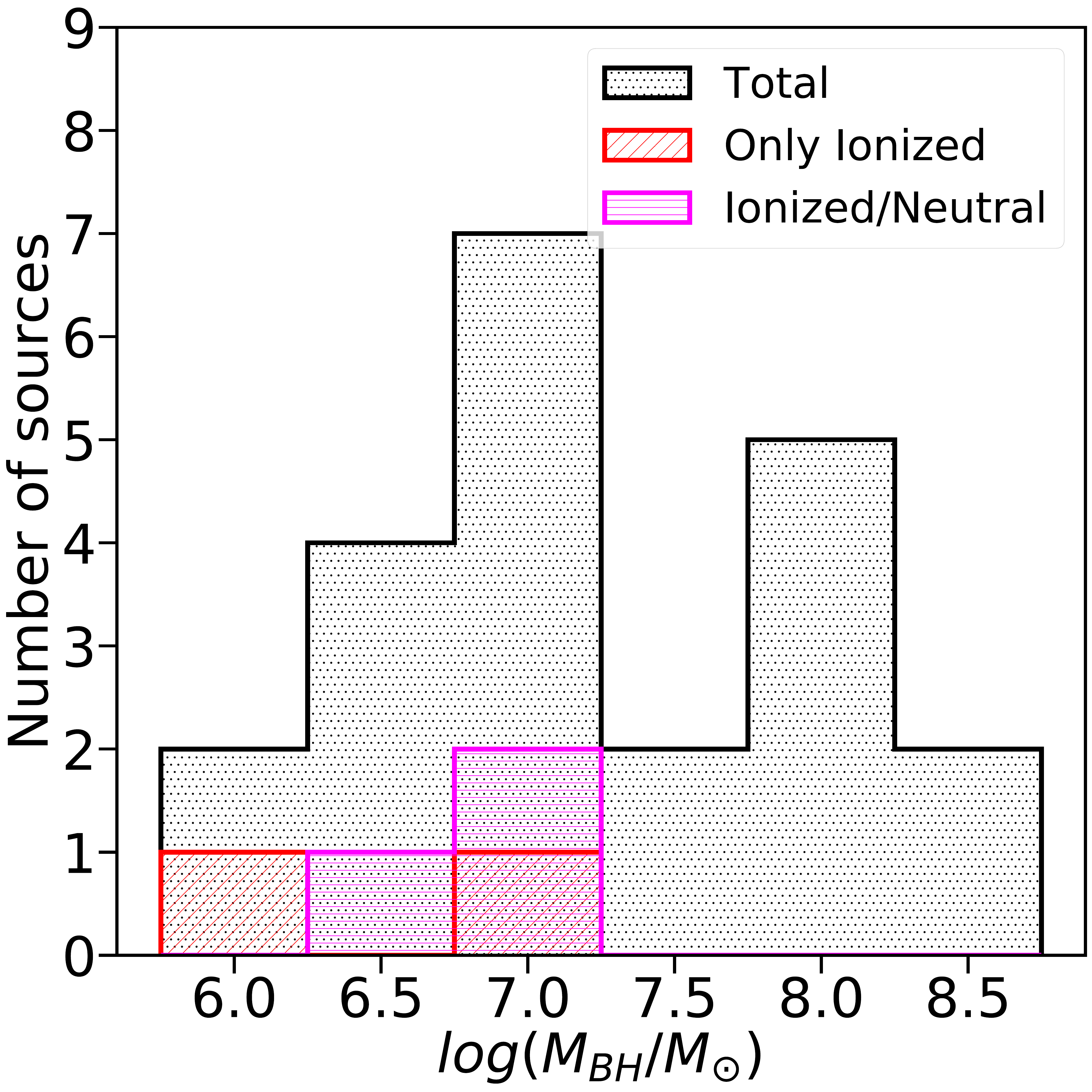} %\hspace{-0.3 cm}
\includegraphics[width=0.63\columnwidth, clip, trim=160 115 0 10]{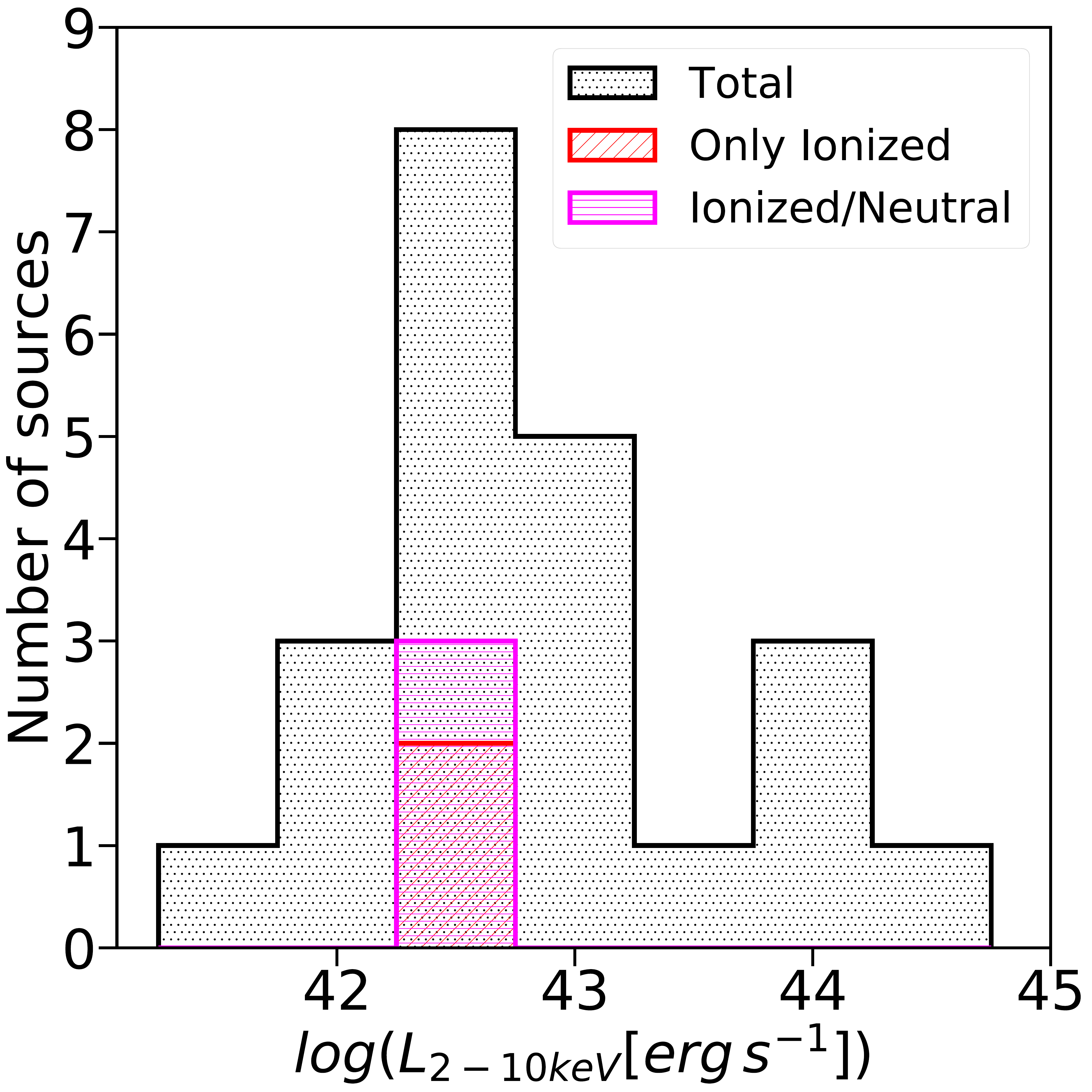} %\hspace{-0.6 cm}
\includegraphics[width=0.63\columnwidth, clip, trim=160 115 0 10]{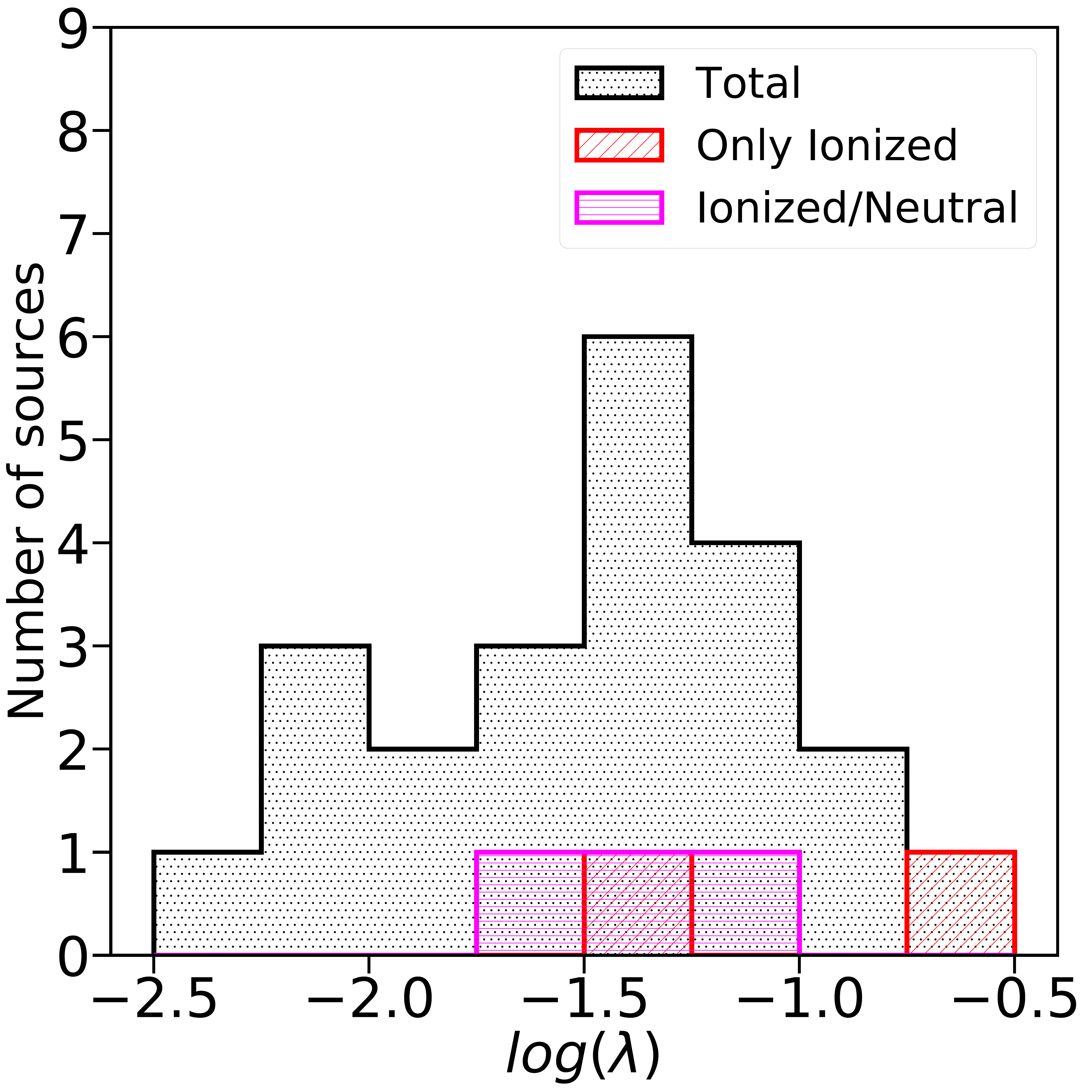} \\ %\hspace{-0.6 cm}
\includegraphics[width=0.69\columnwidth, clip, trim=0 0 0 10]{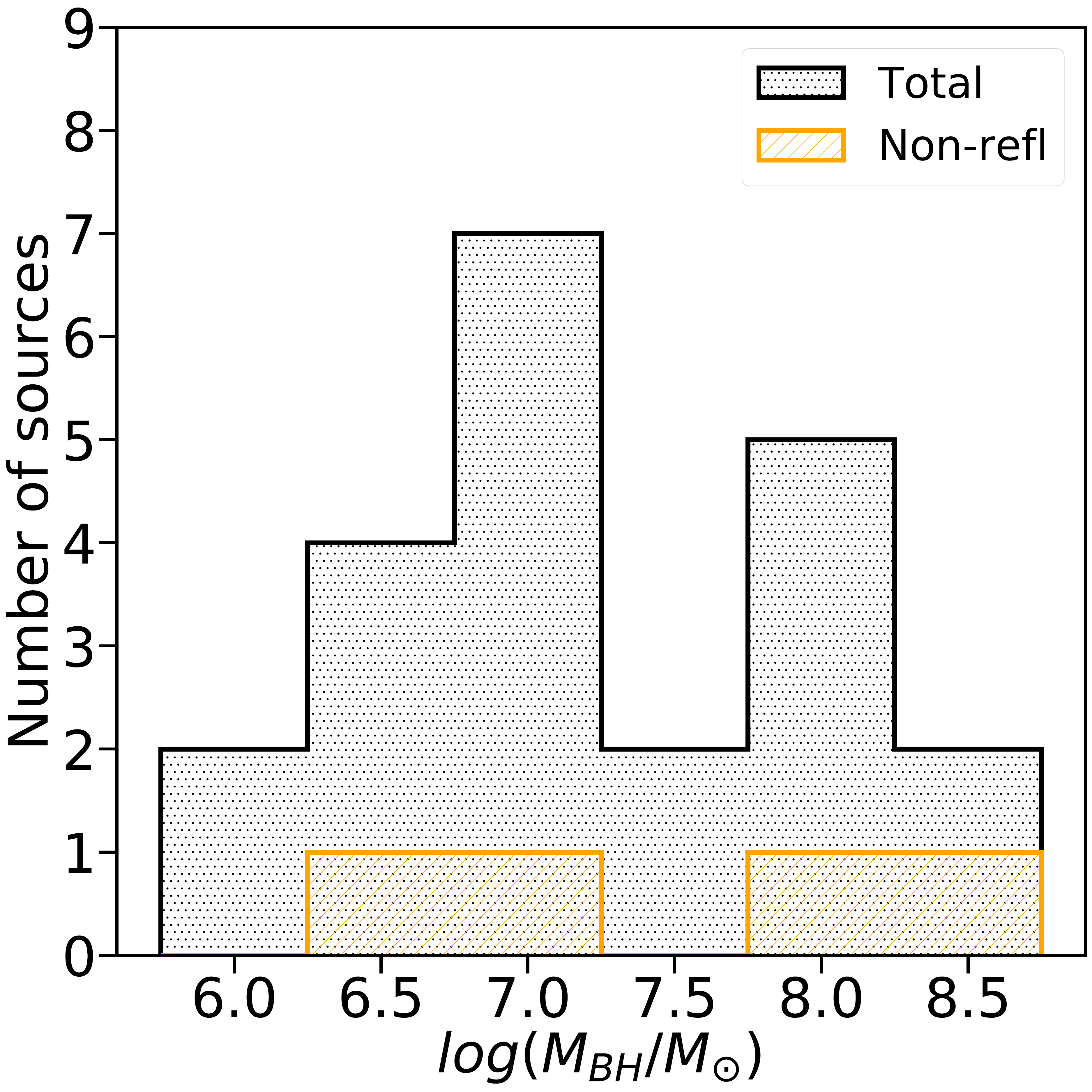}
\includegraphics[width=0.63\columnwidth, clip, trim=160 0 0 10]{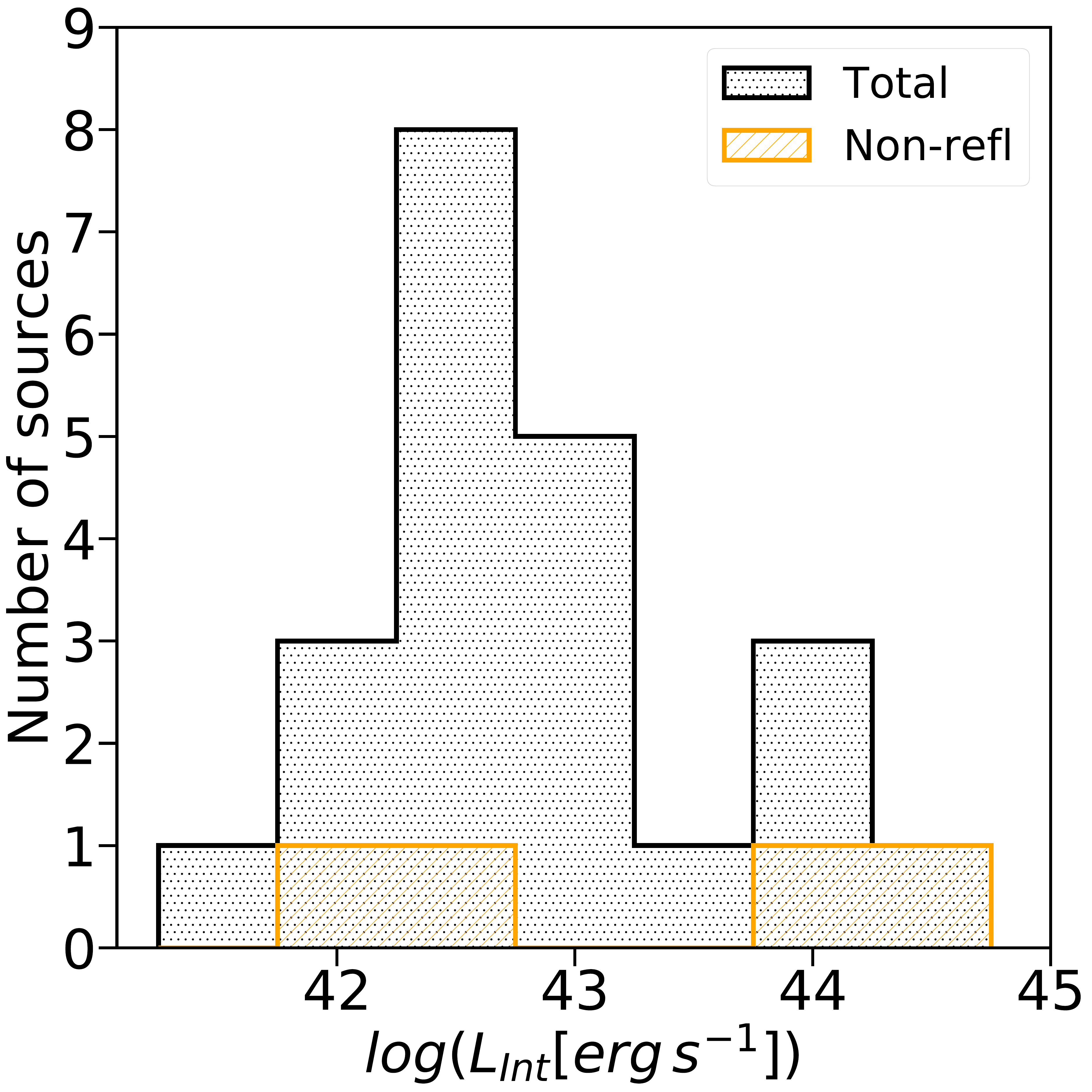}
\includegraphics[width=0.63\columnwidth, clip, trim=160 0 0 10]{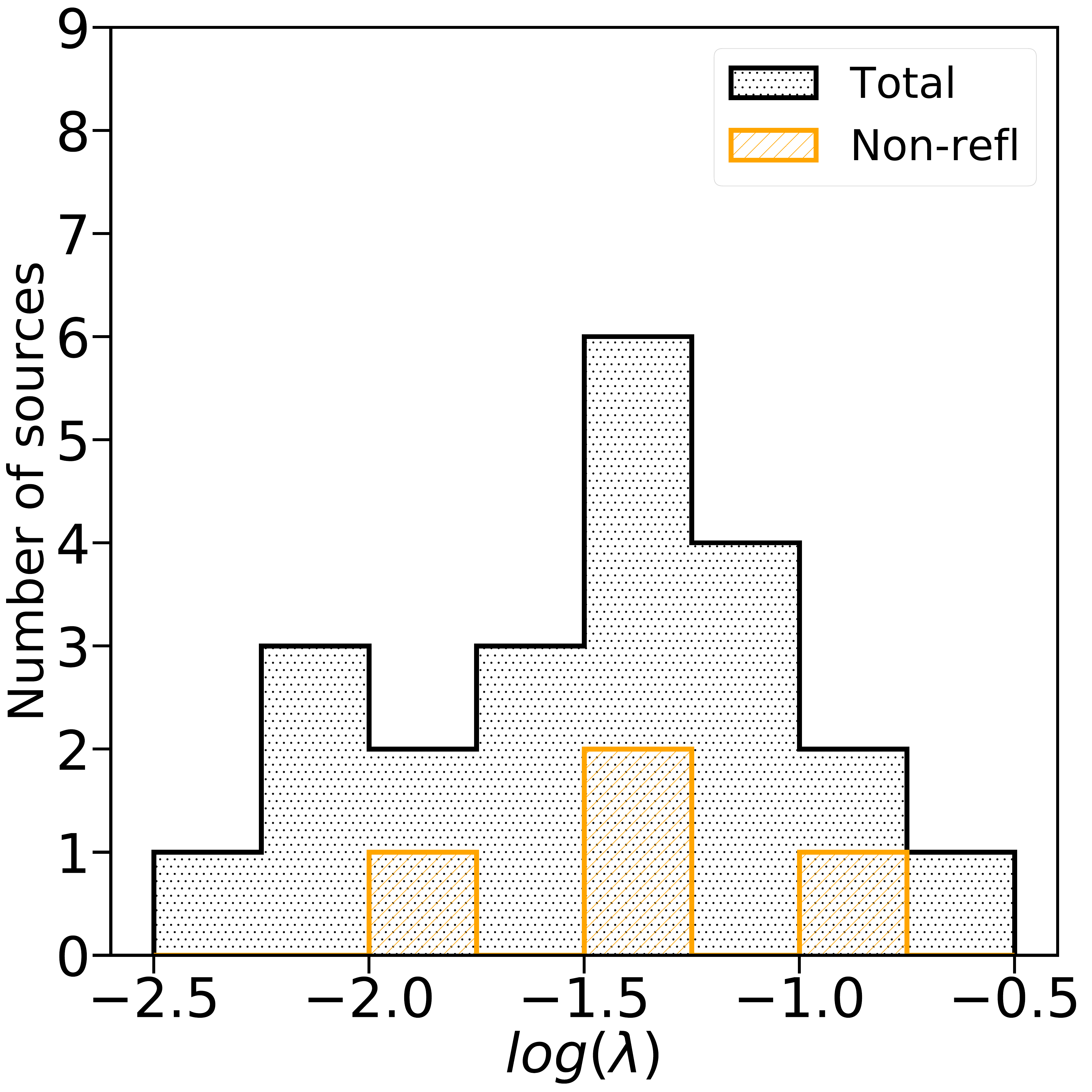}
\caption{Histograms of the distribution of SMBH mass (top), intrinsic 2-10keV luminosity (middle), and Eddington ratio (Bottom) for the full sample (dotted) versus the preferred model (dashed). Bars corresponding to the objects that prefer only the neutral or ionized models are in blue and red colors. Bars of the objects that equally prefer both models are in cyan and magenta. Note that, we do not include the three sources discarded for the analysis on the intrinsic properties of the sample.}
\label{fig:histograms}
\end{center}
\end{figure*}

\begin{figure}[htp]
\begin{center}
\includegraphics[width=1.0\columnwidth]{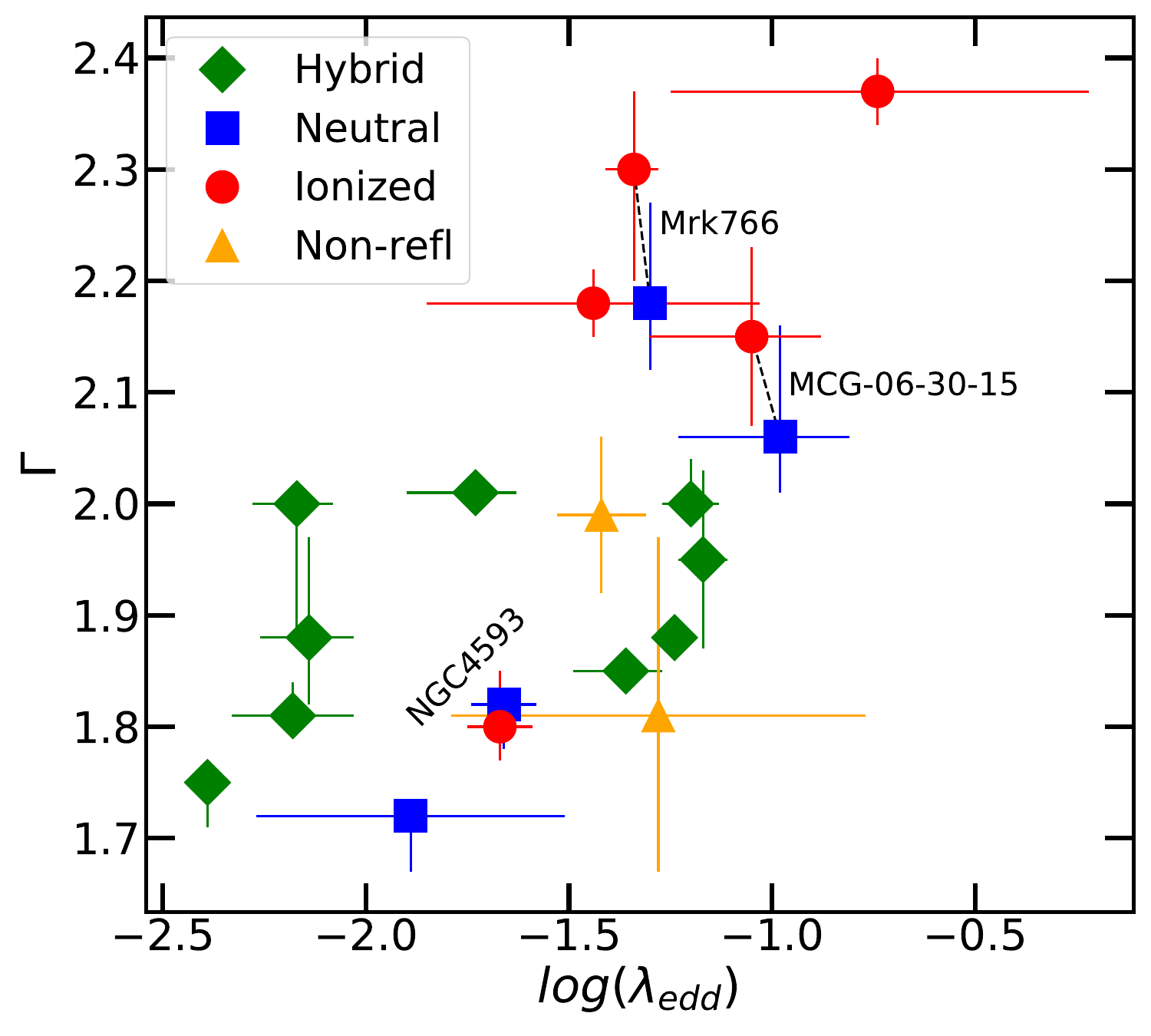}
\caption{Photon index, $\Gamma$, versus Eddington ratio. Green diamonds, blue squares, red circles, and yellow triangles correspond to the hybrid, neutral, ionized and non-reflection models respectively. Symbols of the three sources that equally prefer neutral or ionized models are joined with a dashed black line. Also the name of these sources is written on the side. Note that, we do not include the three sources discarded for the analysis on the intrinsic properties of the sample, and do not include two non-refl sources because we could not get a value for the photon index of these objects.}
\label{fig:gammavsmassvslumvsacc}
\end{center}
\end{figure}

\subsection{AGN intrinsic properties}

In order to study the AGN intrinsic properties we calculated the 2-10 keV band luminosity for each source in our sample, and the contribution of each component to the total luminosity. Also, we compile the SMBH mass from the literature and we calculate the Eddington ratio for each source. We show in Table\,\ref{tab:sample_Lums} such parameters. Figure\,\ref{fig:histograms} shows the histograms of the distribution of SMBH mass, intrinsic 2-10 keV luminosity, and Eddington ratio obtained for the sample. Note that we include those objects which do not require the reflection component (namely non-refl). Also, we excluded NGC\,1365, and NGC\,3783 (marked with an asterisk in Table\,\ref{tab:models_sample}) of the following analysis because the fits obtained with the baseline models used in this work show the worst statistic of the sample ($\chi^{2}_{r} > 1.5$), possibly due to the need to incorporate absorption/emission lines to the baseline model \citep[][]{Rivers15, Mehdipour17}. Considering the contribution of the emission/absorption lines in the baseline model may modify the model parameters, including the estimate of the intrinsic luminosity of the source, and consequently, affect the calculation of the Eddington ratio. Also, we exclude Mrk\,335 because the intrinsic continuum shows smaller contribution than reflection components (see Figure\,\ref{fig:hard_band_fit_HYBRID}), while reflection is expected to represent a fraction of the intrinsic continuum. 

%{\bf We show in Fig.\,\ref{fig:histograms} the histograms of the distribution of SMBH, intrinsic 2-10 keV luminosity, and {\bf Eddington ratio} for the sample.} 

We find that sources which prefer the neutral, hybrid, and non-refl models are in a wide SMBH mass range. Sources showing preferences for ionized reflection are characterised by lower SMBH masses, from $\rm{log(M_{BH})}$ of $\rm{6.23\pm^{0.5}_{0.5}}$ to $\rm{6.91\pm^{0.07}_{0.07}}$. Note that the uncertainties in the luminosities are quite small, so we can well differentiate the most luminous sources from the less luminous ones. The sources that prefer the ionized model shows a narrow range of intrinsic luminosity, around $\rm{log(L_{2-10keV})}$=42.6. The other sources are between $\rm{log(L_{2-10keV})}$=$\rm{41.43\pm^{0.02}_{0.02}}$ and $\rm{log(L_{2-10keV})}$=$\rm{44.68\pm^{0.01}_{0.01}}$.  The sources that do not require the reflection component shows low ($\rm{log(L_{2-10keV})}$=$\rm{42.21\pm^{0.01}_{0.01}}$, $\rm{log(L_{2-10keV})}$=$\rm{42.57\pm^{0.01}_{0.01}}$) and high ($\rm{log(L_{2-10keV})}$=$\rm{43.85\pm^{0.01}_{0.01}}$, $\rm{log(L_{2-10keV})}$= $\rm{44.68\pm^{0.01}_{0.01}}$) luminosities, i.e, these objects do not show any preference for a particular range of luminosities. Finally, the highest Eddington ratios are found for sources that prefer the ionized or non-refl models. However, if we consider the lower limits of Eddington ratio of these sources, they move to lower values. The lower Eddington ratios are observed for the sources that prefer the hybrid model, being Mrk\,140, NGC\,3227, NGC\,4151 and IRAS\,13197-1627 the sources with the lowest Eddington ratios, even considering their uncertainties.

%distributed towards intermediate and lower values are observed for the sources that prefer the neutral and hybrid models.} 

%Also, we study the correlation between the photon index and the Eddington ratio. {\bf We find that the two sources preferring the ionized model shows a trend to have the highest} photon index (see Figure\,\ref{fig:gammavsmassvslumvsacc}). Interestingly, 

Also, we study the correlation between the photon index and the Eddington ratio. We find that the two sources preferring the ionized model have a higher photon index than those preferring the hybrid or non-refl models (see Figure\,\ref{fig:gammavsmassvslumvsacc}). Two sources (Mrk\,766 and MCG\,-06-30-15) of the three sources that equally prefer the neutral and ionized models show the highest photon index in two of the four sources that fit to neutral model and two of the five sources that fit to ionized model. Also, for these sources, the ionized model produces steeper spectra. This could suggest that these two sources have intrinsically steep spectra, however, even excluding these sources, the two objects that prefer only the ionized model show the highest photon index. According to our results, it is difficult to determine if there is any tendency for objects to have a higher photon index or Eddington rate. However, note that previous works have studied the correlation between these parameters. For instance, \cite{Fanali13} study a sample of 71 type-1 AGN. They found that the photon index depends significantly on the Eddington rate, where both parameters have a directly proportional relationship (see their Figure\,5). Note that we also explored the dependence between the photon index and the SMBH mass and 2-10 keV luminosity, however we do not find correlation between these parameters.

\begin{figure*}[t]
\begin{center}
\includegraphics[width=0.622\columnwidth]{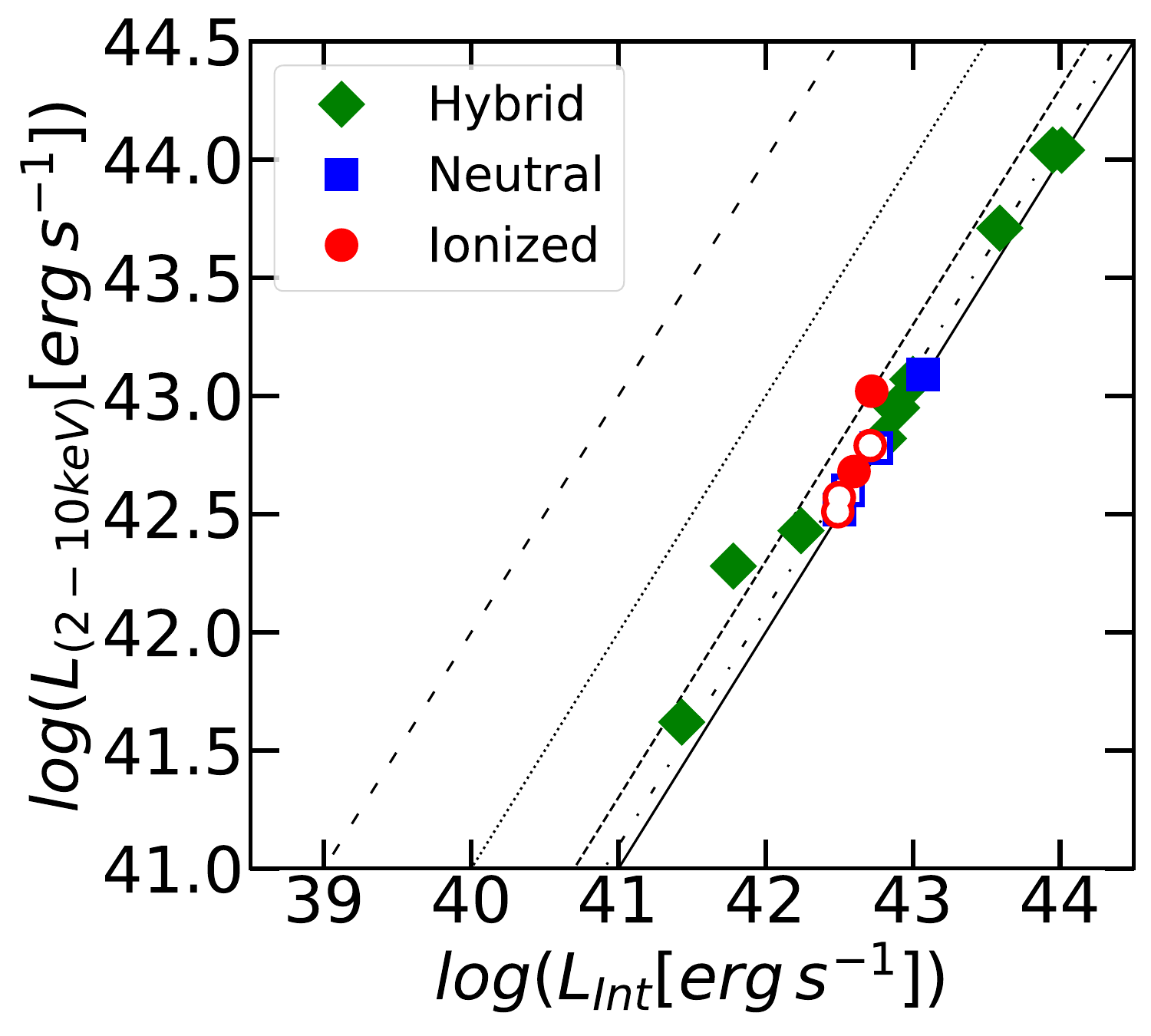} %\hspace{-0.2 cm}
\includegraphics[width=0.483\columnwidth, clip, trim=155 0 5 0]{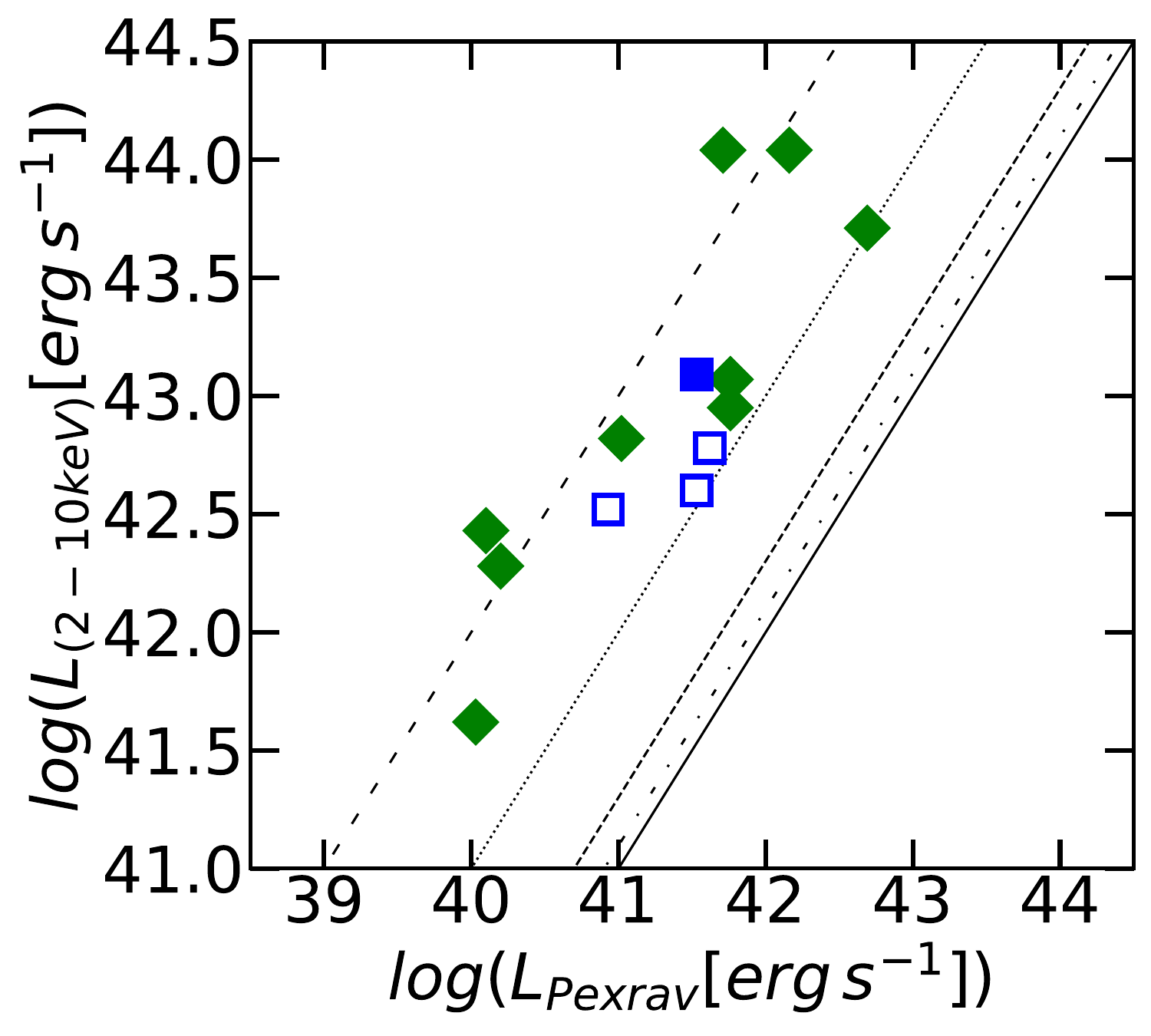} %\hspace{-0.2 cm}
\includegraphics[width=0.483\columnwidth, clip, trim=155 0 5 0]{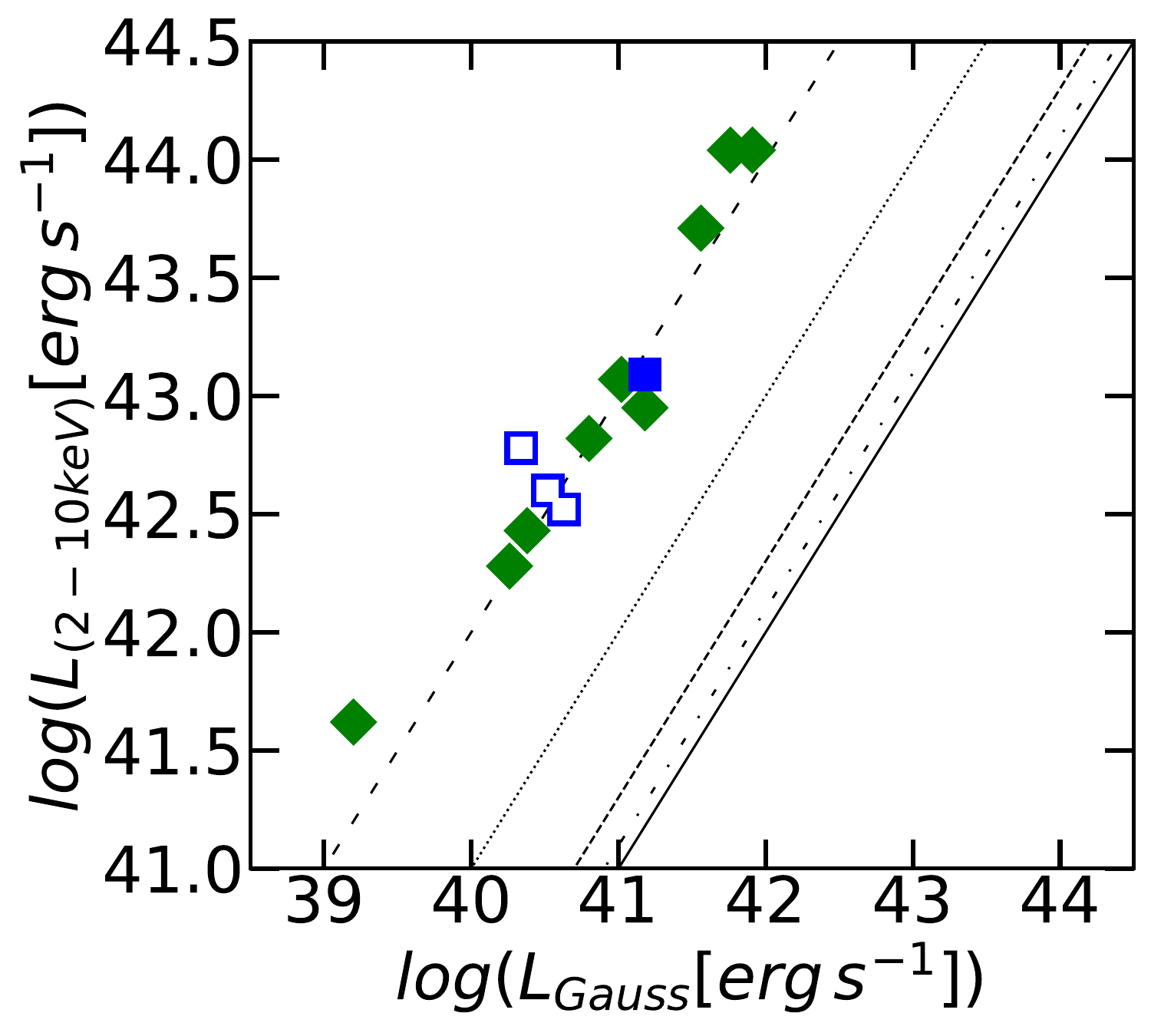} %\hspace{-0.2 cm}
\includegraphics[width=0.483\columnwidth, clip, trim=155 0 5 0]{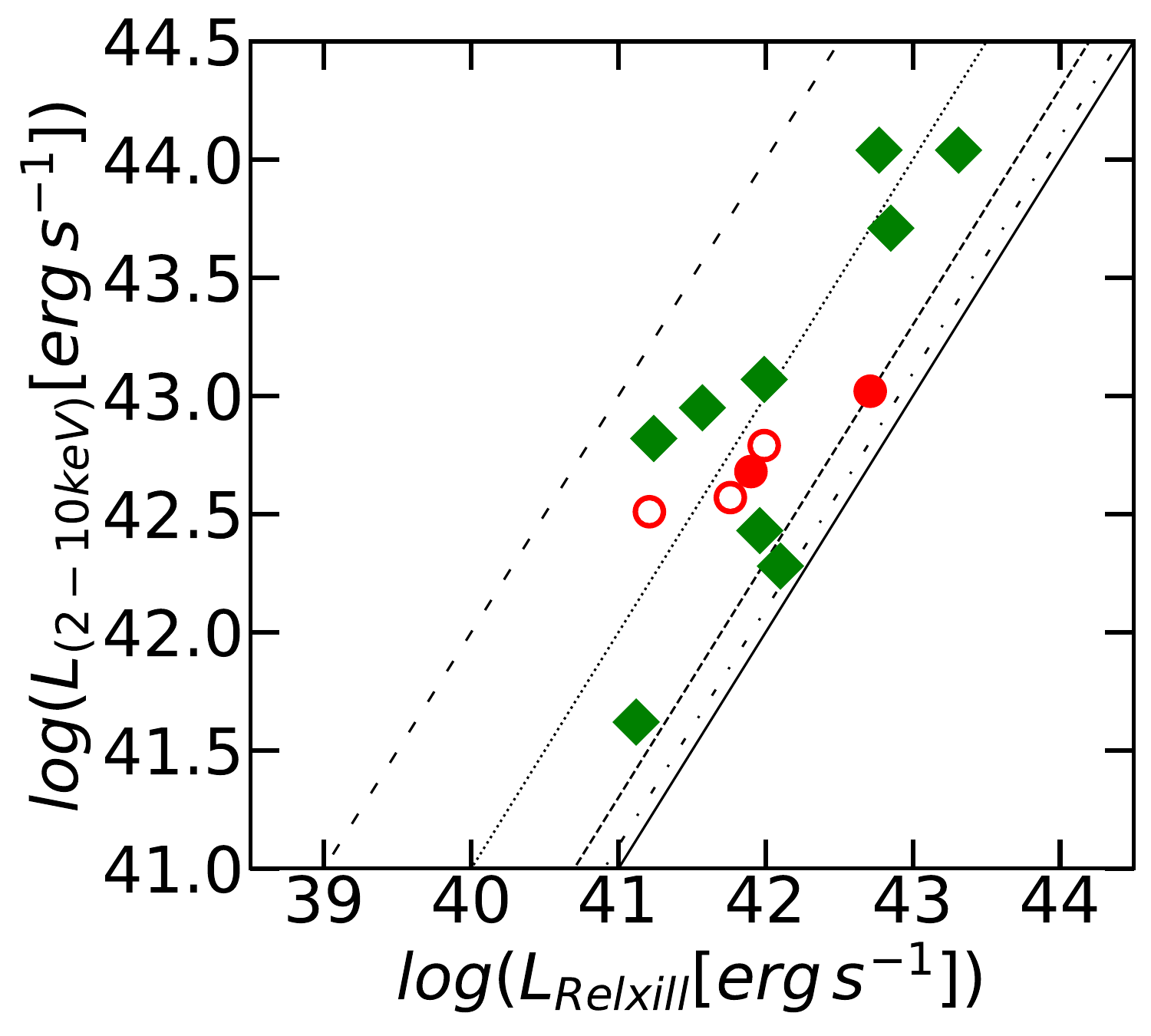}
\caption{2-10 keV band luminosity versus that of intrinsic continuum (left), and neutral (middle) and ionized (right) according to each model. Lines represent the percentage of the quantity in the X-axis with respect to the Y-axis. From bottom to top 100, 80, 50, 10, and 1 per cent of the intrinsic (left panel) and reflection (right panels) luminosities. Green diamonds, blue squares, and red circles correspond to the hybrid, neutral, and ionized models respectively. Empty symbols correspond to the sources that equally prefer neutral or ionized model.}
\label{fig:Lumin_comps_2_10}
\end{center}
\end{figure*}

Finally, we studied the luminosity contribution of each component to the baseline model. Figure\,\ref{fig:Lumin_comps_2_10} shows correlation between the total luminosity versus the intrinsic continuum, and neutral and ionized luminosities according to each model. We find that almost all sources have a contribution of the intrinsic continuum between 80\% and 100\%. This component ranges from $\rm{log(L_{Int})}$ of $\rm{41.43\pm^{0.02}_{0.02}}$ to $\rm{44.68\pm^{0.01}_{0.01}}$, with a mean value of $\rm{log(L_{Int})}$ = $\rm{42.85\pm^{0.01}_{0.01}}$. %NGC\,4151 and Mrk\,1044 has a contribution of the intrinsic around of 50\%. 
The contribution of the neutral reflector is below 10\% for all sources, ranging from $\rm{log(L_{pexrav})}$ of $\rm{40.03\pm^{0.02}_{0.02}}$ to $\rm{42.69\pm^{0.04}_{0.04}}$, with a mean value of $\rm{log(L_{pexrav})}$ = $\rm{41.31\pm^{0.04}_{0.04}}$. Furthermore, the contribution of the $\rm{FeK\alpha}$ to the total luminosity is only around 1\%. This component ranges from $\rm{log(L_{Gauss})}$ of $\rm{39.20\pm^{0.09}_{0.09}}$ to $\rm{41.91\pm^{0.02}_{0.02}}$, with a mean value of $\rm{log(L_{Gauss})}$ = $\rm{40.83\pm^{0.04}_{0.05}}$. For those sources that prefer the ionized model, the reflection component has a contribution close to 10\% except to Mrk\,1044, which has a contribution of the ionized reflector around 50\%. Those sources that prefer the hybrid model have a contribution of the ionized component distributed in a wide range. Four sources below 10\%, four sources between 10 \% and 50\%, and one source above 50\%. The ionized component ranges from $\rm{log(L_{relxill})}$ of $\rm{41.43\pm^{0.03}_{0.03}}$ to $\rm{43.31\pm^{0.01}_{0.01}}$, with a mean value of $\rm{log(L_{relxill})}$ = $\rm{41.43\pm^{0.03}_{0.03}}$.

\section{Discussion}\label{sec:Discussion}

\subsection{Non-reflection component objects}

According to our analysis, four objects (Mrk\,382, IRAS\,13224-3809, Mrk,841, and MR\,2251-178) in our sample do not require a reflection component in order to obtain a good fit. Interestingly, according to previous works, this result is in agreement for only two objects. The absence of reflection features have been observed previously in Mrk\,382 by \cite{Singh92} using EXOSAT observations. They conclude that the observation shows only the unprocessed X-ray emission of the source. They do not find evidence of intrinsic absorption, soft excess or any line feature. \cite{Nardini14} study the spectrum of MR\,2251-178 using \emph{XMM}-Newton observations. They test several physical models to fit the spectrum, however, they conclude that the spectrum features can be ascribed to partial covering of the X-ray source. In the other two objects, reflection features have been observed previously. \citet{Gofford11} also study the spectrum of MR\,2251-178 using \emph{Suzaku} and \emph{Swift}/BAT data. They conclude that the spectrum can be equally well modelled with an absorption-dominated, partially covered, continuum or as a fully covered intrinsic continuum component. \cite{Jiang18} use simultaneous \emph{XMM}-Newton and \emph{NuSTAR} observations to study the spectra of IRAS\,13224-3809. They find that the reflected spectrum is well fit with two relativistic blurred reflection components from the inner accretion disc. Note that they use several observations, 12 of \emph{XMM}-Newton and seven of \emph{NuSTAR}, among which are the two observations used in this work. \cite{Bianchi04} use simultaneous observations of \emph{XMM}-Newton and \emph{BeppoSAX} to study the spectrum of Mrk\,841. They obtain a good fit using an isotropic-illuminated cold slab reflection component.  

A plausible explanation for this discrepancy between our results and those found previously for IRAS\,13224-3809 and Mrk\,841 might be due to the variability of these AGN; i.e. these systems might be in a state of a very high continuum level. \cite{Gallo04} find that high-energy photons in IRAS\,13224-3809 are produced by the combination of the intrinsic continuum associated to the corona and a reflection component associated to the disk. They conclude that, although both processes occur simultaneously, only one dominates at a given time. Mrk841 is known for having a complex, variable iron line \citep{Petrucci02, Longinotti04}, probably indicating emission from the accretion disk but difficult to constrain in our analysis.

Note that, to explore if our four sources have been observed in a lower intrinsic continuum states, we look for other simultaneous observations, however we only find data of IRAS\,13224-3809 and MR\,2251-17. Again, we find that the simple power-law fit is enough to fit the spectra of IRAS\,13224-3809. Tested observations of this source are separated by $\sim$ 9 days. Furthermore, under the interpretation that these four sources are in a high state, we might also expect these sources to occupy the highest values of luminosities and/or Eddington rates. However, although two of them show among the highest luminosities in our sample, the other two show lower values (see Figure\,\ref{fig:histograms}).

\subsection{Reflection scenarios}

Among the sources for which we detect the presence of the reflection component, most objects require the hybrid scenario where both reflection from neutral and the relativistic/ionized media are required. In support of these results, \cite{Nardini11} conclude that the presence of blurred disc reflection as a fundamental component of the X-ray spectra of type 1 AGN. Also, some authors have concluded that it is necessary to consider both neutral/distant and ionized/relativistic reflection components to satisfactorily fit the X-ray spectra above 3 keV of some AGN. \cite{Parker19} use \emph{XMM}-Newton, \emph{NuSTAR}, \emph{Swift} and \emph{HST} observations of Mrk\,335. They conclude that its spectrum is dominated at low energies ($<$ 3 keV) by the photoionized emission, and at high energies ($>$ 5 keV) by distant reflection. They also find that relativistic reflection or partial covering give a similar fit to the data. (Note that one of their \emph{XMM}-Newton and \emph{NuSTAR} observations are the same ones we use for this source). %, and they are necessary to fit the excess flux around 7 keV. 
\citet{Lohfink16} fit the X-ray spectrum of Fairall\,9 using a model which includes distant reflector and relativistically blurred ionized reflection to account for the reflection component. Analyzing \emph{Suzaku} data of Ark\,120, \cite{Nardini11} conclude that in order to obtain a self-consistent interpretation of the broad-band X-ray emission of Ark\,120, a reflection model allowing for both warm/blurred and cold/distant reprocessing is necessary. \cite{Walton18} find that the X-ray spectrum of IRAS\,13197-1627 has contributions from relativistic reflection, absorption and further reprocessing by more distant material, and absorption from an ionized outflow. NGC\,5548 exhibits contributions from cold/distant reflection according to \cite{Brenneman12}. However, \cite{Dehghanian20} find that a translucent wind can contribute a part of the $\rm{FeK\alpha}$ emission line, arguing for a disk wind model which explain the emission lines in the source.

Note that, among the objects that prefer the hybrid model, NGC\,1365 and NGC\,3783 show indications of additional complex spectrum. In particular emission/absorption lines between 7-10 keV are clearly visible (see Figure\,\ref{fig:hard_band_fit_HYBRID}). In agreement with this result, \cite{Rivers15} find multi-layer variable absorbers in NGC\,1365. They find the need of three and two distinct zones of neutral and ionized absorption, respectively. Their spectral fit includes four absorption lines at 6.7, 6.97, 7.88, and 8.27\,keV. Also, \cite{Risalitti05} find that the spectrum of NGC\,1365 switched from reflection-dominated to transmission-dominated and back in timescale of few weeks, which is due the variation in the absorber along the line-of-sight. Also, emission and absorption features in the X-ray spectra of NGC\,3783 have been find by \citet{Mehdipour17} and \citet{Mao19}, associated with an obscuring outflow. 

We find that one reflection component is a good representation of the spectra only for three objects in our sample (Mrk\,915, Mrk\,1044, and IGRJ\,19378-0617). Suporting these results, \citet{Ballo17} find that the reflection component of Mrk\,915 can be explained with a cold reflection from distant matter, and \cite{Mallick18} find that the broad-band spectrum of Mrk\,1044 can be explained through a relativistic reflection from a high-density accretion disc with a broken power-law emissivity profile. %Using \emph{XMM}-Newton, \emph{NuSTAR} and \emph{Swift} observations of Mrk\,1044, 

Thanks to the combination of \emph{NuSTAR} and \emph{XMM}-Newton we could disentangle among the distant/neutral, relativistic/ionized, and hybrid baseline models for the vast majority. However, three sources (named Mrk\,766, NGC\,4593, and MCG\,-06-30-15) equally prefer the neutral and the relativistic/ionized scenarios. %About these sources, some authors have found similar results. 
\cite{Buisson18} test the {\sc pexrav+gauss} and {\sc relxill} models for the combination of \emph{XMM}-Newton, \emph{Swift} and \emph{NuSTAR} data for Mrk\,766, finding no statistical differences among them. Note that we use the same data of \emph{XMM}-Newton and \emph{NuSTAR} and we obtain the same conclusion about the preferred model for this source. %$\rm{\chi^2_{red}}$ of 0.99 and 0.985, respectively. Also, they test a partial covering model, obtaining a $\rm{\chi^2_{red}}$ of 0.996. 
Among them, a strongly debated case is MCG\,-06-30-15, where there is a long discussion of whether pure partial covering or disk reflection could explain the broadening of the emission line \citep{Fabian03, Reynolds09, Miller08, Chiang11}. Among these works, \cite{Marinucci14} test reflection and absorption models using \emph{XMM}-Newton and \emph{NuSTAR} data to study the variability of the source. They statistically disfavor the last model. Their reflection model includes both, cold distant and relativistically blurred reflection. One of the three XMM-Newton observations of \cite{Marinucci14} is the observation used in this work. Interestingly, results of \cite{Marinucci14} favours a hybrid model, while we found that our hybrid model is not required, however, we also found that MCG-06-30-15 equally prefers the neutral and the ionized model. %{\bf X-ray reverberation mapping measurements of \cite{Kara16} do not show reverberation signatures for Mrk\,766 and MCG\,-06-30-15, which suggest that the reflection in these objects occurs in a neutral/distant material instead of the accretion disk (where reverberation signatures would be present).}

Under to the unified model, where the different types of AGN are explain through the viewing angle toward the observer, all objects in our sample should show the contribution of both, neutral/distant and relativistic/ionized reflectors, since our sample contains only type 1 AGN. Supporting this, the hybrid baseline model is the best explanation for most object in our sample. Furthermore, the role of each component to the spectral signatures is quite different; while the relativistic/ionized reflector mostly contribute to the broadening of the $\rm{FeK\alpha}$ emission line and the continuum at lower energies, the distant/neutral reflector is mainly contributing to the narrow component of the $\rm{FeK\alpha}$ emission line and to the shape of the Compton hump above 10\,keV (see Figure\,\ref{fig:hard_band_fit_HYBRID}). However, even though most of sources require these two components to fit their spectra, six objects only require one component, one of them is well fitted with only neutral/distant reflector, and two objects are well fitted with only relativistic/ionized reflector. For the other three objects, only neutral/distant or ionized/relativistic are equally preferred. Additionally, it is also worth to notice that the relativistic/ionized reflector is, in average, contributing more than the distant/neutral reflector to the 2-10\,keV X-ray luminosity for our sample (see Figure\,\ref{fig:Lumin_comps_2_10}). Contrary to the unified model, these results suggest that the difference between AGN types is due to their intrinsic properties rather than their orientation.

Through X-ray reverberation mapping around the FeK$\rm{\alpha}$, which measures the time delays between the photons from the corona and the photons from the accretion disc, it is possible to test if the reflection occurs in the accretion disk \citep{Uttley14}. FeK$\rm{\alpha}$ reverberation signatures has been searched in 16 out of the 22 objects in our sample \citep[compiled in Table\,\ref{tab:sample_Lums}, including Mrk\,335, NGC\,1365, and NGC\,3783,][]{Tripathi11,Kara16,Lobban18,Mallick18}. Among them, 11 show FeK$\rm{\alpha}$ time delays; all but IRAS\,13224-3809 are fitted in this work to the hybrid or the ionized model. On the other hand, five sources do not show signatures of FeK$\rm{\alpha}$ time delays; one do not show a reflection component, one is fitted to the neutral model, one is fitted to the hybrid model, and the other two are equally best fitted to both ionized and neutral models. All these results are compatible with our best fitted models.

The two modes of accretion onto black holes (for stellar mass and supermasive black holes) are either via a geometrically thin, optically thick disk or via a truncated outer disk, and a hot optically thin, geometrically extended advection-dominated flow; being present in low-mass X-ray binaries (LMXBs) and AGN \citep{Meyer09}. Studying the LMXBs, they claim that for high Eddington ratios, the disk reaches inward to the last stable orbit, however, a gap in the disk appears if the Eddington ratio begins to slow (see their Fig.\,1). In a study carried out on a sample of unobscured broad-line, narrow-line and lineless AGN, \cite{Trump11} shows that the Eddington ratio governs their physical properties, arguing the disappearance of the broad emission lines in their sample by an expanding radiatively inefficient accretion flow at the inner radius of the accretion disk. Assuming that the distant and neutral reflection component is always showing the same contribution, which seems to be the case for type-2 and/or low-luminosity AGN \citep{Osorio-Clavijo22}, the preponderance of these two components might be related to this accretion states. In AGN at a high state, the disk is well constructed until the innermost stable orbit, being responsible for most of the reflection component. If that is the case, the signatures of distant and neutral reflector are simply diluted by this prominent disk-reflection component. On the other hand, the objects that are fitted only with the neutral/distant reflection component might have an inner inefficient flow, which makes the accretion disk reflection negligible since the disk start further out from the emitting X-ray corona. Objects that prefer the hybrid model are in a intermediate state. However, note that these objects shows a wide range of Eddington ratios, from $\rm{log(\lambda)}$ = $\rm{-1.17\pm^{0.06}_{0.06}}$ up to $\rm{log(\lambda)}$ = $\rm{-2.39\pm^{0.04}_{0.04}}$.

%the three objects that shows only one reflection medium are consistent with this tentative scenario; i.e. objects preferring the relativistic/ionized model (Mrk\,1044 and IGRJ\,19378-0617) show higher or equal Eddington ratio than that preferring the neutral/distant model (Mrk\,915), even considering the uncertainty (see Figure\,\ref{fig:gammavsmassvslumvsacc}). However, the objects that prefer the hybrid model shows a wide range of Eddington ratios, from $\rm{log(\lambda)}$ = $\rm{-1.17\pm^{0.06}_{0.06}}$ up to $\rm{log(\lambda)}$ = $\rm{-2.39\pm^{0.04}_{0.04}}$. %Moreover, sources that are best (or equally good) fitted to the ionized reflection model, tend to show the highest photon indices and luminosities in our sample (see Fig.\,\ref{fig:gammavsmassvslumvsacc}), {\bf which suggests that the intrinsic component of these sources dominates over the reflected component.} %, changing the broad-line AGN with high accretion rate into narrow-line or lineless AGN with low accretion rate.

\section{Summary}

We have studied the scattered medium of a sample of 22 Seyfert galaxies using simultaneous observations of \emph{XMM}-Newton and \emph{NuSTAR}. For this purpose we selected and tested a set of available reflection models using the {\sc xspec} spectral fitting package. The main results are as follows:\\
1. We find that 18 sources shows evidence of the reflection. Among them, 12 objects prefer a hybrid reflection model, which incorporates neutral and relativistic ionized medium; one object prefer a neutral reflection model; two objects prefer a relativistic ionized reflection model; and three objects equally prefer the neutral and relativistic ionized reflection models.\\
2. We find that four objects do not present the reflection component. We propose the variability of activity of sources as a possible explanation.\\
%3. We constrained the cut-off energy of 12 sources, with $\rm{E_{cut}}$ from $\sim$ 70 keV to $\sim$ 400 keV. \\
%3. We find that the sources that prefer the ionized model shows a trend to have the highest photon index, although a more complete sample is necessary in order to confirm this result.\\
3. For most objects, the intrinsic luminosity represents between 80\% and 100\% of the total luminosity. The neutral reflection component has the smallest contribution to the total luminosity, representing this with less than 10\%. The ionized reflection component contributes over 10\% for almost all objects.

These results suggest that considering a hybrid scenario, in which the reflection from type-1 AGN has a contribution from at least two different media, can satisfactorily explain the observed spectra of most of these objects. %We speculate that exists an intrinsic difference on the reflector for type-1 AGN according to its state of evolution, since in some objects in which we did not find reflection features, other authors have satisfactorily fitted such a component, in addition to the possible trend that we found, where objects that present only ionized reflection have higher photon index than those that present only neutral reflection or both.

\section*{Acknowledgements}

CV-C acknowledge support from a CONACyT scholarship. OG-M acknowledges financial support by the UNAM PAPIIT project IN105720. J-M acknowledges financial support from the State Agency for Research of the Spanish MCIU through the ‘Center of Excellence Severo Ochoa’ award to the Instituto de Astrofísica de Andalucía (SEV-2017-0709) and  the Spanish Ministry of Economy and Competitiveness under grants no. AYA2016-76682-C3 and PID2019-106027GB-C41. A.L.L. acknowledges support from CONAHCyT grant CB-2016-01- 286316 and from DGAPA PAPIIT IA-101623. D.E.-A. also acknowledges financial support from MICINN (Spain) through the programme Juan de la Cierva D.E.-A. acknowledges support from the Spanish Ministry of Science, Innovation, and Universities (MCIU), Agencia Estatal de Investigación (AEI), and the Fondo Europeo de Desarrollo Regional (EU-FEDER) under projects with references AYA2015-68217-P and PID2019- 107010GB-100. This research has made use of dedicated servers (IRyAGN2, Galaxias and Arambolas servers and Calzonzin and Mouruka clusters) maintained by Daniel D\'iaz- Gonz\'alez, Miguel Espejel, Alfonso Ginori Gonz\'alez, and Gilberto Zavala at IRyA-UNAM. All of them are gratefully acknowledged.

%\clearpage

\clearpage
\appendix

\section{Non reflection objects} \label{sec:Non_reflection_objects}

\begin{figure}[!h]
%\begin{center}
\includegraphics[width=0.49\columnwidth]{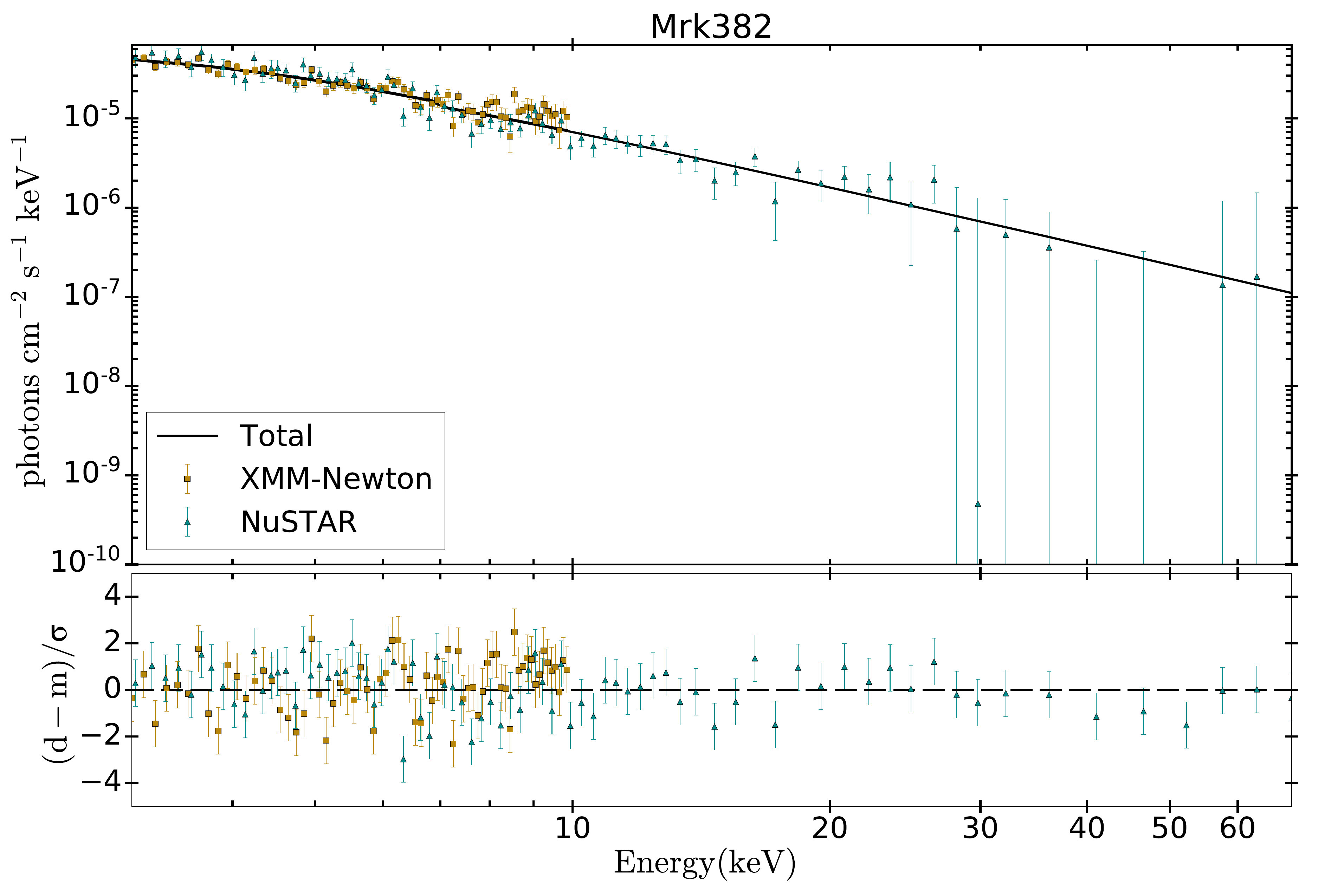}
\includegraphics[width=0.49\columnwidth]{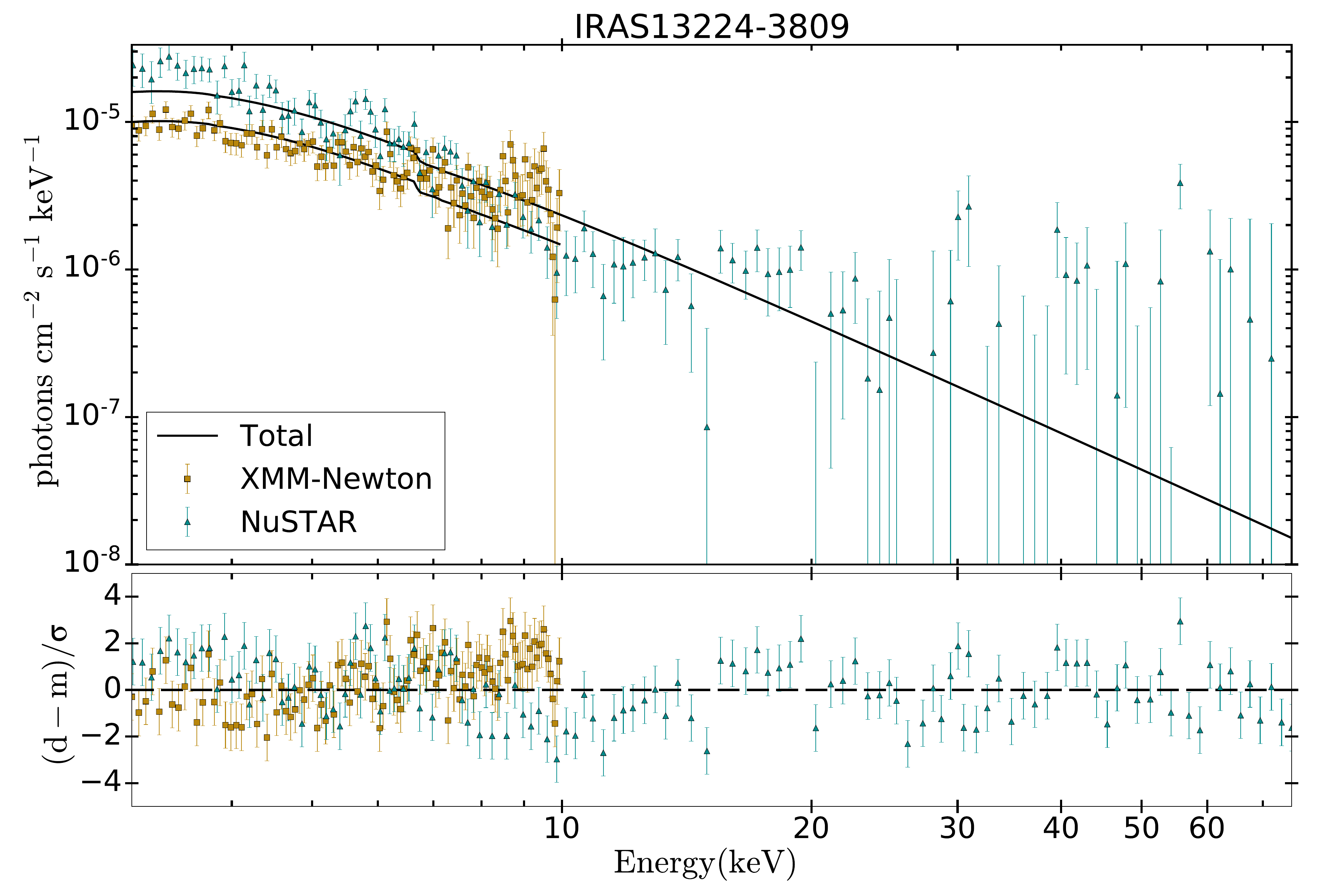}
\includegraphics[width=0.49\columnwidth]{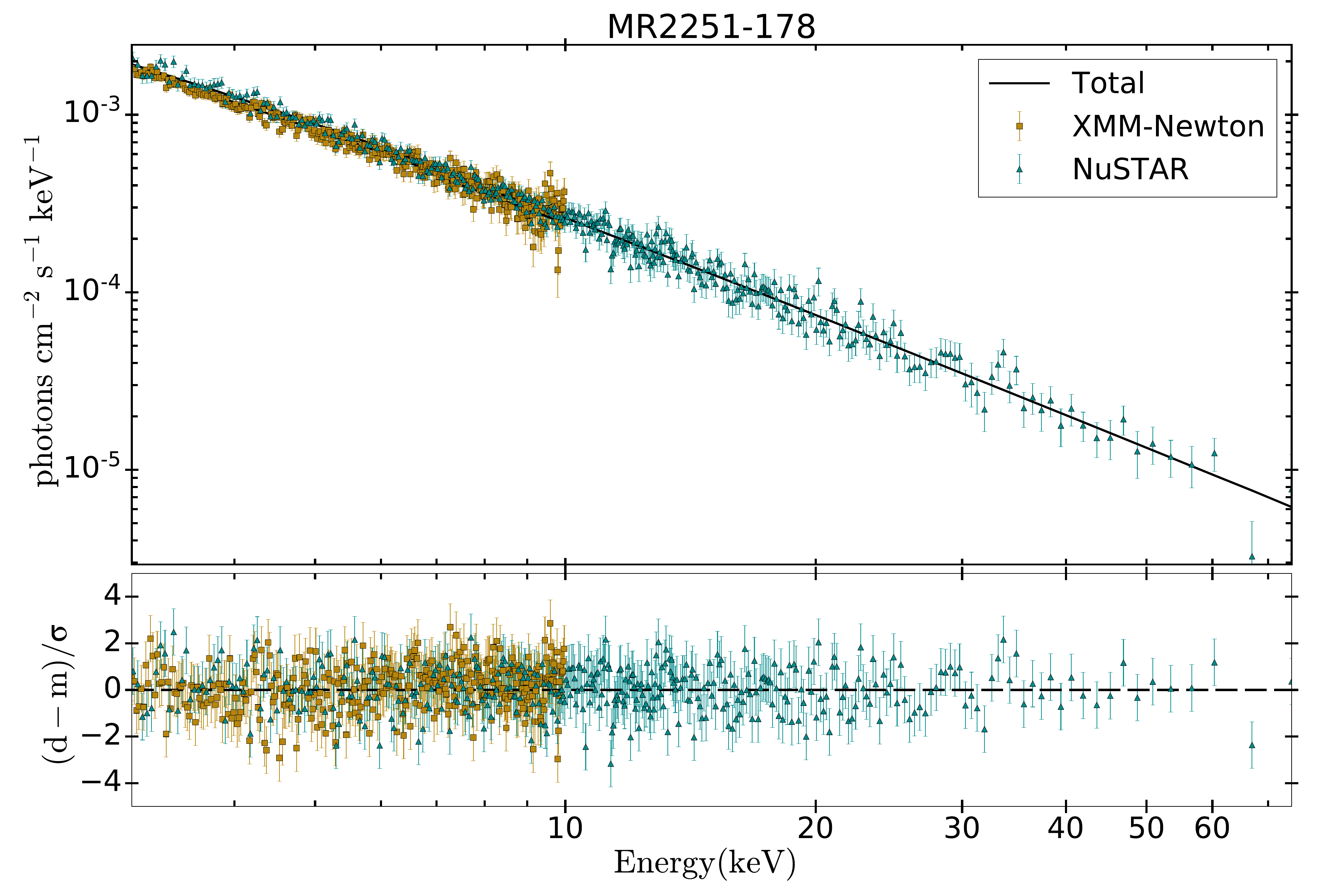}
\caption{Power-law fit to Mrk\,382 (top left), IRAS\,13224-3809 (top right), and MR\,2251-178 (bottom). We show the best fit (black and red solid line) to the data in the top panel and the ratio between model and data in the bottom panel. The gold and dark cyan dots show the data from \emph{XMM}-Newton and \emph{NuSTAR}, respectively.}
\label{fig:non_reflection}
%\end{center}
\end{figure}

Figure\,\ref{fig:non_reflection} shows the power-law fit to Mrk\,382, IRAS\,13224-3809, and MR\,2251-178, which do not require the reflection component according to the F-test statistic (besides Mrk\,841, which is described in Section\,\ref{sec:The existence of the reflection component}. See also Figure\,\ref{fig:Mrk841_non_reflection}). The spectrum of MR\,2251-178 is well fitted with a power-law. The FeK line or compton hump features are absent in the spectrum. More complex is the case of IRAS\,13224-3809, where the spectrum shows poor S/N above $\sim$ 10 keV. Probably in this object the S/N is the reason why we can not find the presence of the reflection component in our analysis.

\begin{table*}[ht]
%\tiny
\scriptsize 
\renewcommand{\tabcolsep}{0.05cm}
\begin{center}
\begin{tabular}{c|cc|ccccc|ccccc}\hline
\hspace{-1 cm} & \multicolumn{2}{c}{{\sc power-law}} & \multicolumn{5}{c}{{\sc pexrav+gauss}} & \multicolumn{5}{c}{{\sc relxill}} \\
 & {\bf $\rm{N_H}$} & $\rm{\Gamma}$ & Z & $\rm{Z_{Fe}}$ & incl & $\rm{E_{line}}$ & $\rm{\sigma}$ & index & a & incl & $\rm{log(\xi)}$ & $\rm{Z_{fe}}$ \\ 
 & & & ($\rm{Z_{\odot}}$) & ($\rm{Z_{\odot}}$) & (cos) & (keV) & (keV) & & & (deg) & & ($\rm{Z_{\odot}}$) \\
 
 \hline 
 
%$\rm{Mrk\,335^H}$ & $\dots$ & $\rm{2.33\pm^{0.01}_{0.16}}$ & $\dots$ & $\rm{2.0\pm^{0.3}_{0.8}}$ & $\dots$ & $\dots$ & $\rm{6.44\pm^{0.03}_{0.01}}$ & $\dots$ & $\rm{4.5\pm^{0.6}_{0.7}}$ & $\rm{0.57\pm^{0.13}_{0.07}}$ & $\rm{48.7\pm^{0.8}_{1.8}}$ & $\rm{2.0\pm^{0.14}_{0.08}}$ & $\rm{3.5\pm^{0.4}_{0.4}}$ \\

$\rm{Mrk\,335^H}$ & $\dots$ & $\dots$ & $\rm{1.74\pm^{0.19}_{0.5}}$ & $\dots$ & $\rm{0.74\pm^{0.11}_{0.55}}$ & $\rm{6.45\pm^{0.01}_{0.01}}$ & $\dots$ & $\rm{4.0\pm^{1.6}_{0.6}}$ & $\rm{0.57\pm^{0.13}_{0.16}}$ & $\rm{49\pm^{1}_{2}}$ & $\rm{2.11\pm^{0.53}_{0.10}}$ & $\dots$ \\

%$\rm{Fairall\,9^H}$ & $\dots$ & $\rm{1.93\pm^{0.02}_{0.07}}$ & $\rm{1.99\pm^{0.02}_{0.07}}$ & $\rm{1.2\pm^{0.4}_{0.3}}$ & $\dots$ & $\dots$ & $\rm{6.44\pm^{0.01}_{0.01}}$ & $\dots$ & $\dots$ & $\rm{0.83\pm^{0.1}_{0.25}}$ & $\dots$ & $\rm{3.0\pm^{0.05}_{0.08}}$ & $\dots$ \\ 

$\rm{Fairall\,9^H}$ & $\dots$ & $\rm{1.85\pm^{0.01}_{0.01}}$ & $\rm{1.09\pm^{0.49}_{0.92}}$ & $\dots$ & $\dots$ & $\rm{6.44\pm^{0.01}_{0.01}}$ & $\dots$ & $\dots$ & $\rm{0.82\pm^{0.07}_{0.29}}$ & $\dots$ & $\rm{3.0\pm^{0.02}_{0.06}}$ & $\dots$ \\ 

%$\rm{Mrk\,1040^H}$ & $\dots$ & $\rm{1.81\pm^{0.06}_{0.02}}$ & $\rm{1.88\pm^{0.07}_{0.01}}$ & $\dots$ & $\dots$ & $\dots$ & $\rm{6.39\pm^{0.01}_{0.01}}$ & $\rm{0.07\pm^{0.02}_{0.02}}$ & $\dots$ & $\dots$ & $\dots$ & $\dots$ & $\dots$ \\

$\rm{Mrk\,1040^H}$ & $\dots$ & $\rm{1.81\pm^{0.03}_{0.01}}$ & $\dots$ & $\dots$ & $\dots$ & $\rm{6.39\pm^{0.01}_{0.01}}$ & $\rm{0.07\pm^{0.02}_{0.02}}$ & $\dots$ & $\dots$ & $\dots$ & $\dots$ & $\dots$ \\

$\rm{NGC\,1365^H}$ & $\rm{1.91\pm^{0.13}_{0.07}}$ & $\rm{1.94\pm^{0.01}_{0.01}}$ & $\dots$ & $\dots$ & $\dots$ & $\rm{6.37\pm^{0.01}_{0.01}}$ & $\dots$ & $\rm{3.50\pm^{0.03}_{0.04}}$ & $\dots$ & $\dots$ & $\rm{2.70\pm^{0.01}_{0.02}}$ & $\dots$ \\ 

$\rm{Ark\,120^H}$ & $\rm{4.04\pm^{0.18}_{1.98}}$ & $\rm{2.0\pm^{0.04}_{0.01}}$ & $\dots$ & $\dots$ & $\rm{0.19\pm^{0.01}_{0.12}}$ & $\rm{6.45\pm^{0.01}_{0.01}}$ & $\rm{0.1\pm^{0.01}_{0.02}}$ & $\dots$ & $\dots$ & $\rm{7\pm^{7}_{3}}$ & $\rm{3.70\pm^{0.14}_{0.01}}$ & $\dots$ \\ 

$\rm{NGC\,3227^H}$ & $\rm{2.63\pm^{0.10}_{0.40}}$ & $\rm{2.0\pm^{0.02}_{0.13}}$ & $\rm{1.20\pm^{0.28}_{0.37}}$ & $\dots$ & $\dots$ & $\rm{6.44\pm^{0.01}_{0.01}}$ & $\rm{0.06\pm^{0.02}_{0.01}}$ & $\dots$ & $\dots$ & $\dots$ & $\rm{2.51\pm^{0.19}_{0.79}}$ & $\rm{2.76\pm^{0.46}_{0.41}}$ \\ 

$\rm{NGC\,3783^H}$ & $\dots$ & $\rm{2.17\pm^{0.09}_{0.03}}$ & $\dots$ & $\dots$ & $\dots$ & $\rm{6.44\pm^{0.01}_{0.01}}$ & $\rm{0.05\pm^{0.01}_{0.01}}$ & $\dots$ & $\rm{0.98\pm^{0.01}_{0.01}}$ & $\rm{75\pm^{1}_{1}}$ & $\dots$ & $\rm{1.84\pm^{0.12}_{0.09}}$ \\ 

%$\rm{NGC\,4051^H}$ & $\rm{1.3\pm^{0.1}_{0.1}}$ & $\rm{1.99\pm^{0.01}_{0.02}}$ & $\rm{2.06\pm^{0.04}_{0.01}}$ & $\rm{3.2\pm^{0.2}_{0.2}}$ & $\dots$ & $\dots$ & $\rm{6.42\pm^{0.03}_{0.02}}$ & $\dots$ & $\rm{1.26\pm^{0.15}_{0.6}}$ & $\dots$ & $\dots$ & $\rm{2.98\pm^{0.03}_{0.01}}$ & $\dots$ \\ 

$\rm{NGC\,4051^H}$ & $\rm{1.25\pm^{0.10}_{0.11}}$ & $\rm{2.01\pm^{0.01}_{0.01}}$ & $\rm{3.01\pm^{0.27}_{0.23}}$ & $\dots$ & $\dots$ & $\rm{6.42\pm^{0.01}_{0.01}}$ & $\dots$ & $\rm{1.28\pm^{0.18}_{0.36}}$ & $\dots$ & $\dots$ & $\rm{3.01\pm^{0.01}_{0.01}}$ & $\dots$ \\ 

$\rm{NGC\,4151^H}$ & $\rm{13.94\pm^{1.11}_{0.96}}$ & $\rm{1.75\pm^{0.02}_{0.04}}$ & $\dots$ & $\dots$ & $\dots$ & $\rm{6.38\pm^{0.01}_{0.01}}$ & $\rm{0.06\pm^{0.03}_{0.03}}$ & $\dots$ & $\rm{0.42\pm^{0.10}_{0.09}}$ & $\rm{33\pm^{3}_{2}}$ & $\rm{3.02\pm^{0.04}_{0.05}}$ & $\dots$ \\

IRAS\,13197-1627${\rm ^H}$ & $\rm{79\pm^{4}_{4}}$ & $\rm{1.88\pm^{0.09}_{0.06}}$ & $\rm{1.1\pm^{0.3}_{0.5}}$ & $\rm{1.0\pm^{0.2}_{0.2}}$ & $\dots$ & $\rm{6.42\pm^{0.01}_{0.01}}$ & $\rm{0.07\pm^{0.01}_{0.01}}$ & $\rm{3.3\pm^{0.3}_{0.3}}$ & $\rm{0.26\pm^{0.16}_{0.17}}$ & $\rm{12\pm^{3}_{3}}$ & $\rm{2.3\pm^{0.1}_{0.2}}$ & $\dots$ \\ 

$\rm{NGC\,5548^H}$ & $\dots$ & $\rm{1.88\pm^{0.01}_{0.01}}$ & $\dots$ & $\dots$ & $\dots$ & $\rm{6.48\pm^{0.01}_{0.01}}$ & $\dots$ & $\rm{0.95\pm^{0.18}_{0.46}}$ & $\dots$ & $\dots$ & $\rm{2.36\pm^{0.02}_{0.06}}$ & $\dots$ \\

$\rm{NGC\,7469^H}$ & $\rm{0.32\pm^{0.26}_{0.24}}$ & $\rm{1.95\pm^{0.08}_{0.08}}$ & $\rm{1.64\pm^{1.0}_{0.65}}$ & $\dots$ & $\dots$ & $\rm{6.43\pm^{0.02}_{0.01}}$ & $\dots$ & $\rm{2.29\pm^{0.31}_{0.31}}$ & $\dots$ & $\rm{61\pm^{7}_{4}}$ & $\rm{2.99\pm^{0.10}_{0.55}}$ & $\dots$ \\

\hline

%$\rm{NGC\,4593^N}$ & $\dots$ & $\rm{1.74\pm^{0.05}_{0.01}}$ & $\rm{1.83\pm^{0.06}_{0.03}}$ & $\dots$ & $\rm{2.3\pm^{1.4}_{0.8}}$ & $\dots$ & $\rm{6.48\pm^{0.01}_{0.01}}$ & $\dots$ & - & - & - & - & - \\ 

%$\rm{Mrk\,915^N}$ & $\rm{2.4\pm^{0.5}_{0.3}}$ & $\rm{1.7\pm^{0.09}_{0.01}}$ & $\rm{1.82\pm^{0.13}_{0.06}}$ & $\dots$ & $\dots$ & $\rm{0.08\pm^{0.25}_{0.01}}$ & $\rm{6.42\pm^{0.02}_{0.01}}$ & $\rm{0.07\pm^{0.02}_{0.02}}$ & - & - & - & - & - \\ 

$\rm{Mrk\,915^N}$ & $\rm{2.97\pm^{0.42}_{0.21}}$ & $\rm{1.72\pm^{0.01}_{0.05}}$ & $\dots$ & $\dots$ & $\dots$ & $\rm{6.43\pm^{0.01}_{0.02}}$ & $\rm{0.08\pm^{0.02}_{0.02}}$ & - & - & - & - & - \\ 

%$\rm{Mrk\,766^N}$ & $\dots$ & $\rm{2.19\pm^{0.1}_{0.07}}$ & $\dots$ & $\dots$ & $\rm{0.99\pm^{0.06}_{0.3}}$ & $\dots$ & $\dots$ & $\dots$ & - & - & - & - & - \\

$\rm{Mrk\,766^N}$ & $\rm{0.39\pm^{0.62}_{0.33}}$ & $\rm{2.18\pm^{0.09}_{0.06}}$ & $\dots$ & $\rm{1.03\pm^{0.55}_{0.16}}$ & $\dots$ & $\dots$ & $\dots$ & - & - & - & - & - \\

$\rm{NGC\,4593^N}$ & $\rm{0.53\pm^{0.14}_{0.22}}$ & $\rm{1.82\pm^{0.01}_{0.04}}$ & $\dots$ & $\rm{2.75\pm^{1.51}_{1.06}}$ & $\dots$ & $\rm{6.49\pm^{0.01}_{0.02}}$ & $\dots$ & - & - & - & - & - \\ 

MCG\,-06-30-15${\rm ^N}$ & $\rm{2.08\pm^{0.10}_{0.26}}$ & $\rm{2.06\pm^{0.10}_{0.05}}$ & $\dots$ & $\rm{0.81\pm^{0.15}_{0.27}}$ & $\dots$ & $\rm{6.44\pm^{0.02}_{0.03}}$ & $\dots$ & - & - & - & - & - \\

\hline

$\rm{Mrk\,1044^I}$ & $\rm{0.79\pm^{0.37}_{0.35}}$ & $\rm{2.37\pm^{0.03}_{0.03}}$ & - & - & - & - & - & $\dots$ & $\rm{0.93\pm^{0.01}_{0.04}}$ & $\dots$ & $\rm{3.18\pm^{0.06}_{0.06}}$ & $\rm{0.81\pm^{0.14}_{0.12}}$ \\ 

IGRJ\,19378-0617${\rm ^I}$ & $\rm{0.53\pm^{0.22}_{0.17}}$ & $\rm{2.18\pm^{0.03}_{0.03}}$ & - & - & - & - & - & $\rm{2.19\pm^{0.17}_{0.37}}$ & $\dots$ & $\rm{44\pm^{7}_{3}}$ & $\rm{3.13\pm^{0.06}_{0.06}}$ & $\dots$ \\

%$\rm{Mrk\,766^I}$ & $\dots$ & $\rm{2.3\pm^{0.1}_{0.1}}$ & $\rm{2.29\pm^{0.08}_{0.05}}$ & - & - & - & - & - & $\rm{1.96\pm^{0.05}_{0.86}}$ & $\dots$ & $\rm{49\pm^{10}_{12}}$ & $\rm{1.7\pm^{0.2}_{0.4}}$ & $\dots$  \\

$\rm{Mrk\,766^I}$ & $\dots$ & $\rm{2.30\pm^{0.07}_{0.10}}$ & - & - & - & - & - & $\rm{1.93\pm^{0.38}_{0.83}}$ & $\dots$ & $\rm{50\pm^{6}_{12}}$ & $\rm{1.70\pm^{0.16}_{0.39}}$ & $\rm{0.68\pm^{0.17}_{0.18}}$ \\

$\rm NGC\,4593^I$ & $\rm 0.39\pm^{0.33}_{0.30}$ & $\rm 1.80\pm^{0.05}_{0.03}$ & - & - & - & - & - & $\dots$ & $\dots$ & $\dots$ & $\rm 2.47\pm^{0.12}_{0.11}$ & $\dots$ \\ 

MCG\,-06-30-15${\rm ^I}$ & $\rm{2.37\pm^{0.22}_{0.30}}$ & $\rm{2.15\pm^{0.08}_{0.08}}$ & - & - & - & - & - & $\rm{1.45\pm^{0.37}_{0.81}}$ & $\dots$ & $\rm{31\pm^{10}_{5}}$ & $\rm{1.86\pm^{0.11}_{0.17}}$ & $\rm{0.71\pm^{0.23}_{0.14}}$ \\

\hline

$\rm{Mrk\,382^{NR}}$ & $\dots$ & $\rm{1.81\pm^{0.16}_{0.14}}$ & - & - & - & - & - & - & - & - & - & - \\

IRAS\,13224-3809${\rm ^{NR}}$ & $\dots$ & $\dots$ & - & - & - & - & - & - & - & - & - & - \\

$\rm{Mrk\,841^{NR}}$ & $\dots$ & $\rm{1.99\pm^{0.07}_{0.07}}$ & - & - & - & - & - & - & - & - & - & - \\

MR\,2251-178${\rm ^{NR}}$ & $\rm{1.61\pm^{0.19}_{0.19}}$ & $\dots$ & - & - & - & - & - & - & - & - & - & - \\

\hline

\end{tabular}
\end{center}
\caption{Spectral fit parameters obtained with the preferred model for each source. Superscript in the object name indicates the model corresponding to the parameters (H for hybrid, N for neutral, I for ionized, and NR for non-refl). Dots indicate when the parameter could not be constrained. Dash indicate when the model does not contain that parameter. Column density is in units of $\rm{10^{22}cm^{-2}}$, ionization parameter in log erg cm $\rm s^{-1}$}.
%The cutoff energy was linked between power-law and reflection components.}
\label{tab:sample_params}
\end{table*}

\begin{table*}[ht]
%\tiny
\scriptsize 
\renewcommand{\tabcolsep}{0.05cm}
\begin{center}
\begin{tabular}{c|ccccc|cccccccc}\hline
Source & type & z & $\rm{log(M_{BH})}$ & $\rm{log(\lambda)}$ & FeK$\rm{\alpha_{rev}}$ & \multicolumn{5}{c}{$\rm{L_{2-10 keV}}$}  \\
&  &  &  &  &  & $\rm{log(L_{Int})}$ & $\rm{log(L_{Pexrav})}$ & $\rm{log(L_{Gauss})}$ & $\rm{log(L_{Relxill})}$ & $\rm{log(L_{total})}$ \\ 
(1) & (2) & (3) & (4) & (5) & (6) & (7) & (8) & (9) & (10) & (11) \\\hline
 
%Mrk\,335 & Sy1.2 & 0.035 & 7.23$^a$ & -1.43 & 42.46 & 37.66 & 42.05 & 41.21 & 42.21 \\ $\rm{0.89\pm^{0.06}_{0.28}}$

Fairall\,9 & Sy1.2 & 0.047 & $\rm{8.299\pm^{0.078}_{0.116}}^a$ & $\rm{-1.36\pm^{0.09}_{0.13}}$ & - & $\rm{44.01\pm^{0.01}_{0.01}}$ & $\rm{42.16\pm^{0.09}_{0.10}}$ & $\rm{41.91\pm^{0.03}_{0.04}}$ & $\rm{42.77\pm^{0.04}_{0.05}}$ & $\rm{44.04\pm^{0.01}_{0.01}}$ \\ 

Mrk\,1040 & Sy1.5 & 0.011 & $\rm{7.77\pm^{0.14}_{0.14}}^c$ & $\rm{-2.18\pm^{0.15}_{0.15}}$ & Yes$^2$ & $\rm{42.80\pm^{0.01}_{0.01}}$ & $\rm{41.02\pm^{0.06}_{0.02}}$ & $\rm{40.80\pm^{0.03}_{0.04}}$ & $\rm{41.24\pm^{0.04}_{0.05}}$ & $\rm{42.82\pm^{0.01}_{0.01}}$ \\

%NGC\,1365 & Sy1.8 & 0.004 & 7.97$^c$ & -3.42 & 41.88 & 41.82 & -8.04 & 39.9 & 41.24 \\ 

Ark\,120 & Sy1 & 0.032 & $\rm{8.068\pm^{0.048}_{0.063}}^a$ & $\rm{-1.20\pm^{0.07}_{0.07}}$ & Yes$^3$ & $\rm{43.95\pm^{0.02}_{0.01}}$ & $\rm{41.71\pm^{0.02}_{0.02}}$ & $\rm{41.76\pm^{0.03}_{0.03}}$ & $\rm{43.31\pm^{0.01}_{0.01}}$ & $\rm{44.04\pm^{0.02}_{0.01}}$ \\ 

NGC\,3227 & Sy1.5 & 0.004 & $\rm{6.684\pm^{0.081}_{0.102}}^a$ & $\rm{-2.17\pm^{0.09}_{0.11}}$ & No$^1$ & $\rm{41.78\pm^{0.01}_{0.01}}$ & $\rm{40.20\pm^{0.03}_{0.04}}$ & $\rm{40.26\pm^{0.03}_{0.03}}$ & $\rm{42.10\pm^{0.05}_{0.07}}$ & $\rm{42.28\pm^{0.03}_{0.05}}$ \\ 

%NGC\,3783 & Sy1 & 0.011 & 7.08$^a$ & -1.59 & 42.83 & 42.72 & 40.27* & 41.25 & 42.48 \\ 

NGC\,4051 & Sy1.2 & 0.003 & $\rm{5.891\pm^{0.084}_{0.145}}^a$ & $\rm{-1.73\pm^{0.10}_{0.17}}$ & Yes$^1$ & $\rm{41.43\pm^{0.02}_{0.02}}$ & $\rm{40.03\pm^{0.02}_{0.02}}$ & $\rm{39.20\pm^{0.09}_{0.09}}$ & $\rm{41.12\pm^{0.03}_{0.03}}$ & $\rm{41.62\pm^{0.02}_{0.02}}$ \\ 

NGC\,4151 & Sy1.5 & 0.002 & $\rm{7.374\pm^{0.027}_{0.032}}^a$ & $\rm{-2.39\pm^{0.04}_{0.04}}$ & Yes$^1$ & $\rm{42.24\pm^{0.01}_{0.01}}$ & $\rm{40.10\pm^{0.01}_{0.01}}$ & $\rm{40.38\pm^{0.02}_{0.01}}$ & $\rm{41.96\pm^{0.02}_{0.01}}$ & $\rm{42.43\pm^{0.01}_{0.01}}$ \\

IRAS\,13197-1627 & Sy1.8 & 0.020 & $\rm{7.81\pm^{0.10}_{0.10}}^d$ & $\rm{-2.14\pm^{0.11}_{0.12}}$ & - & $\rm{42.89\pm^{0.01}_{0.02}}$ & $\rm{41.76\pm^{0.01}_{0.02}}$ & $\rm{41.18\pm^{0.01}_{0.03}}$ & $\rm{41.57\pm^{0.02}_{0.02}}$ & $\rm{42.95\pm^{0.01}_{0.02}}$ \\ 

NGC\,5548 & Sy1.5 & 0.025 & $\rm{7.692\pm^{0.016}_{0.016}}^a$ & $\rm{-1.24\pm^{0.03}_{0.03}}$ & Yes$^1$ & $\rm{43.59\pm^{0.01}_{0.01}}$ & $\rm{42.69\pm^{0.04}_{0.04}}$ & $\rm{41.56\pm^{0.04}_{0.03}}$ & $\rm{42.85\pm^{0.06}_{0.07}}$ & $\rm{43.71\pm^{0.01}_{0.01}}$ \\

NGC\,7469 & Sy1.2 & 0.014 & $\rm{6.956\pm^{0.048}_{0.050}}^a$ & $\rm{-1.17\pm^{0.06}_{0.06}}$ & Yes$^1$ & $\rm{43.00\pm^{0.01}_{0.01}}$ & $\rm{41.76\pm^{0.03}_{0.03}}$ & $\rm{41.02\pm^{0.03}_{0.04}}$ & $\rm{41.99\pm^{0.04}_{0.05}}$ & $\rm{43.07\pm^{0.01}_{0.01}}$ \\

\hline

Mrk\,915 & Sy1 & 0.024 & $\rm{7.76\pm^{0.37}_{0.37}}^e$ & $\rm{-1.89\pm^{0.38}_{0.38}}$ & - & $\rm{43.07\pm^{0.01}_{0.01}}$ & $\rm{41.53\pm^{0.08}_{0.11}}$ & $\rm{41.18\pm^{0.03}_{0.04}}$ & - & $\rm{43.09\pm^{0.01}_{0.01}}$ \\ 

Mrk\,766 & Sy1.5 & 0.013 & $\rm{6.822\pm^{0.050}_{0.057}}^a$ & $\rm{-1.30\pm^{0.06}_{0.07}}$ & No$^1$ & $\rm{42.75\pm^{0.01}_{0.01}}$ & $\rm{41.62\pm^{0.03}_{0.03}}$ & $\rm{40.34\pm^{0.13}_{0.17}}$ & - & $\rm{42.78\pm^{0.01}_{0.01}}$ \\

NGC\,4593 & Sy1 & 0.008 & $\rm{6.912\pm^{0.069}_{0.068}}^a$ & $\rm{-1.66\pm^{0.08}_{0.08}}$ & No$^1$ & $\rm{42.50\pm^{0.01}_{0.01}}$ & $\rm{40.93\pm^{0.05}_{0.06}}$ & $\rm{40.63\pm^{0.04}_{0.04}}$ & - & $\rm{42.52\pm^{0.01}_{0.01}}$ \\ 

MCG\,-06-30-15 & Sy1.2 & 0.008 & $\rm{6.295\pm^{0.157}_{0.237}}^a$ & $\rm{-0.98\pm^{0.17}_{0.25}}$ & No$^1$ & $\rm{42.56\pm^{0.01}_{0.01}}$ & $\rm{41.53\pm^{0.01}_{0.01}}$ & $\rm{40.52\pm^{0.04}_{0.04}}$ & - & $\rm{42.60\pm^{0.01}_{0.01}}$ \\

\hline

Mrk\,1044 & Sy1 & 0.016 & $\rm{6.23\pm^{0.50}_{0.50}}^b$ & $\rm{-0.74\pm^{0.52}_{0.51}}$ & Yes$^4$ & $\rm{42.72\pm^{0.02}_{0.01}}$ & - & - & $\rm{42.71\pm^{0.01}_{0.02}}$ & $\rm{43.02\pm^{0.01}_{0.01}}$ \\ 

IGRJ\,19378-0617 & Sy1 & 0.010 & $\rm{6.8\pm^{0.40}_{0.40}}^g$ & $\rm{-1.44\pm^{0.41}_{0.41}}$ & - & $\rm{42.60\pm^{0.01}_{0.01}}$ & - & - & $\rm{41.90\pm^{0.02}_{0.02}}$ & $\rm{42.68\pm^{0.01}_{0.01}}$ \\

Mrk\,766 & Sy1.5 & 0.013 & $\rm{6.822\pm^{0.050}_{0.057}}^a$ & $\rm{-1.34\pm^{0.06}_{0.07}}$ & No$^1$ & $\rm{42.71\pm^{0.01}_{0.01}}$ & - & - & $\rm{41.99\pm^{0.01}_{0.02}}$ & $\rm{42.79\pm^{0.01}_{0.01}}$ \\

NGC\,4593 & Sy1 & 0.008 & $\rm{6.912\pm^{0.069}_{0.068}}^a$ & $\rm{-1.67\pm^{0.08}_{0.08}}$ & No$^1$ & $\rm{42.49\pm^{0.01}_{0.01}}$ & - & - & $\rm{41.21\pm^{0.03}_{0.03}}$ & $\rm{42.51\pm^{0.01}_{0.01}}$ \\ 

MCG\,-06-30-15 & Sy1.2 & 0.008 & $\rm{6.295\pm^{0.157}_{0.237}}^a$ & $\rm{-1.05\pm^{0.17}_{0.25}}$ & No$^1$ & $\rm{42.50\pm^{0.01}_{0.01}}$ & - & - & $\rm{41.76\pm^{0.01}_{0.01}}$ & $\rm{42.57\pm^{0.01}_{0.01}}$ \\

\hline 

%RXJ & NLSy1 & 0.237 & ? &  &  &  & - & - & - \\

Mrk\,382 & Sy1 & 0.027 & $\rm{6.61\pm^{0.50}_{0.50}}^b$ & $\rm{-1.28\pm^{0.51}_{0.51}}$ & - & $\rm{42.57\pm^{0.01}_{0.01}}$ & - & - & - & $\rm{42.57\pm^{0.01}_{0.01}}$ \\

IRAS\,13224-3809 & NLSy1 & 0.066 & $\rm{6.8\pm^{0.50}_{0.50}}^f$ & $\rm{-1.85\pm^{0.51}_{0.51}}$ & Yes$^1$ & $\rm{42.21\pm^{0.01}_{0.01}}$ & - & - & - & $\rm{42.21\pm^{0.01}_{0.01}}$ \\

Mrk\,841 & Sy1.5 & 0.036 & $\rm{8.17\pm^{0.10}_{0.10}}^d$ & $\rm{-1.42\pm^{0.11}_{0.11}}$ & No$^1$ & $\rm{43.85\pm^{0.01}_{0.01}}$ & - & - & - & $\rm{43.85\pm^{0.01}_{0.01}}$ \\

MR\,2251-178 & Sy1 & 0.064 & $\rm{8.71\pm^{0.11}_{0.11}}^c$ & $\rm{-0.97\pm^{0.12}_{0.12}}$ & - & $\rm{44.68\pm^{0.01}_{0.01}}$ & - & - & - & $\rm{44.68\pm^{0.01}_{0.01}}$ \\

\hline
\end{tabular}
\end{center}
\caption{Parameters of the sample. (1) Name of the source; (2) AGN classification; (3) Redshift; (4) log ($M_{BH}/M_{\odot}$); (5) Eddington ratio; (6) FeK$\rm{\alpha}$ reverberation
signatures; (7)-(11) 2-10 keV band, intrinsic, {\sc pexrav}, {\sc gauss}, and {\sc relxill} luminosities (in erg/s) . References: a)\cite{Bentz15}, b)\cite{Wang01}, c)\cite{Khorunzhev12} , d) \cite{Vasudevan10}, e) \cite{Hinkle21}, f) \cite{Waddell20}, g) \cite{Chang21}, 1) \cite{Kara16}, 2) \cite{Tripathi11}, 3) \cite{Lobban18}, 4) \cite{Mallick18}.}
\label{tab:sample_Lums}
\end{table*}

%Tripathi11,Kara16,Lobban18,Mallick18


\begin{thebibliography}{}

%\bibitem[Ananna et al.(2022)]{Ananna22} Ananna, T.~T., Urry, C.~M., Ricci, C., et al.\ 2022, \apjl, 939, L13. doi:10.3847/2041-8213/ac9979

%\bibitem[Ananna et al.(2022)]{Ananna22b} Ananna, T.~T., Weigel, A.~K., Trakhtenbrot, B., et al.\ 2022, \apjs, 261, 9. doi:10.3847/1538-4365/ac5b64

%\bibitem[Antonucci(1993)]{Antonucci93} Antonucci, R.\ 1993, \araa, 31, 473 

\bibitem[Ballo et al.(2017)]{Ballo17} Ballo, L., Severgnini, P., Della Ceca, R., et al.\ 2017, \mnras, 470, 3924. doi:10.1093/mnras/stx1360

\bibitem[Balokovi{\'c} et al.(2018)]{Balokovic18} Balokovi{\'c}, M., Brightman, M., Harrison, F.~A., et al.\ 2018, \apj, 854, 42

\bibitem[Bauer et al.(2015)]{Bauer15} Bauer, F.~E., Ar{\'e}valo, P., Walton, D.~J., et al.\ 2015, \apj, 812, 116. doi:10.1088/0004-637X/812/2/116

\bibitem[Bentz \& Katz(2015)]{Bentz15} Bentz, M.~C. \& Katz, S.\ 2015, \pasp, 127, 67. doi:10.1086/679601

\bibitem[Bianchi et al.(2004)]{Bianchi04} Bianchi, S., Matt, G., Balestra, I., et al.\ 2004, \aap, 422, 65. doi:10.1051/0004-6361:20047128

%\bibitem[Blandford \& K{\"o}nigl(1979)]{Blandford19} Blandford, R.~D. \& K{\"o}nigl, A.\ 1979, \apj, 232, 34. doi:10.1086/157262

\bibitem[Blustin et al.(2002)]{Blustin02} Blustin, A.~J., Branduardi-Raymont, G., Behar, E., et al.\ 2002, \aap, 392, 453. doi:10.1051/0004-6361:20020914

%\bibitem[Brenneman \& Reynolds(2006)]{Brenneman06} Brenneman, L.~W. \& Reynolds, C.~S.\ 2006, \apj, 652, 1028. doi:10.1086/508146

%\bibitem[Brenneman et al.(2007)]{Brenneman07} Brenneman, L.~W., Reynolds, C.~S., Wilms, J., et al.\ 2007, \apj, 666, 817. doi:10.1086/520763

\bibitem[Brenneman et al.(2011)]{Brenneman11} Brenneman, L.~W., Reynolds, C.~S., Nowak, M.~A., et al.\ 2011, \apj, 736, 103

\bibitem[Brenneman et al.(2012)]{Brenneman12} Brenneman, L.~W., Elvis, M., Krongold, Y., et al.\ 2012, \apj, 744, 13. doi:10.1088/0004-637X/744/1/13

\bibitem[Brightman \& Nandra(2011)]{Brightman11} Brightman, M. \& Nandra, K.\ 2011, \mnras, 413, 1206. doi:10.1111/j.1365-2966.2011.18207.x

\bibitem[Brightman \& Ueda(2012)]{Brightman12} Brightman, M. \& Ueda, Y.\ 2012, \mnras, 423, 702. doi:10.1111/j.1365-2966.2012.20908.x

%\bibitem[Brightman et al.(2013)]{Brightman13} Brightman, M., Silverman, J.~D., Mainieri, V., et al.\ 2013, \mnras, 433, 2485. doi:10.1093/mnras/stt920

\bibitem[Brightman et al.(2015)]{Brightman15} Brightman, M., Balokovi{\'c}, M., Stern, D., et al.\ 2015, \apj, 805, 41. doi:10.1088/0004-637X/805/1/41

\bibitem[Buisson et al.(2018)]{Buisson18} Buisson, D.~J.~K., Parker, M.~L., Kara, E., et al.\ 2018, \mnras, 480, 3689. doi:10.1093/mnras/sty2081

%\bibitem[Chainakun et al.(2022)]{Chainakun22} Chainakun, P., Fongkaew, I., Hancock, S., et al.\ 2022, \mnras, 513, 648. doi:10.1093/mnras/stac924

\bibitem[Chang et al.(2021)]{Chang21} Chang, N., Xie, F.~G., Liu, X., et al.\ 2021, \mnras, 503, 1987. doi:10.1093/mnras/stab521

\bibitem[Chiang \& Fabian(2011)]{Chiang11} Chiang, C.-Y. \& Fabian, A.~C.\ 2011, \mnras, 414, 2345. doi:10.1111/j.1365-2966.2011.18553.x

%\bibitem[Dadina \& Cappi(2004)]{Dadina04} Dadina, M. \& Cappi, M.\ 2004, \aap, 413, 921. doi:10.1051/0004-6361:20031561

\bibitem[Dauser et al.(2010)]{Dauser10} Dauser, T., Wilms, J., Reynolds, C.~S., et al.\ 2010, \mnras, 409, 1534

\bibitem[Dehghanian et al.(2020)]{Dehghanian20} Dehghanian, M., Ferland, G.~J., Kriss, G.~A., et al.\ 2020, \apj, 898, 141. doi:10.3847/1538-4357/ab9cb2

\bibitem[Dewangan et al.(2007)]{Dewangan07} Dewangan, G.~C., Griffiths, R.~E., Dasgupta, S., et al.\ 2007, \apj, 671, 1284. doi:10.1086/523683

\bibitem[Diaz et al.(2020)]{Diaz20} Diaz, Y., Ar{\'e}valo, P., Hern{\'a}ndez-Garc{\'\i}a, L., et al.\ 2020, \mnras, 496, 5399. doi:10.1093/mnras/staa1762

\bibitem[Emmanoulopoulos et al.(2011)]{Emmanoulopoulos11} Emmanoulopoulos, D., Papadakis, I.~E., McHardy, I.~M., et al.\ 2011, \mnras, 415, 1895. doi:10.1111/j.1365-2966.2011.18834.x

\bibitem[Emmanoulopoulos et al.(2016)]{Emmanoulopoulos16} Emmanoulopoulos, D., Papadakis, I.~E., Epitropakis, A., et al.\ 2016, \mnras, 461, 1642

\bibitem[Fabian et al.(1989)]{Fabian89} Fabian, A.~C., Rees, M.~J., Stella, L., et al.\ 1989, \mnras, 238, 729

\bibitem[Fabian \& Vaughan(2003)]{Fabian03} Fabian, A.~C. \& Vaughan, S.\ 2003, \mnras, 340, L28. doi:10.1046/j.1365-8711.2003.06465.x

\bibitem[Falocco et al.(2014)]{Falocco14} Falocco, S., Carrera, F.~J., Barcons, X., et al.\ 2014, \aap, 568, A15. doi:10.1051/0004-6361/201322812

\bibitem[Fanali et al.(2013)]{Fanali13} Fanali, R., Caccianiga, A., Severgnini, P., et al.\ 2013, \mnras, 433, 648. doi:10.1093/mnras/stt757

\bibitem[Gallo et al.(2004)]{Gallo04} Gallo, L.~C., Boller, T., Tanaka, Y., et al.\ 2004, \mnras, 347, 269. doi:10.1111/j.1365-2966.2004.07196.x

\bibitem[Garc{\'\i}a et al.(2014)]{Garcia14} Garc{\'\i}a, J., Dauser, T., Lohfink, A., et al.\ 2014, \apj, 782, 76

\bibitem[George \& Fabian(1991)]{George91} George, I.~M. \& Fabian, A.~C.\ 1991, \mnras, 249, 352. doi:10.1093/mnras/249.2.352

%\bibitem[Gierli{\'n}ski \& Done(2004)]{Gierlinski04} Gierli{\'n}ski, M., \& Done, C.\ 2004, \mnras, 349, L7

\bibitem[Gofford et al.(2011)]{Gofford11} Gofford, J., Reeves, J.~N., Turner, T.~J., et al.\ 2011, \mnras, 414, 3307. doi:10.1111/j.1365-2966.2011.18634.x

\bibitem[Gondoin et al.(2001)]{Gondoin01} Gondoin, P., Lumb, D., Siddiqui, H., et al.\ 2001, \aap, 373, 805. doi:10.1051/0004-6361:20010582

%\bibitem[Goosmann et al.(2016)]{Goosmann16} Goosmann, R.~W., Holczer, T., Mouchet, M., et al.\ 2016, \aap, 589, A76

%\bibitem[Haardt \& Maraschi(1991)]{Haardt91} Haardt, F. \& Maraschi, L.\ 1991, \apjl, 380, L51. doi:10.1086/186171

\bibitem[Haardt \& Maraschi(1993)]{Haardt93} Haardt, F. \& Maraschi, L.\ 1993, \apj, 413, 507. doi:10.1086/173020

\bibitem[Harrison et al.(2013)]{Harrison13} Harrison, F.~A., Craig, W.~W., Christensen, F.~E., et al.\ 2013, \apj, 770, 103

\bibitem[Hinkle \& Mushotzky(2021)]{Hinkle21} Hinkle, J.~T. \& Mushotzky, R.\ 2021, \mnras, 506, 4960. doi:10.1093/mnras/stab1976

\bibitem[Inaba et al.(2022)]{Inaba22} Inaba, K., Ueda, Y., Yamada, S., et al.\ 2022, \apj, 939, 88. doi:10.3847/1538-4357/ac97ec

\bibitem[Jansen et al.(2001)]{Jansen01} Jansen, F., Lumb, D., Altieri, B., et al.\ 2001, \aap, 365, L1

\bibitem[Jiang et al.(2018)]{Jiang18} Jiang, J., Parker, M.~L., Fabian, A.~C., et al.\ 2018, \mnras, 477, 3711. doi:10.1093/mnras/sty836

\bibitem[Kalberla et al.(2005)]{Kalberla05} Kalberla, P.~M.~W., Burton, W.~B., Hartmann, D., et al.\ 2005, \aap, 440, 775. doi:10.1051/0004-6361:20041864

%\bibitem[Kang \& Wang(2022)]{Kang22} Kang, J.-L. \& Wang, J.-X.\ 2022, \apj, 929, 141. doi:10.3847/1538-4357/ac5d49

\bibitem[Kara et al.(2016)]{Kara16} Kara, E., Alston, W.~N., Fabian, A.~C., et al.\ 2016, \mnras, 462, 511. doi:10.1093/mnras/stw1695

\bibitem[Khorunzhev et al.(2012)]{Khorunzhev12} Khorunzhev, G.~A., Sazonov, S.~Y., Burenin, R.~A., et al.\ 2012, Astronomy Letters, 38, 475. doi:10.1134/S1063773712080026

%\bibitem[Komossa \& Meerschweinchen(2000)]{Komossa00} Komossa, S. \& Meerschweinchen, J.\ 2000, \aap, 354, 411

\bibitem[Krongold et al.(2003)]{Krongold03} Krongold, Y., Nicastro, F., Brickhouse, N.~S., et al.\ 2003, \apj, 597, 832. doi:10.1086/378639

\bibitem[Krongold et al.(2005)]{Krongold05} Krongold, Y., Nicastro, F., Brickhouse, N.~S., et al.\ 2005, \apj, 622, 842. doi:10.1086/427621

\bibitem[Krongold et al.(2021)]{Krongold21} Krongold, Y., Longinotti, A.~L., Santos-Lle{\'o}, M., et al.\ 2021, \apj, 917, 39. doi:10.3847/1538-4357/ac0977

\bibitem[Laha \& Ghosh(2021)]{Laha21} Laha, S. \& Ghosh, R.\ 2021, \apj, 915, 93. doi:10.3847/1538-4357/abfc56

\bibitem[Laor(1991)]{Laor91} Laor, A.\ 1991, \apj, 376, 90. doi:10.1086/170257

%\bibitem[Liu(2016)]{Liu16} Liu, J.\ 2016, \mnras, 463, L108. doi:10.1093/mnrasl/slw164

\bibitem[Liu et al.(2020)]{Liu20} Liu, H., Wang, H., Abdikamalov, A.~B., et al.\ 2020, \apj, 896, 160. doi:10.3847/1538-4357/ab917a

\bibitem[Lobban et al.(2018)]{Lobban18} Lobban, A.~P., Porquet, D., Reeves, J.~N., et al.\ 2018, \mnras, 474, 3237. doi:10.1093/mnras/stx2889

\bibitem[Lohfink et al.(2016)]{Lohfink16} Lohfink, A.~M., Reynolds, C.~S., Pinto, C., et al.\ 2016, \apj, 821, 11. doi:10.3847/0004-637X/821/1/11

\bibitem[Longinotti et al.(2004)]{Longinotti04} Longinotti, A.~L., Nandra, K., Petrucci, P.~O., et al.\ 2004, \mnras, 355, 929. doi:10.1111/j.1365-2966.2004.08369.x

\bibitem[Longinotti et al.(2007)]{Longinotti07} Longinotti, A.~L., Sim, S.~A., Nandra, K., et al.\ 2007, \mnras, 374, 237. doi:10.1111/j.1365-2966.2006.11138.x

\bibitem[Magdziarz \& Zdziarski(1995)]{Magdziarz95} Magdziarz, P., \& Zdziarski, A.~A.\ 1995, \mnras, 273, 837

%\bibitem[Malizia et al.(2014)]{Malizia14} Malizia, A., Molina, M., Bassani, L., et al.\ 2014, \apjl, 782, L25. doi:10.1088/2041-8205/782/2/L25

\bibitem[Mallick et al.(2018)]{Mallick18} Mallick, L., Alston, W.~N., Parker, M.~L., et al.\ 2018, \mnras, 479, 615. doi:10.1093/mnras/sty1487

\bibitem[Mao et al.(2019)]{Mao19} Mao, J., Mehdipour, M., Kaastra, J.~S., et al.\ 2019, \aap, 621, A99

\bibitem[Marchesi et al.(2018)]{Marchesi18} Marchesi, S., Ajello, M., Marcotulli, L., et al.\ 2018, \apj, 854, 49. doi:10.3847/1538-4357/aaa410

\bibitem[Marchesi et al.(2022)]{Marchesi22} Marchesi, S., Zhao, X., Torres-Alb{\`a}, N., et al.\ 2022, \apj, 935, 114. doi:10.3847/1538-4357/ac80be

\bibitem[Marconi et al.(2004)]{Marconi04} Marconi, A., Risaliti, G., Gilli, R., et al.\ 2004, \mnras, 351, 169. doi:10.1111/j.1365-2966.2004.07765.x

\bibitem[Marinucci et al.(2014)]{Marinucci14} Marinucci, A., Matt, G., Miniutti, G., et al.\ 2014, \apj, 787, 83. doi:10.1088/0004-637X/787/1/83

\bibitem[Matsuoka et al.(1990)]{Matsuoka90} Matsuoka, M., Piro, L., Yamauchi, M., et al.\ 1990, \apj, 361, 440. doi:10.1086/169209

\bibitem[Mehdipour et al.(2015)]{Mehdipour15} Mehdipour, M., Kaastra, J.~S., Kriss, G.~A., et al.\ 2015, \aap, 575, A22. doi:10.1051/0004-6361/201425373

\bibitem[Mehdipour et al.(2017)]{Mehdipour17} Mehdipour, M., Kaastra, J.~S., Kriss, G.~A., et al.\ 2017, \aap, 607, A28. doi:10.1051/0004-6361/201731175

\bibitem[Meyer-Hofmeister et al.(2009)]{Meyer09} Meyer-Hofmeister, E., Liu, B.~F., \& Meyer, F.\ 2009, \aap, 508, 329. doi:10.1051/0004-6361/200913044

%\bibitem[Middei et al.(2018)]{Middei18} Middei, R., Bianchi, S., Cappi, M., et al.\ 2018, \aap, 615, A163

\bibitem[Miller et al.(2008)]{Miller08} Miller, L., Turner, T.~J., \& Reeves, J.~N.\ 2008, \aap, 483, 437. doi:10.1051/0004-6361:200809590

\bibitem[Miller et al.(2009)]{Miller09} Miller, L., Turner, T.~J., \& Reeves, J.~N.\ 2009, \mnras, 399, L69. doi:10.1111/j.1745-3933.2009.00726.x

%\bibitem[Nandra(2006)]{Nandra06} Nandra, K.\ 2006, \mnras, 368, L62. doi:10.1111/j.1745-3933.2006.00158.x

\bibitem[Nandra et al.(2007)]{Nandra07} Nandra, K., O'Neill, P.~M., George, I.~M., et al.\ 2007, \mnras, 382, 194

\bibitem[Nardini et al.(2011)]{Nardini11} Nardini, E., Fabian, A.~C., Reis, R.~C., et al.\ 2011, \mnras, 410, 1251. doi:10.1111/j.1365-2966.2010.17518.x

\bibitem[Nardini et al.(2014)]{Nardini14} Nardini, E., Reeves, J.~N., Porquet, D., et al.\ 2014, \mnras, 440, 1200. doi:10.1093/mnras/stu333

%\bibitem[Nishimura et al.(1986)]{Nishimura86} Nishimura, J., Mitsuda, K., \& Itoh, M.\ 1986, \pasj, 38, 819

%\bibitem[Noda \& Done(2018)]{NodaDone18} Noda, H. \& Done, C.\ 2018, \mnras, 480, 3898. doi:10.1093/mnras/sty2032

\bibitem[O'Neill et al.(2007)]{ONeill07} O'Neill, P.~M., Nandra, K., Cappi, M., et al.\ 2007, \mnras, 381, L94. doi:10.1111/j.1745-3933.2007.00376.x

%\bibitem[Osorio-Clavijo et al.(2020)]{Osorio-Clavijo20} Osorio-Clavijo, N., Gonz{\'a}lez-Mart{\'\i}n, O., Papadakis, I.~E., et al.\ 2020, \mnras, 491, 29

\bibitem[Osorio-Clavijo et al.(2022)]{Osorio-Clavijo22} Osorio-Clavijo, N., Gonz{\'a}lez-Mart{\'\i}n, O., S{\'a}nchez, S.~F., et al.\ 2022, \mnras, 510, 5102. doi:10.1093/mnras/stab3752

%\bibitem[Padovani et al.(2017)]{Padovani17} Padovani, P., Alexander, D.~M., Assef, R.~J., et al.\ 2017, \aapr, 25, 2. doi:10.1007/s00159-017-0102-9

\bibitem[Panagiotou \& Walter(2019)]{Panagiotou19} Panagiotou, C. \& Walter, R.\ 2019, \aap, 626, A40. doi:10.1051/0004-6361/201935052

\bibitem[Parker et al.(2019)]{Parker19} Parker, M.~L., Longinotti, A.~L., Schartel, N., et al.\ 2019, \mnras, 490, 683. doi:10.1093/mnras/stz2566

\bibitem[Patrick et al.(2011)]{Patrick11} Patrick, A.~R., Reeves, J.~N., Porquet, D., et al.\ 2011, \mnras, 411, 2353. doi:10.1111/j.1365-2966.2010.17852.x

\bibitem[Petrucci et al.(2002)]{Petrucci02} Petrucci, P.~O., Henri, G., Maraschi, L., et al.\ 2002, \aap, 388, L5. doi:10.1051/0004-6361:20020534

%\bibitem[Piro et al.(1990)]{Piro90} Piro, L., Yamauchi, M., \& Matsuoka, M.\ 1990, \apjl, 360, L35. doi:10.1086/185806

\bibitem[Porquet et al.(2018)]{Porquet18} Porquet, D., Reeves, J.~N., Matt, G., et al.\ 2018, \aap, 609, A42. doi:10.1051/0004-6361/201731290

\bibitem[Pounds et al.(1990)]{Pounds90} Pounds, K.~A., Nandra, K., Stewart, G.~C., et al.\ 1990, \nat, 344, 132. doi:10.1038/344132a0

%\bibitem[Reeves et al.(2008)]{Reeves08} Reeves, J., Done, C., Pounds, K., et al.\ 2008, \mnras, 385, L108

\bibitem[Reis et al.(2012)]{Reis12} Reis, R.~C., Fabian, A.~C., Reynolds, C.~S., et al.\ 2012, \apj, 745, 93. doi:10.1088/0004-637X/745/1/93

\bibitem[Reynolds et al.(2009)]{Reynolds09} Reynolds, C.~S., Fabian, A.~C., Brenneman, L.~W., et al.\ 2009, \mnras, 397, L21. doi:10.1111/j.1745-3933.2009.00676.x

\bibitem[Ricci et al.(2011)]{Ricci11} Ricci, C., Walter, R., Courvoisier, T.~J.-L., et al.\ 2011, \aap, 532, A102. doi:10.1051/0004-6361/201016409

\bibitem[Ricci et al.(2015)]{Ricci15} Ricci, C., Ueda, Y., Koss, M.~J., et al.\ 2015, \apjl, 815, L13. doi:10.1088/2041-8205/815/1/L13

\bibitem[Risaliti et al.(2005)]{Risalitti05} Risaliti, G., Elvis, M., Fabbiano, G., et al.\ 2005, \apjl, 623, L93. doi:10.1086/430252

\bibitem[Rivers et al.(2015)]{Rivers15} Rivers, E., Risaliti, G., Walton, D.~J., et al.\ 2015, \apj, 804, 107. doi:10.1088/0004-637X/804/2/107

\bibitem[Ross et al.(1999)]{Ross99} Ross, R.~R., Fabian, A.~C., \& Young, A.~J.\ 1999, \mnras, 306, 461

\bibitem[Ross \& Fabian(2005)]{Ross05} Ross, R.~R., \& Fabian, A.~C.\ 2005, \mnras, 358, 211

%\bibitem[Ruan et al.(2019)]{Ruan19} Ruan, J.~J., Anderson, S.~F., Eracleous, M., et al.\ 2019, \apj, 883, 76. doi:10.3847/1538-4357/ab3c1a

%\bibitem[Saez et al.(2008)]{Saez08} Saez, C., Chartas, G., Brandt, W.~N., et al.\ 2008, \aj, 135, 1505. doi:10.1088/0004-6256/135/4/1505

%\bibitem[Scott et al.(2011)]{Scott11} Scott, A.~E., Stewart, G.~C., Mateos, S., et al.\ 2011, \mnras, 417, 992. doi:10.1111/j.1365-2966.2011.19325.x

\bibitem[Singh et al.(1992)]{Singh92} Singh, K.~P., Rao, A.~R., \& Vahia, M.~N.\ 1992, \aap, 262, 49

\bibitem[Str{\"u}der et al.(2001)]{Struder01} Str{\"u}der, L., Briel, U., Dennerl, K., et al.\ 2001, \aap, 365, L18. doi:10.1051/0004-6361:20000066

%\bibitem[Tortosa et al.(2018)]{Tortosa18} Tortosa, A., Bianchi, S., Marinucci, A., et al.\ 2018, \aap, 614, A37. doi:10.1051/0004-6361/201732382

\bibitem[Traina et al.(2021)]{Traina21} Traina, A., Marchesi, S., Vignali, C., et al.\ 2021, \apj, 922, 159. doi:10.3847/1538-4357/ac1fee

\bibitem[Tripathi et al.(2011)]{Tripathi11} Tripathi, S., Misra, R., Dewangan, G., et al.\ 2011, \apjl, 736, L37. doi:10.1088/2041-8205/736/2/L37

\bibitem[Trump et al.(2011)]{Trump11} Trump, J.~R., Impey, C.~D., Kelly, B.~C., et al.\ 2011, \apj, 733, 60. doi:10.1088/0004-637X/733/1/60

%\bibitem[Urry \& Padovani(1995)]{Urry-Padovani95} Urry, C.~M. \& Padovani, P.\ 1995, \pasp, 107, 803. doi:10.1086/133630

\bibitem[Ursini et al.(2016)]{Ursini16} Ursini, F., Petrucci, P.-O., Matt, G., et al.\ 2016, \mnras, 463, 382. doi:10.1093/mnras/stw2022

\bibitem[Uttley et al.(2014)]{Uttley14} Uttley, P., Cackett, E.~M., Fabian, A.~C., et al.\ 2014, \aapr, 22, 72. doi:10.1007/s00159-014-0072-0

\bibitem[Vasudevan et al.(2010)]{Vasudevan10} Vasudevan, R.~V., Fabian, A.~C., Gandhi, P., et al.\ 2010, \mnras, 402, 1081. doi:10.1111/j.1365-2966.2009.15936.x

\bibitem[Waddell \& Gallo(2020)]{Waddell20} Waddell, S.~G.~H. \& Gallo, L.~C.\ 2020, \mnras, 498, 5207. doi:10.1093/mnras/staa2783

%\bibitem[Walton et al.(2014)]{Walton14} Walton, D.~J., Risaliti, G., Harrison, F.~A., et al.\ 2014, \apj, 788, 76. doi:10.1088/0004-637X/788/1/76

\bibitem[Walton et al.(2018)]{Walton18} Walton, D.~J., Brightman, M., Risaliti, G., et al.\ 2018, \mnras, 473, 4377. doi:10.1093/mnras/stx2659

\bibitem[Wang \& Lu(2001)]{Wang01} Wang, T. \& Lu, Y.\ 2001, \aap, 377, 52. doi:10.1051/0004-6361:20011071

%\bibitem[Williams et al.(2022)]{Williams22} Williams, D.~R.~A., Pahari, M., Baldi, R.~D., et al.\ 2022, \mnras, 510, 4909. doi:10.1093/mnras/stab3310

\bibitem[Zdziarski et al.(1995)]{Zdziarski95} Zdziarski, A.~A., Johnson, W.~N., Done, C., et al.\ 1995, \apjl, 438, L63. doi:10.1086/187716

%\bibitem[Zdziarski et al.(1996)]{Zdziarski96} Zdziarski, A.~A., Johnson, W.~N., \& Magdziarz, P.\ 1996, \mnras, 283, 193

%\bibitem[Zhang et al.(2018)]{Zhang18} Zhang, J.-X., Wang, J.-X., \& Zhu, F.-F.\ 2018, \apj, 863, 71. doi:10.3847/1538-4357/aacf92

%\bibitem[{\.Z}ycki et al.(1999)]{Zycki99} {\.Z}ycki, P.~T., Done, C., \& Smith, D.~A.\ 1999, \mnras, 309, 561



\end{thebibliography}
\end{document}